\documentclass{aa}
\usepackage{graphicx}
\usepackage{txfonts}
\usepackage{hyperref}
\usepackage[normalem]{ulem}
\hypersetup{colorlinks = true, citecolor = {blue}, urlcolor= {blue}}
\def\gapprox{\;\rlap{\lower 3.0pt                       % approximately smaller
        \hbox{$\sim$}}\raise 2.5pt\hbox{$>$}\;}
\def\lapprox{\;\rlap{\lower 3.1pt                       % approximately smaller
        \hbox{$\sim$}}\raise 2.7pt\hbox{$<$}\;}

\newcommand{\be}{ \begin{equation} }
\newcommand{\ee}{\end{equation}}

\newcommand{\ben}{\begin{enumerate}}
\newcommand{\een}{\end{enumerate}}

\renewenvironment{thebibliography}[1]{\begin{oldthebibliography}{#1}\setlength{\parskip}{0ex}\setlength{\itemsep}{0ex}}{\end{oldthebibliography}}

\usepackage{diagbox}
\usepackage{siunitx}
\newcommand{\orcid}[1]{\href{https://orcid.org/#1}{\protect\includegraphics[width=8pt]{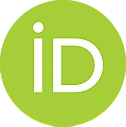}}}
\makeatletter
\renewcommand*\aa@pageof{, page \thepage{} of \pageref*{LastPage}}
\makeatother

\usepackage[dvipsnames]{xcolor}

\definecolor{darkgreen}{RGB}{31, 207, 31}

\begin{document}
   
\title{Milky Way globular clusters on cosmological timescales. II. Interaction with the Galactic centre}

\author{Maryna~Ishchenko
\inst{1,2}\orcid{0000-0002-6961-8170}
\and
Margaryta~Sobolenko
\inst{1}\orcid{0000-0003-0553-7301}
\and
Dana~Kuvatova
\inst{2}\orcid{0000-0002-5937-4985}
\and
Taras~Panamarev
\inst{3,2}\orcid{0000-0002-1090-4463}
\and
Peter~Berczik
\inst{4,5,2,1}\orcid{0000-0003-4176-152X}
}

\institute{Main Astronomical Observatory, National 
           Academy of Sciences of Ukraine,
           27 Akademika Zabolotnoho St, 03143 Kyiv, Ukraine  \email{\href{mailto:marina@mao.kiev.ua}{marina@mao.kiev.ua}}
           \and
           Fesenkov Astrophysical Institute, Observatory 23, 050020 Almaty, Kazakhstan
           \and
           Rudolf Peierls Centre for Theoretical Physics, Parks Road, OX1 3PU, Oxford, UK
           \and
           Astronomisches Rechen-Institut, Zentrum f\"ur Astronomie, University of Heidelberg, M\"onchhofstrasse 12-14, 69120 Heidelberg, Germany
           \and
           Konkoly Observatory, Research Centre for Astronomy and Earth Sciences, E\"otv\"os Lor\'and Research Network (ELKH), MTA Centre of Excellence, Konkoly Thege Mikl\'os \'ut 15-17, 1121 Budapest, Hungary
           }
   
\date{Received xxx / Accepted xxx}

% \abstract{}{}{}{}{}
% 5 {} token are mandatory
\abstract
{}
{ We estimate the dynamical evolution of the Globular Clusters interaction with the Galactic centre that dynamically changed in the past.}
{We simulated the orbits of 147 globular clusters over 10~Gyr lookback time using the parallel $N$-body code "$\varphi$-GPU". For each globular cluster, we generated 1000 sets of initial data with random proper motions and radial velocities based on the observed values. 
To distinguish globular clusters interacting with the galactic centre, we used the criterion of a relative distance of less than 100~pc. We used four external potentials from the IllustrisTNG-100 database, which were selected for their similarity to the present-day Milky Way, to simulate the structure of the Galaxy at different times.}
{We obtained $\sim$3--4 globular cluster interactions per Gyr at distances of less than 50 pc and $\sim$5--6 interactions per Gyr at distances of less than 80 pc among the studied 147 globular clusters that had close passages near the Galactic centre. We selected 10 of them for detailed study and found almost 100\% probability of interaction with the Galactic centre for six of them.}
{According to our results, the maximum interaction frequency of globular clusters with the Galactic centre in the Milky Way is likely to be a few dozens of passages per Gyr within a central zone of 100~pc. This low frequency may not be sufficient to fully explain the relatively high mass (of order 10$^7$ M$_\odot$) of the nuclear star cluster in the Milky Way, if we consider only the periodic capture of stars from globular clusters during close encounters. Therefore, we must also consider the possibility that some early globular clusters were completely tidally disrupted during interactions with the forming nuclear star cluster and the Galactic centre.}

\keywords{Galaxy: globular clusters: general - Galaxy: center - Methods: numerical}

\titlerunning{Part II. Interaction with the Galactic centre}
\authorrunning{M.~Ishchenko et al.}
\maketitle

%%%%%%%%%%%%%%%%%%%%%%%%%%%%%%%%%%%%%%%%%%%%%%%%%%%%%%%%%%%%%%%%%%%%
\section{Introduction}\label{sec:Intr}
%%%%%%%%%%%%%%%%%%%%%%%%%%%%%%%%%%%%%%%%%%%%%%%%%%%%%%%%%%%%%%%%%%%%

According to the standard $\Lambda$CDM model, the Milky Way (MW) Globular Clusters (GCs) are the first bound stellar systems with a typical age of about 10-12~Gyr that formed in the early Universe \citep{VandenBerg2013, Valcin2020}. The GCs are quite common objects: at the beginning of 2020, 150 of them were discovered in the Milky Way \citep{Gaia2018}, at the 2022 -- about 10 more GCs were discovered by \cite{Gaia2021} and there are more than 10 new candidates for the GCs. In larger galaxies, there may be more of them: for example, in the Andromeda galaxy, their number can reach up to 500 \citep{Barmby2001}. Some giant elliptical galaxies, especially those at the centre of galaxy clusters such as M 87, may have up to 13 thousand GCs \citep{McLaughlin1994}.

The GCs are a very useful tool in the study of merging and interaction history of the galaxies \citep{Ashman1992}. It is known that in our Galaxy there are even swallowed up stellar streams \citep{Ibata2021} and satellite galaxies \citep{McConnachie2012, McConnachie2021} with their own GCs. Therefore, the study of the GCs kinematic characteristics together with their chemical properties helps in understanding the global evolution of the Galaxy itself.

Centres of the most galaxies host a supermassive black hole (SMBH).
Observations of the the so-called S-stars \citep{Genzel2010, Gillessen2017} in the centre of our Galaxy showed the presence of a central object with a mass ~4 million solar masses, what is the direct evidence of the existence of the SMBH in the Milky Way centre \citep{Ghez2005}. As a result of recent direct observations of the Galactic centre (GalC) with Event Horizon Telescope, even the direct image of the Milky Way's SMBH was obtained \citep{EHT2022}.

According to our preliminary research \citep{Ishchenko2021, Ishchenko2023b}, we have found that the orbits of some MW GCs can pass close to the Galactic centre . This idea is confirmed in the papers: \cite{Burkert2010, Harris2011, Gonzalez-Lopezlira2017, Harris2014}, where the authors show a correlation between the mass of the central SMBH and the number of GCs in elliptical and spiral galaxies.

In addition to the SMBH, there is a nuclear star cluster (NSC) in the centre of the Galaxy, which is a very dense star system \citep{Neumayer2020}. Observations show that for a large number of galaxies, the formation of the NSC could be the result the GCs in-spiralling \citep{Lotz2004} toward the centre of the Galaxy due to dynamical friction and their merging into a compact dense star system. It is important to note, that the masses of GCs must be large enough for this phenomenon to occur \citep{Sedda2019}.
We consider that the destroyed GCs contributed to the formation and growth of the NSC \citep[for a review see][]{Neumayer2020}. The NSC can also grow by capturing stars from the GCs as they pass through the pericentre. Indeed, observations supported by numerical modelling suggest the presence of a distinct metal poor population of stars in the NSC \citep{Do2020}, that may have originated from infalling GC \citep{Arca-Sedda2020, Wang2022}. Moreover, the NSC and SMBHs may not have formed simultaneously at early times, as recent observations suggest a challenge to coevolutionary models \citep{Zhuo2023, Just2012}.
In this regard, it would be interesting to study such a possibility of the growth of the NSC. For example, one can analyse the GCs orbits passing near the centre of the Galaxy \citep{Ishchenko2021, Ishchenko2023a}.

The first attempts to analyse in detail the MW GCs orbital evolution based on the \textit{Gaia} DR2 data were made by \cite{Baumgardt2019, Bajkova2020, Bajkova2021}. The authors investigated the orbital evolution of MW GCs subsystem by integrating back in time up to 5 Gyr. In continuation of these studies we have already done a similar type of orbital integration in a fixed MW potential, analysing in detail the GCs close passages with each other and with the Galactic centre \citep{Ishchenko2021, Ishchenko2023b}. 

In order to increase the realism of our simulations, i.e. the realism of the Galaxy structure evolution, in a current investigation we applied time-dependent Galactic gravitational potentials extracted from IllustrisTNG-100 cosmological simulation database. Currently, the IllustrisTNG data is one of the best publicly available databases for such investigation~\citep{2018MNRAS.473.4077P, 2018MNRAS.475..624N, 2018MNRAS.475..676S, 2018MNRAS.480.5113M, 2018MNRAS.477.1206N, Nelson2019}. For example, the simulation's cosmological box sizes and physical mass resolution are currently one of the best. We adopted a multi-dimensional (mass and spatial) fit for the basic Galactic structures, such as halo and disk (masses and sizes) for each TNG100 simulation snapshot \citep{Mardini2020}.

The main idea of this work is to carry out the dynamic evolution of the orbits of Globular Clusters subsystem sample in lookback time up to 10 Gyr. This allows us to estimate in the common statistical way the average probability and the possibility of GCs close interaction with the Galactic centre (that dynamically changed in the past).

The paper is organised as follows. In Section \ref{sec:met} we present the GCs initial data with the integration procedure in time-variable potentials including the influence of measurement uncertainties on the GCs orbits. %Also here present errors effects.
In \ref{subsec:tempo} we present the GCs interactions rate with the Galactic centre and the statistical analysis of such events. In the Sections \ref{sec:phys} and \ref{sec:conc} we present the physical characteristics of the selected GCs and summarize our findings.

%%%%%%%%%%%%%%%%%%%%%%%%%%%%%%%%%%%%%%%%%%%%%%%%%%%%%%%%%%%%%%%%%%%%%
\section{Method}\label{sec:met}
%%%%%%%%%%%%%%%%%%%%%%%%%%%%%%%%%%%%%%%%%%%%%%%%%%%%%%%%%%%%%%%%%%%%

%%%%%%%%%%%%%%%%%%%%%%%%%%%%%%%%%%%%%%%%%%%%%%%%%%%%%%%%%%%%%%%%%%%%%
\subsection{Globular Clusters initial data}\label{subsec:DR3}
%%%%%%%%%%%%%%%%%%%%%%%%%%%%%%%%%%%%%%%%%%%%%%%%%%%%%%%%%%%%%%%%%%%%

We have selected GCs from catalogues compiled based on \textit{Gaia} Data Release 3 observations \citep{Baumgardt2021,VasBaum2021}. The catalogues are up-to-date and contain information on more than 160~objects. In particular, the catalogues contain GCs masses and their 6D phase space coordinates which are used as initial conditions for our numerical simulations: 3 position coordinates \citep{Baumgardt2021}, proper motion in right ascension (PMRA), proper motion in declination (PMDEC) and radial velocity \citep[RV;][]{VasBaum2021}. To minimize the influence of observational uncertainties on numerical simulations, we excluded the GCs with relative errors in PMRA, PMDEC and RV larger than 30\% \footnote{Error values for PMRA, PMDEC and RV \url{https://people.smp.uq.edu.au/HolgerBaumgardt/globular/orbits_table.txt}}. 12 objects do not satisfy our selection criteria and were therefore excluded from further modelling.
%Because during the dynamical simulation of the GCs, we used the clusters with the self-gravity (for more details, see the next subsection) together with the Galaxy external potential, we excluded the GC Mercer 5 due to the absence of mass information for this GC in the catalogues above.
During the dynamical simulation of the GCs, we used the clusters with the self-gravity (for more details, see the next subsection) in conjunction with the Galaxy's external potential. However, we excluded the GC Mercer 5 from our analysis due to the absence of mass information for this cluster in the above catalogues. Thus, we finally got the sample of the 147 GCs for the future integration and analysis.

%%%%%%%%%%%%%%%%%%%%%%%%%%%%%%%%%%%%%%%%%%%%%%%%%%%%%%%%%%%%%%%%%%%%%
\subsection{Time variable potentials and integration procedure}\label{sybsec:integr-TNG}
%%%%%%%%%%%%%%%%%%%%%%%%%%%%%%%%%%%%%%%%%%%%%%%%%%%%%%%%%%%%%%%%%%%%

For the GCs orbital integration, we used a high-order parallel dynamical $N$-body code $\varphi$-GPU.  This code is based on the fourth-order Hermite integration scheme with hierarchical individual block time steps \citep{Berczik2011,BSW2013}. Each GC was integrated as one physical particle with the fixed mass from the catalogue \citep{Baumgardt2021}. All 147 GCs which we investigated in this paper were integrated together taking into account the GCs self interactions and the interactions with the external potential. 

As an integration time step parameter \citep{MA1992} we decided to use $\eta=0.01$ as a good compromise between the speed of calculation and accuracy of integration. Using $\eta=0.01$ we have obtained the total relative energy drift ($\Delta E_{\rm tot}/ E_{{\rm tot}, ~t=0}$) over a 10~Gyr backward integration below $\approx2.5\times10^{-13}$. Typical integration time (on a desktop system AMD Ryzen Threadripper 3960X 24 core with 10 parallel threads) takes approximately 21~minutes. 

To be more physically motivated we performed our integration of GCs evolution in time-varying Milky Way-like potentials. We used the fitted data from IllustrisTNG-100 cosmological modelling database \citep{Nelson2019} for the external potentials. 

The IllustrisTNG-100 is characterised by a simulation box $\sim100$~Mpc$^3$. In a box of such size, each simulation can provide us a sufficient number of the MW-mass size disk galaxies with the mass resolution of $7.5\times10^{6}\rm\;M_{\odot}$ for dark matter and $1.4\times10^{6}\rm\;M_{\odot}$ for the baryonic particles, respectively. For our analysis, we identified the MW-like galaxy candidates from the Illustris simulations,  with at least $10^{5}$ dark matter particles and at least $10^{3}$ baryonic particles (stars and gas) at redshift zero.

Based on the IllustrisTNG-100 for each snapshot-time we constructed the five parameters fitting of particle distribution with the Navarro-Frenk-White (NFW) halo and Miyamoto-Nagai (MN) disk profiles. To obtain the spatial scales of the disks and dark matter haloes, we decomposed the mass distribution using the Miyamoto-Nagai~$\Phi_{\rm d} (R,z)$ \citep{Miyamoto1975} and  NFW~$\Phi_{\rm h} (R,z)$ \citep{NFW1997} potentials:
\begin{equation}
\begin{split}
\Phi_{\rm tot} &= \Phi_{\rm d} (R,z) + \Phi_{\rm h} (R,z) = \\
&= - \frac{GM_{\rm d}}{\sqrt{R^{2}+\Bigl(a_{\rm d}+\sqrt{z^{2}+b^{2}_{\rm d}}\Bigr)^{2}}} - 
\frac{GM_{\rm h}\cdot{\rm ln}\Bigr(1+\frac{\sqrt{R^{2}+z^{2}}}{b_{\rm h}}\Bigl)}{\sqrt{R^{2}+z^{2}}},
\end{split}
\end{equation}
where 
$R=\sqrt{x^{2}+y^{2}}$ is the planar Galactocentric radius, 
$z$ is the distance above the plane of the disc, 
$G$ is the gravitational constant, 
$a_{\rm d}$ is the disk scale length, 
$b_{\rm d,h}$ are the disk and halo scale heights, respectively, 
$M_{\rm d}$ and 
$M_{\rm h}=4\pi\rho_{0}b^{3}_{\rm h}$ ($\rho_{0}$ is the central mass density of the halo) are masses of the disk and halo, respectively. 

For our investigation we used four of the pre-selected IllustrisTNG time-variable potentials (TNG-TVPs) that have a maximally similar parameters to our Galaxy at present: {\tt \#411321, \#441327, \#451323} and {\tt \#462077}, see Table~\ref{tab:pot-val}
\citep{Ishchenko2023a, Ishchenko2023b, Mardini2020}, see TNG-TVP potential on line \footnote{Milky Way-like TNG-TVP potentials \url{https://sites.google.com/view/mw-type-sub-halos-from-illustr/TNG-MWl}}. 

%-------------------------------------------------------------------------%
\begin{table*}[htbp]
\caption{Parameters of the time-varying potentials selected from the IllustrisTNG-100 simulation at redshift zero. The last column shows the parameters of the corresponding MW components according to \cite{BBH2022} at present.}
\centering
\begin{tabular}{llccccccc}
\hline
\hline
\multicolumn{1}{c}{Parameter} & Unit & {\tt \#411321} & {\tt \#441327} & {\tt \#451323} & {\tt \#462077} & Milky Way \\
\hline
\hline
Disk mass, $M_{\rm d}$         & $10^{10}~\rm M_{\odot}$ & 7.110 & 7.970 & 7.670 & 7.758 & 6.788 \\
Halo mass, $M_{\rm h}$         & $10^{12}~\rm M_{\odot}$ & 1.190 & 1.020 & 1.024 & 1.028 & 1.000 \\
Disk scale length, $a_{\rm d}$ & 1~kpc                     & 2.073 & 2.630 & 2.630 & 1.859 & 3.410 \\
Disk scale height, $b_{\rm d}$ & 1~kpc                     & 1.126 & 1.356 & 1.258 & 1.359 & 0.320 \\
Halo scale height, $b_{\rm h}$ & 10 kpc                  & 2.848 & 1.981 & 2.035 & 2.356 & 2.770 \\
\hline
\end{tabular}
\label{tab:pot-val}
\end{table*} 
%-------------------------------------------------------------------------%

We also used a circular velocity value at the solar distance~($\approx 8$~kpc) in the model as an extra parameter to select the best TNG galaxies which represent the MW-type systems. This value indicates the position of the Sun at present. According to the age and chemical compositions of the stars in the solar neighbourhood, we know that over the past few Gyrs, there were no big changes in the radial motion of masses. This means that the circular velocity at the distance of the Sun in the Galactic disk should remain approximately constant during the last few Gyrs near the $V_{\odot}\approx235$~km~s$^{-1}$ \citep{Mardini2020}.  

The TNG-TVPs are obtained from large scale cosmological simulations involving millions of galaxies\footnote{IllustrisTNG-100: \url{ https://www.tng-project.org/data/}} \citep{Nelson2019}. The morphology of individual galaxies central region cannot be accurately resolved since these simulations are focused on large-scale structure. Taking into account these numerical limitations of our external potentials we specially added the extra SMBH into simulations. A SMBH was added as one special particle with a fixed position and mass equal to $4.1\times10^{6}~\rm\;M_{\odot}$ \citep{Ghez2008}. The SMBH mass was fixed throughout all time of the integration. So, in total we got four TNG-TVPs plus one modified potential {\tt \#411321} (hereafter {\tt \#411321-m}).

Based on our earlier study of dynamical friction in GalC 
\citep[see section 6.1,][]{Just2011} we can conclude that in the case of MW GCs fast (250-500 km~s$^{-1}$) near central passages these forces will be not dominant. We can also mention that the average orbital distance of GCs from GalC are well beyond few kpc, so the densities at these distances are significantly lower compared to the GalC near central densities (< 100 pc).    

%%%%%%%%%%%%%%%%%%%%%%%%%%%%%%%%%%%%%%%%%%%%%%%%%%%%%%%%%%%%%%%%%%%%%
\subsection{Influence of the measurement errors on Globular Cluster orbits}\label{sybsec:error-effect}
%%%%%%%%%%%%%%%%%%%%%%%%%%%%%%%%%%%%%%%%%%%%%%%%%%%%%%%%%%%%%%%%%%%%

%-------------------------------------------------------------------------%
\begin{figure*}[htbp!]
\centering
\includegraphics[width=0.9\linewidth]{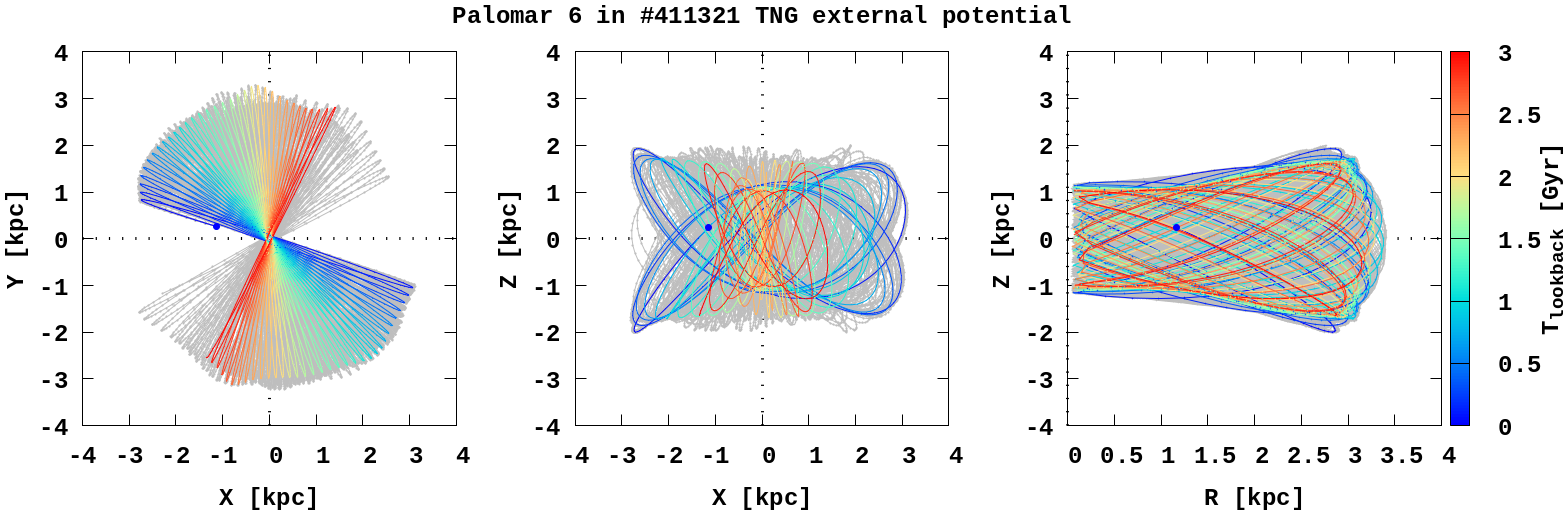}
\includegraphics[width=0.9\linewidth]{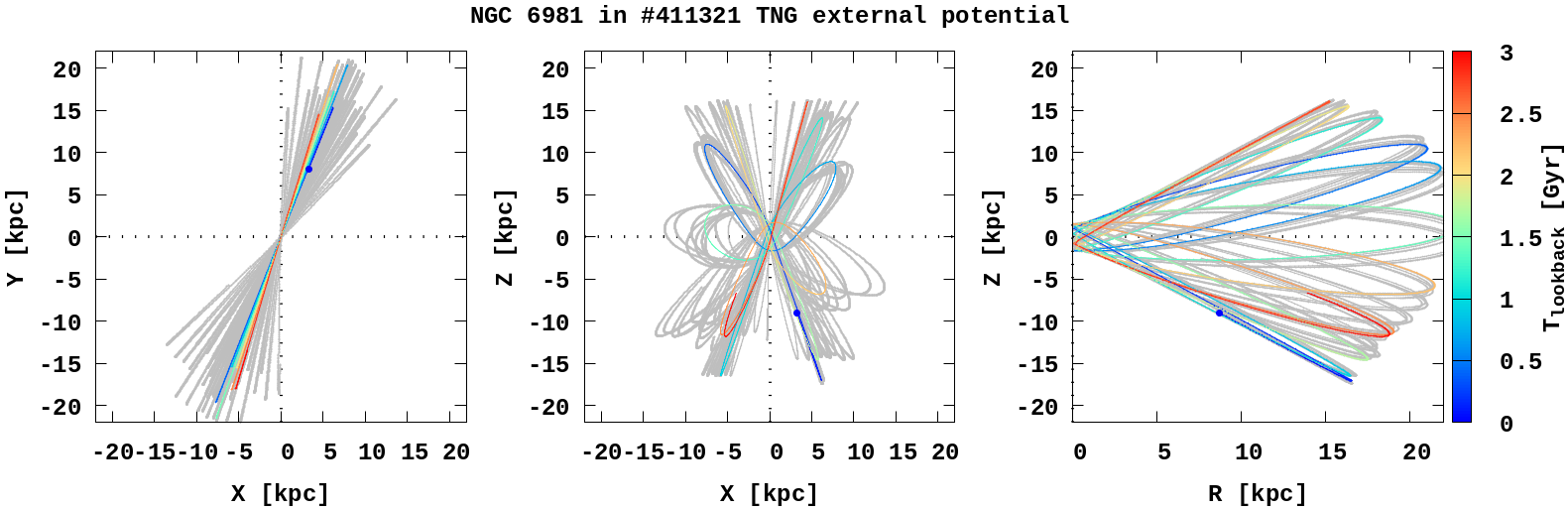}
\caption{The orbital evolution of two selected clusters: Pal~6 and NGC~6981. On the plot we show only the first 3 Gyr evolution in {\tt \#411321} TNG-TVP external potential. Colour coded line presents the orbit based on catalogue initial positions and velocities. Grey colour represents the orbits for ten different random realisations of initial data.}
\label{fig:orb-err}
\end{figure*}
%-------------------------------------------------------------------------%

For most of the selected GCs the measurement uncertainties in velocities are on the order of several percent. To check the possible influence of these uncertainties on orbital integration, we performed 10 test simulations. The positions were fixed and PMRA, PMDEC and RA were varied within the range of the observational errors (ePMRA, ePMDEC and eRV) obtained from the \cite{VasBaum2021} catalogue. Figure~\ref{fig:orb-err} illustrates the orbital evolution of two randomly selected clusters (Pal~6 and NGC~6981) with their orbits affected by the errors. The influence of the measurement errors (radial velocity eRV, proper motions in right ascension ePMRA and in declination ePMDEC) leads to the moderate variations in orbits. As can be seen, the orbits are qualitatively similar, therefore we can conclude that the velocity errors do not significantly affect orbital properties of the selected GCs.

Thus, we selected the GCs from the \cite{VasBaum2021} catalogues with relative errors in velocities of less than 30\% and used the provided positions and velocities as the initial conditions for our $N$-body simulations. In order to obtain statistically significant results, we performed 1000 simulations varying initial velocities of the GCs within $\pm1\sigma$ of the measurement errors taken from the normal distribution.

So, at the end we carried out 5000 simulations: 4000 for the four TNG-TVPs and 1000 for the {\tt \#411321-m} potential. The wall clock time calculation for the set of 1000 runs in one external potential takes more than 20~days of continuous calculations.

%%%%%%%%%%%%%%%%%%%%%%%%%%%%%%%%%%%%%%%%%%%%%%%%%%%%%%%%%%%%%%%%%%%%%%%%%%%%%%%
\section{Globular Clusters interaction with the Galactic centre}\label{sec:gc-bh}
%%%%%%%%%%%%%%%%%%%%%%%%%%%%%%%%%%%%%%%%%%%%%%%%%%%%%%%%%%%%%%%%%%%%%%%%%%%%%%%

%%%%%%%%%%%%%%%%%%%%%%%%%%%%%%%%%%%%%%%%%%%%%%%%%%%%%%%%%%%%%%%%%%%%%
\subsection{Global Globular Clusters interactions rate with the Galactic centre}\label{subsec:tempo}
%%%%%%%%%%%%%%%%%%%%%%%%%%%%%%%%%%%%%%%%%%%%%%%%%%%%%%%%%%%%%%%%%%%%

We adopt the following simple criterion to define the close passages of the GCs with the GalC, namely, if the minimum orbital distance between the GalC and the GC become less then 100 pc. We chose this value because it corresponds well with 
%outside the SMBH influence radius and with outside the size of the MW NSC. 
the outer influence radius of the SMBH and the outer size of the MW NSC. We simply define the actual distance between the GalC and GC as: $D_{\rm G} = \sqrt{X_{\rm GC}^{2} + Y_{\rm GC}^{2}+ Z_{\rm GC}^{2}}$, and the minimum separation between the GC and GalC over the time (pericentre distance) as $D_{\rm m}$. 

This approach allows us to analyse all possible statistically significant close interactions between the GCs and GalC. The GCs interactions with GalC during their orbital evolution were analysed for all 1000 sets of randomizations and for all the TNG-TVP external potentials.

\begin{figure}[htbp!]
\centering
\includegraphics[width=0.98\linewidth]{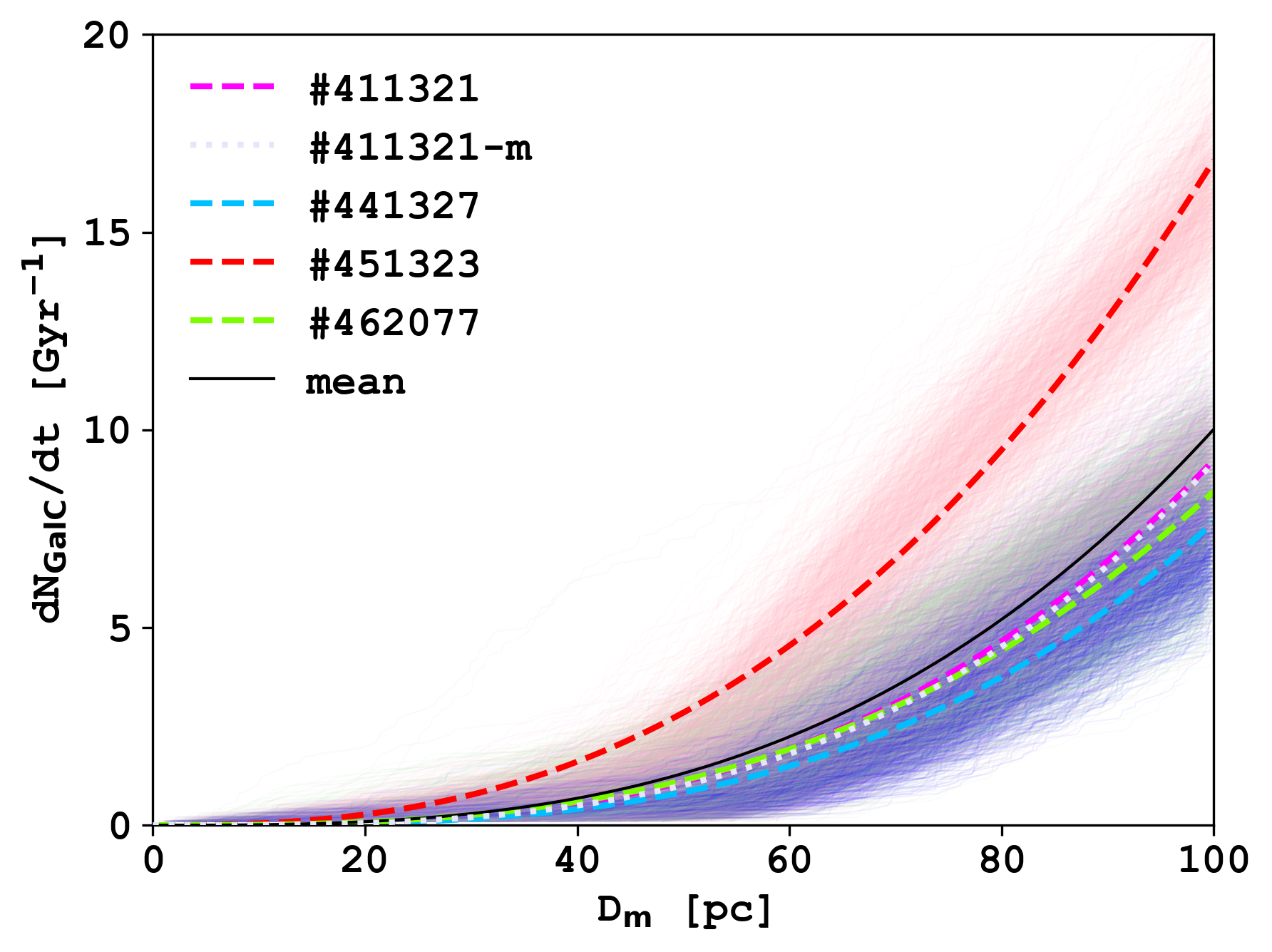}
\caption{Interaction rate of GCs with Galactic centre as a function of the relative distance {\bf from the centre} for the five TNG-TVP external potentials and 1000 random realisations (thin solid coloured lines). Thick dashed lines are a power-law fit in TNG potentials. Solid black line is a mean fitting line (see Table~\ref{tab:bh-fit}). The pale violet dotted line is a power-law fit for potential {\tt \#411321-m} with SMBH mass accounting.}
\label{fig:statR-BH}
\end{figure}

We estimated the global interaction rate of GCs with GalC (total number of close passages for all GCs per 1 Gyr) as a function of minimum relative distance from the centre $D_{\rm m}$ (see Fig.~\ref{fig:statR-BH}). This relation can be fitted by a simple power-law function:
\begin{equation}
\frac{dN_{\rm GalC} (D_{\rm m})}{dt}=10^{{\rm a}\cdot\lg(D_{\rm m})+{\rm b}},
\label{eq:fit}
\end{equation}
where the best-fit slope parameters {\bf a} and {\bf b} for all the TNG-TVP external potentials are compiled in the Table~\ref{tab:bh-fit}. 

%-------------------------------------------------------------------------%
%-------------------------------------------------------------------------%

%-------------------------------------------------------------------------%
\begin{table}[htbp!]
\caption{Fitting parameters for interaction rate GC with GalC as a function of the relative distance from the centre for four TNG-TVP external potentials.}
\centering
\begin{tabular}{lll}
\hline
\hline
\multicolumn{1}{c}{Potential} & \multicolumn{1}{c}{a} & \multicolumn{1}{c}{b}  \\
\hline
\hline
{\tt \#411321}          & 3.073\;$\pm$\;0.588 & -5.181\;$\pm$\;1.205 \\
{\tt \#411321-m}$^{*}$  & 3.183\;$\pm$\;0.609 & -5.404\;$\pm$\;1.247 \\
{\tt \#441327}          & 3.202\;$\pm$\;0.584 & -5.519\;$\pm$\;1.187 \\
{\tt \#451323}          & 2.563\;$\pm$\;0.328 & -3.900\;$\pm$\;0.670 \\
{\tt \#462077}          & 2.897\;$\pm$\;0.594 & -4.867\;$\pm$\;1.218 \\
Mean                    & 2.934\;$\pm$\;0.524 & -4.867\;$\pm$\;1.070 \\
\hline
\end{tabular}
\tablefoot{$^{*}$ simulations with additional SMBH mass accounting. This value was not used for the mean calculation.}
\label{tab:bh-fit}
\end{table}
%-------------------------------------------------------------------------%

From general physical assumptions (the probability is proportional to GalC volume), we can estimate the GCs average interaction rate with GalC as a simple power relation $dN_{\rm GalC}/dt \sim (D_{\rm m})^{3}$. As we can see from Fig.~\ref{fig:statR-BH} and from the Table~\ref{tab:bh-fit}, the GCs interaction rate with GalC that averaged for the four TNG-TVPs (excluding the model with the central SMBH) {\bf a} and {\bf b} slope parameters of the simple power-law function (equation \ref{eq:fit}) have a quite small variance: ${\rm \bar{a}} = 2.93\pm0.52$ and ${\rm \bar{b}} = -4.87\pm1.07$. For example, analysing the Fig.~\ref{fig:statR-BH} we can conclude, that at the relative distance from GalC less then 50~pc we can expect on average about 3--4 GCs close passages during each 1~Gyr and at the relative distance less than 80~pc we can expect $\sim$5--6 GCs close passages near the GalC.

\begin{figure*}[htbp!]
\centering
\includegraphics[width=0.43\linewidth]{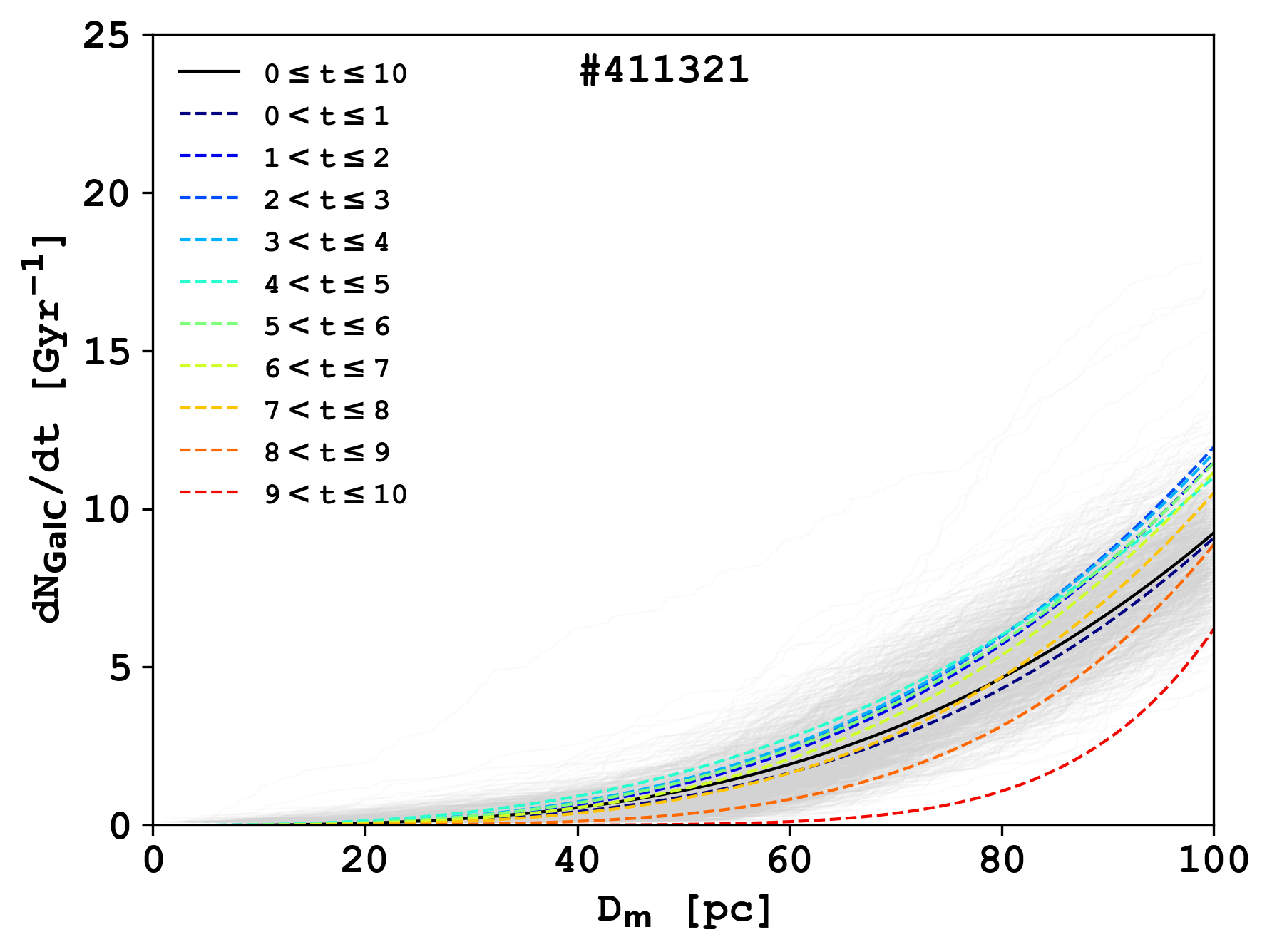}
\includegraphics[width=0.46\linewidth]{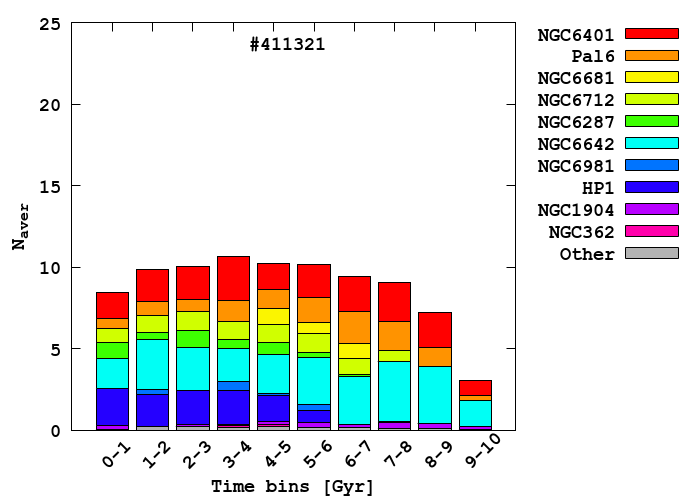}
\includegraphics[width=0.43\linewidth]{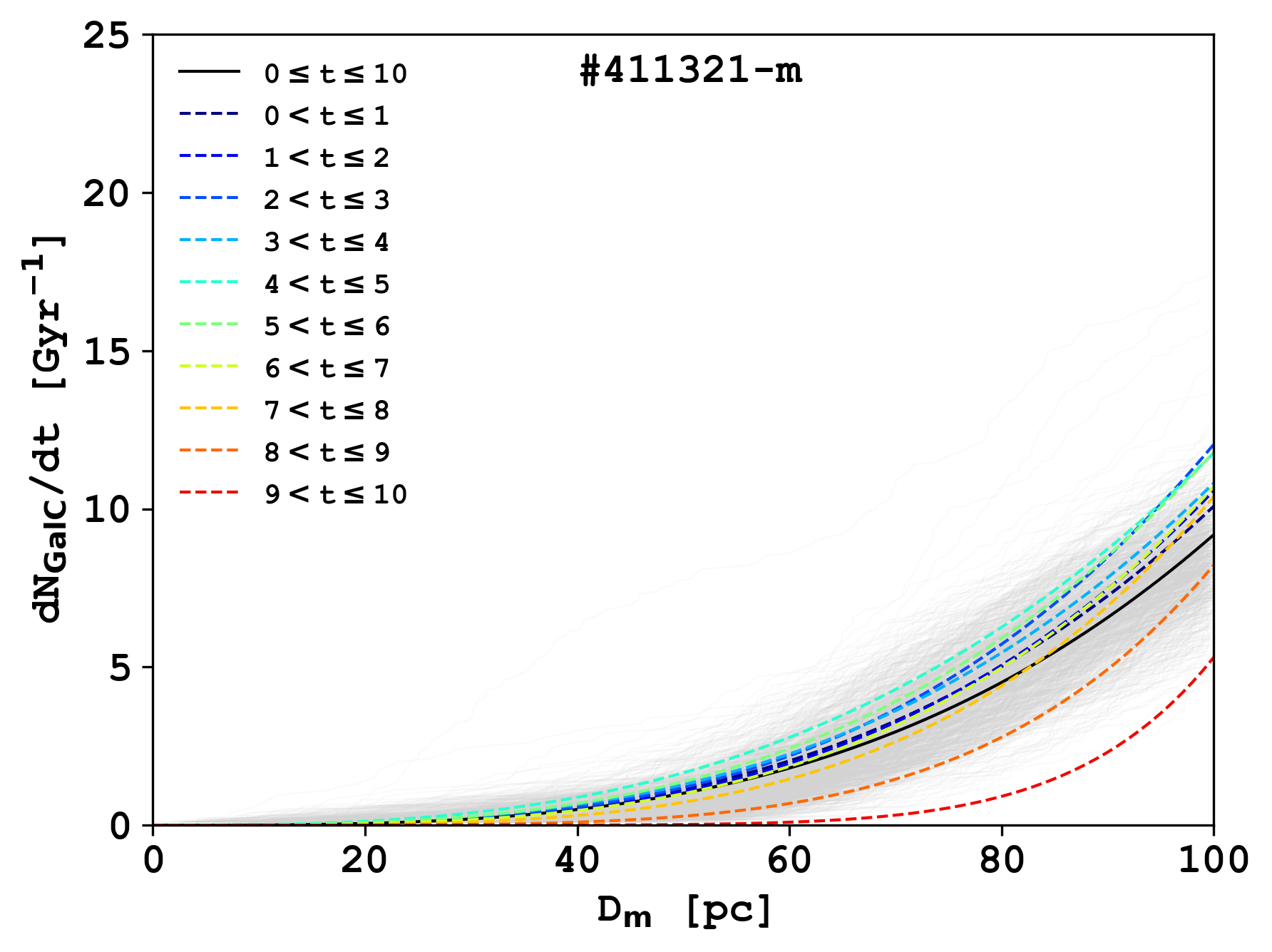}
\includegraphics[width=0.46\linewidth]{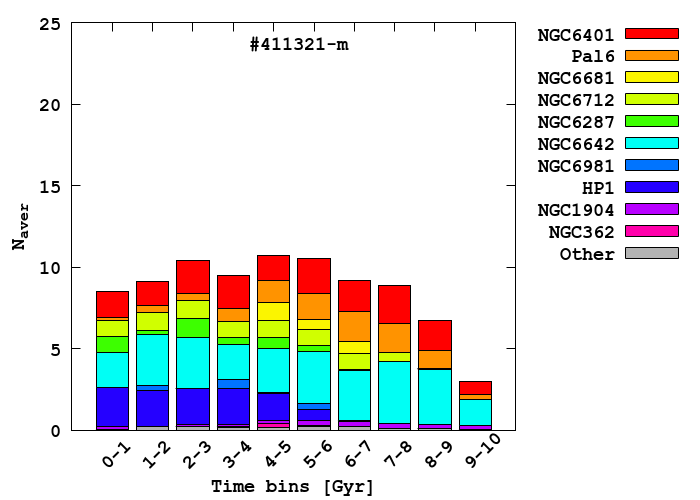}
\caption{Interaction rate of GCs with GalC as a function of the relative distance from the centre in different time intervals (colour dashed lines) for {\tt \#411321} and {\tt \#411321-m} TNG-TVPs \textit{left panels}. Each time interval have a length of 1~Gyr. The solid black line is a global close passages rate for a whole time interval 10~Gyr. The grey solid lines are results for 1000 simulations with different random realisations. 
 Contribution of individual GCs into global collision rate at different time ranges for {\tt \#411321} and {\tt \#411321-m} TNG-TVPs, \textit{right panels}.}
\label{fig:time-clip-bh}
\end{figure*}

We also estimated the interaction rate between the GCs and GalC as a function of the relative distance from the centre at different time intervals (1~Gyr) for each of four TNG-TVP: {\tt \#411321}, {\tt \#441327}, {\tt \#451323} and {\tt \#462077} (see Fig.~\ref{fig:time-clip-bh} and Appendix~\ref{app:hist-gc-bh}). In general, we can see that the collision rates are lower in the early stages of the evolution (8--10 Gyr ago). But the individual behaviour of the collision rates depends on the specific TNG-TVPs. For comparison, the solid black line in the upper left panel of the figure shows the global cumulative interaction rate as a function of the relative distance from the Galactic centre. In the right panels of the Fig.~\ref{fig:time-clip-bh} and Appendix~\ref{app:hist-gc-bh} we present the contribution from individual GCs to the interaction rates divided in to time bins. As we can see, there are several clusters that dominate in the global interaction rates. We analyse their properties in details in the Section~\ref{sec:phys}.

The lower panels in the Fig.~\ref{fig:time-clip-bh} show interaction rates in time bins for the potential that include the SMBH {\tt \#411321-m} (left) and the contribution from individual GCs (right). Comparing upper and lower panels of the figure, we can conclude that the effects of the SMBH on the global interaction rates are marginal. 

%-------------------------------------------------------------------------%
%-------------------------------------------------------------------------%

We also present the orbital evolution with and without the SMBH mass influence in {\tt \#411321} TNG-TVP external potential for the GC NGC 362 (Fig. \ref{fig:orb1}), PH 1 (Fig. \ref{fig:orb4}) and for the NGC 6401 (Fig. \ref{fig:orb5}). As we can see, significant changes of the orbits start only after 8~Gyr of lookback time. We can understand this as a result of the significant halo and disk mass loss of the model galaxy after 8~Gyr. However, as we can see in Fig.~\ref{fig:time-clip-bh}, the influence of the central SMBH mass are generally not significant on the GC vs. GalC interaction rates. 

For example, on {\tt \#411321} and {\tt \#451323} TNG-TVPs we see the similar initial behaviour of the interaction rates as a function of time. At the beginning of our integration (from 0 to 2 Gyr) we have growing interaction rates in total around $\sim$10--15 events per Gyr. Around 4--6 Gyr we have a maximum of interaction rates 12--20 events per Gyr. In a case of {\tt \#441327} TNG-TVP we have a completely different shape -- almost constant interaction rates in time (8 event per Gyr). In potential {\tt \#462077} we have an interesting feature, see Appendix~\ref{app:hist-gc-bh}. Here the local minimum happens in between 2--3 Gyr. So, we first have a drop to 8 events per Gyr and only after that it starts to grow up to 15 events per Gyr. As we can see from Fig.~\ref{fig:time-clip-bh} and Appendix~\ref{app:hist-gc-bh} on the right panels the most ``active'' (in a sense of interaction) GCs are: NGC 6642 (cyan), NGC 6401 (red) and HP 1 (blue) independently from the TNG-TVP external potentials.  

%%%%%%%%%%%%%%%%%%%%%%%%%%%%%%%%%%%%%%%%%%%%%%%%%%%%%%%%%%%%%%%%%%%%%
\subsection{Statistical analysis of the Globular Clusters interaction rates}\label{subsec:stat-prob}
%%%%%%%%%%%%%%%%%%%%%%%%%%%%%%%%%%%%%%%%%%%%%%%%%%%%%%%%%%%%%%%%%%%%

Our analysis of the GCs interaction rates presented in the previous subsection (Fig.~\ref{fig:time-clip-bh}) shows that there are several GCs that play a role of major contributors to the interaction rates with GalC (such as NGC~6401, Pal~6, NGC~6681, NGC~6712, NGC~6287, NGC~4462, NGC~6981, HP~1, NGC~1904 and NGC~362). While others (e.g. NGC 6638, NGC 5946, UKS-1, Pal 14, Pal 15, NGC 6229, ESO-452, IC-1257, Pal 2) have far fewer potential encounters. In this subsection we investigate each of the individual GCs and quantify their probability of their close encounters.   

To calculate the probability that each individual GC experienced the close encounter with the GalC at least once during its lifetime we search for the close passages in all random realisations for each of the four TNG potentials. Thus, the probability is $P_{\rm GC} = (M_{\rm rand}/1000)\cdot100\%$, where $M_{\rm rand}$ is the number of models in which the GC approaches GalC ($D_{\rm m}$ < 100 pc) at least once from the 1000 random realisations in a particular potential.

In total, in our four TNG-TVP potentials we have 36657 cases, while we have at least one individual interactions between GalC and 98 GCs (from all 147 GCs) from all (4$\times$1000 = 4000) random realisations (see the end of Subsection~\ref{subsec:DR3}). We selected the top ten GCs with the highest chance of interaction and summarised their probabilities $P_{\rm GC}$ in Table~\ref{tab:bh-prob}. 

%-------------------------------------------------------------------------%
\begin{table*}[htbp]
\caption{The percent of probability the GCs interaction with Galactic centre in all 1000 sets of randomisation for four TNG time-variable potentials.}
\centering
\sisetup{separate-uncertainty}
\begin{tabular}{c
S[table-format=3.1]
S[table-format=3.1]
S[table-format=3.1]
S[table-format=3.1]
S[table-format=3.1]
S[table-format=3.1(2)]}
\hline
\hline
GC & \multicolumn{1}{c}{{\tt \#411321}} & \multicolumn{1}{c}{{\tt \#411321-m}} & \multicolumn{1}{c}{{\tt \#441327}} & \multicolumn{1}{c}{{\tt \#451323}} & \multicolumn{1}{c}{{\tt \#462077}} & \multicolumn{1}{c}{Mean} \\
{(1)} & {(2)} & {(3)} & {(4)} & {(5)} & {(6)} & {(7)} \\ 
\hline
\hline
NGC~6401     & 100.0 & 99.9  & 100.0 & 100.0 & 100.0 &  100.0 \\
Pal~6        & 100.0 & 99.6  &  99.9 & 100.0 & 100.0 &  99.9\pm0.1 \\
NGC~6681     &  99.9 & 99.9  & 100.0 & 100.0 & 100.0 &  99.9\pm0.1 \\
NGC~6712     &  99.9 & 100.0 &  99.9 &  99.8 & 100.0 &  99.9\pm0.1 \\
NGC~6287     & 100.0 & 100.0 & 100.0 &  97.0 &  92.1 &  97.3\pm3.7 \\
NGC~6642     &  99.8 & 99.8  &  99.3 & 100.0 &  99.5 &  99.7\pm0.3 \\
NGC~6981     &  83.9 & 84.9  &  90.2 &  93.5 &  87.8 &  88.9\pm4.0 \\
HP~1         &  98.7 & 99.0  &  70.7 &  99.5 &  83.0 &  89.9\pm13.8 \\
NGC~1904     &  72.4 & 73.0  &  73.6 &  83.2 &  76.7 &  76.5\pm4.8 \\
NGC~362      &  24.4 & 27.9  &  30.7 &  41.2 &  12.9 &  27.3\pm11.8 \\
\hline
\end{tabular}
\tablefoot{
Column~(1) -- name of a GC. Columns~(2)--(6) -- interaction probabilities for GCs with GalC for each TNG-TVP external potentials in percent. {\tt \#411321-m} -- the special potential with the MW SMBH mass. Column~(7) -- the average probability value with error over all potentials, excluding the {\tt \#411321-m} potential.}
\label{tab:bh-prob}
\end{table*}
%-------------------------------------------------------------------------%

As we can see from Table~\ref{tab:bh-prob}, the six GCs: NGC~6401, Pal~6, NGC~6681, NGC~6712, NGC~6287 and NGC~6642 have very reliable close passages near the GalC in all four TNG external potentials with the probability of almost 100\%. This fact can already be stated as a strong conclusion about the dynamical evolution of these GCs. The other four GCs: NGC~6981, HP~1, NGC~1904 and NGC~362 have interactions probability values in the range from 90\% to 27\%. 

We additionally estimated the impact of the random sample sizes on the probability results for the selected GCs in Table~\ref{tab:bh-prob}. For this investigation from all 1000 random realisations for selected {\tt \#411321} TNG-TVP potential, we randomly constructed sub-samples with sizes 200, 400, 600, 800 and compared the probability of interaction $P_{\rm GC}$ with the full 1000 randomisation results. Starting from a few hundred realisations, the interaction numbers are saturated. As an example, we can show GC HP~1 interaction probabilities in a different sub-samples: 99\% (200), 98\% (400), 98\% (600), 98\% (800) and 99\% (1000). For the GC NGC~6401, we get 100\% probability for all sample sizes. As we expected, the interaction probability results $P_{\rm GC}$ practically do not depend on the sample size starting from a sub-sample size 200.

%%%%%%%%%%%%%%%%%%%%%%%%%%%%%%%%%%%%%%%%%%%%%%%%%%%%%%%%%%%%%%%%%%%%%
\section{Physical characteristics of the selected Globular Clusters}\label{sec:phys}
%%%%%%%%%%%%%%%%%%%%%%%%%%%%%%%%%%%%%%%%%%%%%%%%%%%%%%%%%%%%%%%%%%%%

\begin{figure}[htbp]
\centering
\includegraphics[width=0.99\linewidth]{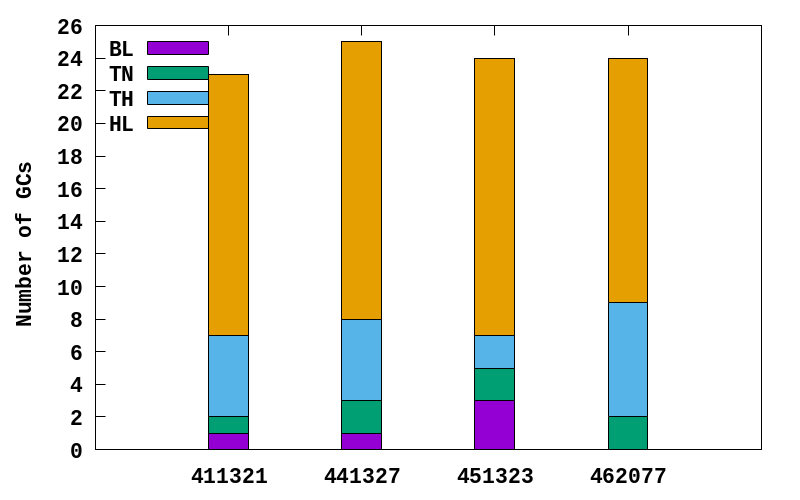}
\caption{Distribution of the GCs by their origin: bulge (BL), thin disk (TN), thick disk (TH), and halo (HL) in four Galactic components. 
} 
\label{fig:gisogal-orb}
\end{figure}

For all our selected GCs, we determined the regions of the Galaxy where they are generally located. For selection we apply the criteria according to \cite{Bland-Hawthorn2016}. In the Fig.~\ref{fig:gisogal-orb} we present the distribution of the GCs that have at least one individual interaction with the GalC according to the location in different Galactic regions for the four TNG-TVP potentials. 

As a main conclusion from the Fig.~\ref{fig:gisogal-orb}, we can argue that the majority of the GCs in our sample that interact with the GalC come from the halo (HL) and thick disk (TH) regions. As an interesting fact, we can note that for the TNG-TVP {\tt \#462077}, we do not have any objects from the bulge (BL) region that could potentially come into close vicinity of the GalC. 

%-------------------------------------------------------------------------%
%-------------------------------------------------------------------------%

\begin{figure}[htbp]
\centering
\includegraphics[width=0.95\linewidth]{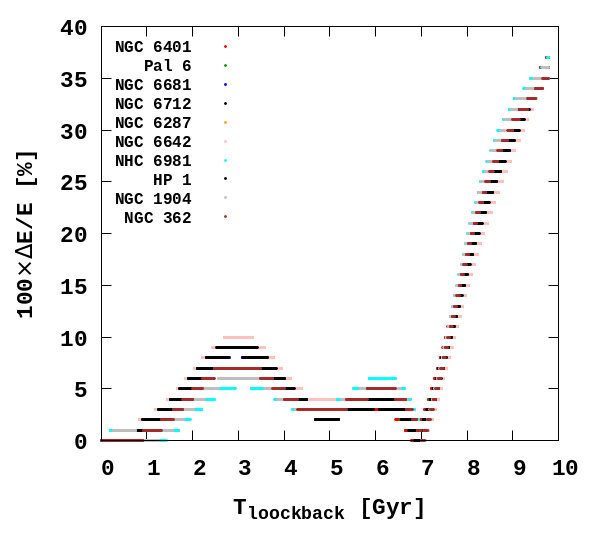}
\caption{Relative energy $\Delta E/E$ changes of the selected GCs in percent during orbital evolution for the {\tt \#411321} TNG-TVP external potential.}
\label{fig:de}
\end{figure}

We calculated the changes in the relative energy $\Delta E/E$ of our selected sample of 10 GCs during the full time evolution up to ~10 Gyr. As an example, the energy changes of individual GCs in the TNG-TVP {\tt \#411321} are presented in Fig.~\ref{fig:de}. The maximum changes in energy reaches $\sim$40\% but only at the end of the integration. From the Fig.~\ref{fig:de}, we can conclude that our selected 10 GCs are not strongly affected by the time evolution of the external gravitational potential up to $\sim$7 Gyr. We see some significant changes only for the last two Gyrs.

%-------------------------------------------------------------------------%
%-------------------------------------------------------------------------%

\begin{figure*}[htbp!]
\centering
\includegraphics[width=0.99\linewidth]{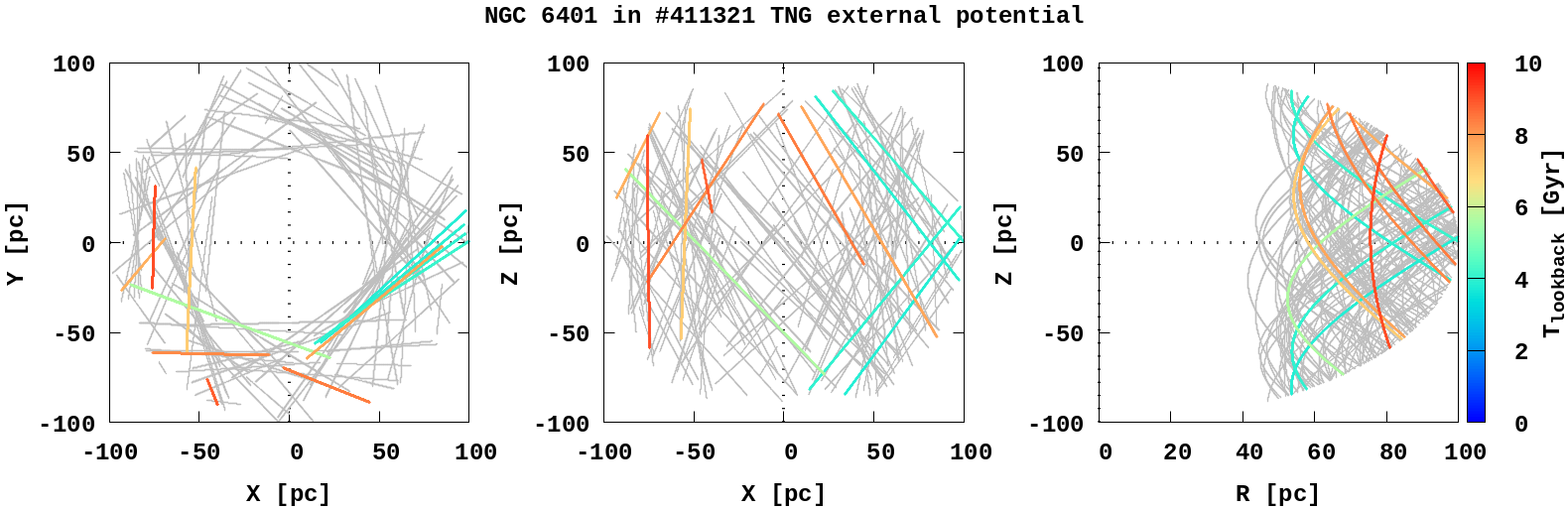}
\includegraphics[width=0.98\linewidth]{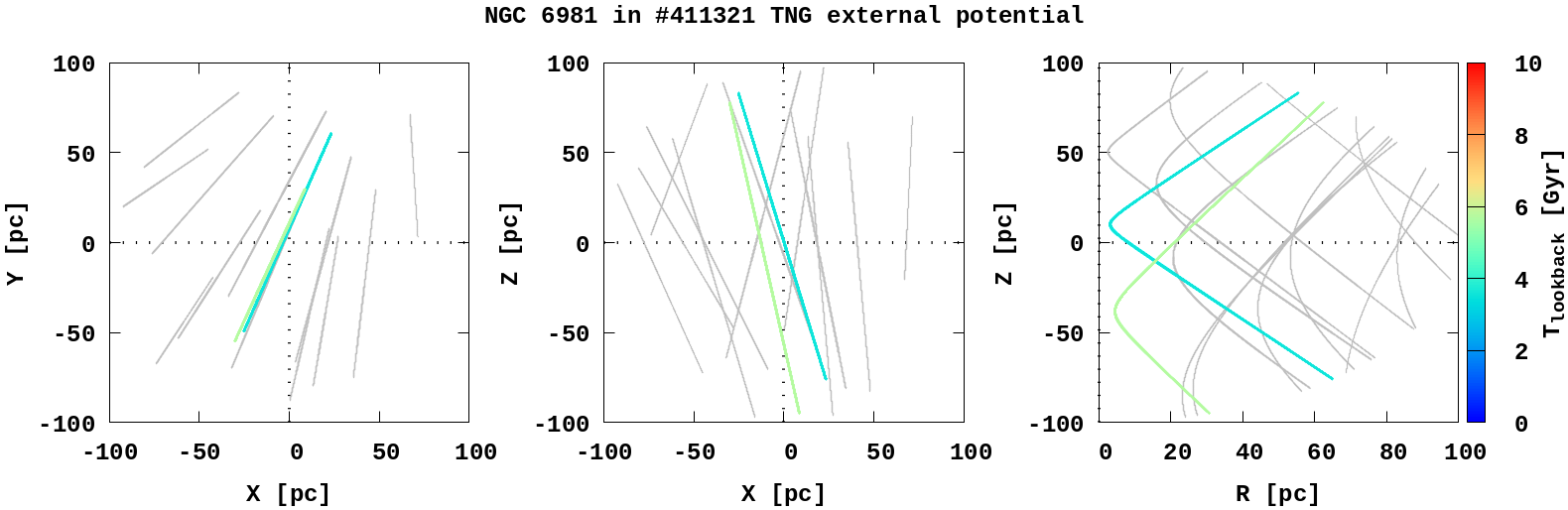}
\caption{Detailed trajectories that have interactions with the Galactic centre in three planes ($X$, $Y$), ($X$, $Z$) and ($R$, $Z$) (where $R$ is the planar Galactocentric radius) inside the box 100$\times$100 pc. The GCs \textit {from top to bottom panels}: NGC 6401 and NGC 6981 in {\tt \#411321} TNG-TVP potential. The colour line presents the trajectory based on the data from the catalogues. Gray lines -- taking into account the measurement errors.}
\label{fig:box-orb}
\end{figure*}

In Fig.~\ref{fig:box-orb}, we present the orbital trajectories of the two selected GCs (NGC 6401 and NGC 6981) as an example. The figure shows all the different realisations (grey lines) that reflect the assumed velocity measurement errors (eRV, ePMRA and eMDEC). The colour line (coded by the lookback time) represents the trajectory based on the data from the catalogues, without taking into account the measurements errors. The trajectories are presented in three planes ($X$, $Y$), ($X$, $Z$) and ($R$, $Z$) (where $R$ is the planar Galactocentric radius) inside the box 100$\times$100 pc. The plots for other eight GCs, such as Pal 6, NGC 6712, NGC 6287, NGC 6642, HP 1, NGC 6681 and NGC 1904, are presented in Appendix~\ref{app:rand-orb-box}. 

%-------------------------------------------------------------------------%
%-------------------------------------------------------------------------%

As an additional item in Table~\ref{tab:bh-prob}, in column (3), we present the interaction probability of the GCs with the GalC, taking into account the presence of the MW SMBH. As we can see from columns (2) and (3), the differences are minimal for the {\tt \#411321} TNG-TVP -- not more than few percent. The similar behaviour we see also in the Table~\ref{tab:bh-phis-param}. The statistics of the GCs' interactions with the GalC are almost identical in the potentials {\tt \#411321} and {\tt \#411321-m}. 

We also checked in detail, the parameters of close Galactic centre encounters for the selected 10 GCs for all 1000 random sample realisations. The minimum distance in pc between the GC and GalC in our sample realisations is denoted as $<D_{\rm m}>$, the corresponding value of the relative velocity is denoted as $<dV>$ and the average number of individual interactions for each randomisation for each TNG-TVPs is denoted as $<N_{\rm int}>$. We present all these values in Table~\ref{tab:bh-phis-param}. The ($\pm$) values in each column represent the standard deviation from these average values based on our 1000 random realisations. 

%-------------------------------------------------------------------------%
\begin{table*}[htbp]
\caption{Statistics of characteristics of the GCs interaction with the Galactic centre in all four TNG-TVP external potentials.}
\centering
%\small
\sisetup{separate-uncertainty}
%\resizebox{0.98\textwidth}{!}{
\begin{tabular}{
c|
S[table-format=2.0(2)] 
S[table-format=3.0(2)] 
S[table-format=2.0(2)]|
S[table-format=2.0(2)]
S[table-format=3.0(2)]
S[table-format=2.0(2)]|
S[table-format=2.0(2)]
S[table-format=3.0(2)]
S[table-format=2.0(2)]
}
%\hline
%\hline
\cline{1-7}
\vspace{-1.1em}\\
\cline{1-7}
\multicolumn{1}{r|}{\diagbox[width=2cm]{}{Pot}} & \multicolumn{3}{c|}{{\tt \#411321}} & \multicolumn{3}{c|}{{\tt \#411321-m}} &  \\
\cline{2-7}
GC & {$<D_{\rm m}>$} & {$<dV>$} & {$<N_{\rm int}>$} &  
     {$<D_{\rm m}>$} & {$<dV>$} & {$<N_{\rm int}>$} &  
      &  & \\
     & pc & km~s$^{-1}$ & 
     & pc & km~s$^{-1}$ &  
     &  &  \\
\cline{1-7}
\vspace{-1.1em}\\
\cline{1-7}
NGC~6401  & 56\pm10 & 331\pm18 & 19\pm2  & 58\pm10 & 333\pm17 & 18\pm4 & & & \\
Pal~6     & 58\pm12 & 340\pm10 & 11\pm6  & 58\pm13 & 341\pm10 & 10\pm5 & & & \\
NGC~6681  & 42\pm16 & 410\pm4  &  3\pm1  & 45\pm17 & 411\pm1  & 3\pm1  & & & \\
NGC~6712  & 52\pm8  & 379\pm10 &  8\pm2  & 53\pm8  & 380\pm10 & 8\pm2  & & & \\
NGC~6287  & 82\pm5  & 405\pm10 &  4\pm1  & 81\pm4  & 398\pm11 & 4\pm1  & & & \\
NGC~6642  & 62\pm16 & 262\pm19 & 27\pm10 & 63\pm14 & 260\pm19 & 28\pm9 & & & \\
NGC~6981  & 57\pm24 & 545\pm8  &  2\pm1  & 57\pm24 & 547\pm8  & 2\pm1  & & & \\
HP~1      & 50\pm21 & 304\pm10 & 11\pm4  & 50\pm22 & 306\pm11 & 16\pm4 & & & \\
NGC~1904  & 53\pm25 & 517\pm17 &  3\pm2  & 54\pm24 & 518\pm18 & 3\pm1  & & & \\
NGC~362   & 88\pm10 & 484\pm6  &  1\pm1  & 87\pm10 & 481\pm5  & 1\pm1  & & & \\
\hline
\hline
\multicolumn{1}{r|}{\diagbox[width=2cm]{}{Pot}} & \multicolumn{3}{c|}{{\tt \#441327}} & \multicolumn{3}{c|}{{\tt \#451323}} & \multicolumn{3}{c}{{\tt \#462077}} \\
\cline{2-10}
GC & {$<D_{\rm m}>$} & {$<dV>$} & {$<N_{\rm int}>$} &  
     {$<D_{\rm m}>$} & {$<dV>$} & {$<N_{\rm int}>$} &
     {$<D_{\rm m}>$} & {$<dV>$} & {$<N_{\rm int}>$} \\
     & pc & km~s$^{-1}$ &
     & pc & km~s$^{-1}$ &
     & pc & km~s$^{-1}$  \\
\hline
\hline
NGC~6401  & 48\pm7  & 352\pm12 & 38\pm6 & 50\pm7  & 366\pm19 & 20\pm3  & 54\pm10 & 305\pm12 & 14\pm3 \\
Pal~6     & 58\pm9  & 359\pm15 & 22\pm4 & 70\pm8  & 351\pm29 & 10\pm4  & 65\pm9  & 299\pm7  & 10\pm3 \\
NGC~6681  & 25\pm10 & 440\pm14 & 13\pm2 & 34\pm13 & 439\pm14 &  7\pm1  & 21\pm11 & 368\pm5  &  7\pm2 \\
NGC~6712  & 49\pm10 & 427\pm16 & 13\pm3 & 54\pm6  & 382\pm25 &  8\pm3  & 59\pm10 & 342\pm6  &  7\pm2 \\
NGC~6287  & 94\pm4  & 409\pm11 &  2\pm1 & 84\pm7  & 418\pm20 &  2\pm1  & 77\pm3  & 375\pm4  &  4\pm2 \\
NGC~6642  & 51\pm13 & 300\pm10 & 44\pm7 & 59\pm22 & 276\pm27 & 25\pm13 & 66\pm13 & 226\pm13 & 20\pm6 \\
NGC~6981  & 46\pm24 & 590\pm14 &  3\pm1 & 51\pm24 & 540\pm28 &  3\pm1  & 48\pm24 & 511\pm12 &  3\pm1 \\
HP~1      & 46\pm21 & 327\pm14 & 16\pm6 & 65\pm23 & 290\pm2  &  4\pm2  & 61\pm25 & 282\pm5  &  7\pm4 \\
NGC~1904  & 51\pm25 & 588\pm19 &  4\pm2 & 54\pm25 & 539\pm47 &  3\pm1  & 57\pm25 & 500\pm9  &  3\pm1 \\
NGC~362   & 85\pm11 & 530\pm12 &  2\pm1 & 88\pm10 & 482\pm10 &  1\pm1  & 87\pm10 & 457\pm6  &  1\pm1 \\
\hline
\end{tabular}
%}
\label{tab:bh-phis-param}
\end{table*}
%-------------------------------------------------------------------------%

From Table~\ref{tab:bh-phis-param}, one can conclude that close encounters between the GCs and GalC statistically, on average, occur with the minimum relative distance of around $\sim60$~pc and with the relative velocity -- $\sim$400 km~s$^{-1}$. We can see, that for individual GCs these numbers are highly similar even for different TNG-TVPs potentials.

Analysing the first six GCs that have the highest statistical probability of interaction with GalC, we found that NGCs 6642 and 6401 have more passes than the others. The average values of interaction over all four TNG-TVP potentials are estimated to be $\sim$36 and $\sim$27, respectively. In addition, the NGC 6642 also has the lowest relative velocity at the moment of close passage to the GalC from the entire sample. The next GC with high interaction numbers among the four potentials is Pal 6, which has $\sim$15 passes. However, Pal 6 has much higher uncertainties. 

In the Table~\ref{tab:phis-gc}, we present the orbits main characteristics of the selected 10 clusters, such as the values of pericenter {\tt ''per''}, orbital eccentricity {\tt ''ecc''} and the maximum height above the Galactic plane $z_{\rm max}$. Also we present here the belonging of each cluster to our Galaxy, as well as the possible progenitors according to the classification of \cite{Malhan2022}. Here M-B -- Main Bulge of our Galaxy, G-E -- Gaia-Enceladus, remains of a dwarf galaxy and Pontus -- ancient satellite galaxy. Also here we present some other physical characteristics of the GCs: age, current masses, half-mass radii and Galactocentric distance. In Appendix \ref{app:det-orb} we present orbital evolution up to 10 Gyr for these GCs in four TNG-TVP external potentials {\tt \#411321}, {\tt \#441327}, {\tt \#451323} and {\tt \#462077}. 

%-------------------------------------------------------------------------%
\begin{table*}[htbp]
\caption{Main kinematic characteristics of the selected GCs and their orbits in {\tt \#411321} TNG-TVP external potential.}
\centering
\sisetup{separate-uncertainty}
%\resizebox{0.49\textwidth}{!}{
\begin{tabular}{
rc
S[table-format=2.2] 
cc
S[table-format=2.2] 
S[table-format=2.1] 
cc
S[table-format=2.1] 
c}
\hline
\hline
\multicolumn{1}{c}{ID} & Name & {Mass} & Age & $R_{\rm hm}$  & {$D_{\rm G}$} & GR & {\tt per} & {\tt ecc} & {$z_{\rm max}$} & Progenitor \\
                       &      & {$10^{4}\rm\;M_{\odot}$} & Gyr & pc & {kpc} &    & pc       &           & {kpc}           &            \\
(1) & (2) & {(3)} & (4) & (5) & {(6)} & {(7)} & {(8)} & {(9)} & {(10)} & {(11)}\\
\hline
\hline
 1 & NGC~6401  & 14.50 & 13.20$^{\rm a}$ & 3.28 & 0.75  & TH & 56 & 0.70 &  1.3 & M-B \\
 2 & Pal~6     &  9.45 & 12.40$^{\rm b}$ & 2.89 & 1.19  & TH & 58 & 0.70 &  2.0 & M-B \\
 3 & NGC~6681  & 11.60 & 12.80$^{\rm c}$ & 2.89 & 2.29  & HL & 42 & 0.83 &  5.0 & --  \\
 4 & NGC~6712  &  9.63 & 10.40$^{\rm c}$ & 3.21 & 3.55  & HL & 52 & 0.81 &  3.0 & --  \\
 5 & NGC~6287  & 14.50 & 13.57$^{\rm c}$ & 3.62 & 1.57  & HL & 82 & 0.76 &  4.0 & --  \\
 6 & NGC~6642  &  3.44 & 13.80$^{\rm d}$ & 1.51 & 1.66  & TN & 62 & 0.52 &  0.9 & M-B \\
 7 & NGC~6981  &  6.89 & 10.88$^{\rm c}$ & 5.96 & 12.53 & HL & 57 & 0.94 & 20.0 & G-E \\
 8 & HP~1      & 20.00 & 12.80$^{\rm e}$ & 6.06 & 1.26  & TH & 50 & 0.54 &  2.5 & M-B \\
 9 & NGC~1904  & 13.90 & 11.14$^{\rm c}$ & 3.21 & 19.09 & HL & 53 & 0.94 & 15.0 & G-E \\
10 & NGC~362   & 28.40 & 10.37$^{\rm c}$ & 3.89 & 9.62  & HL & 88 & 0.86 & 10.0 & Pontus \\
\hline
\end{tabular}
%}
\tablefoot{
Columns (1) and (2) -- the GCs index numbers and their names.  
Column (3) -- the GCs mass in $\rm M_{\odot}$ according to \cite{Baumgardt2021} at present time.
Column (4) -- the GCs age in Gyr according to the: 
$^{\rm a}$\cite{Cohen2021}
$^{\rm b}$\cite{Souza2021},
$^{\rm c}$\cite{Forbes2010},
$^{\rm d}$\cite{Balbinot2009},
$^{\rm e}$\cite{Ortolani2011}.
Column (5) -- the GCs half-mass radius in pc at present day according to  \cite{Baumgardt2021} .
Column (6) -- GCs distance from the Galactic centre in kpc at present day according to \cite{Baumgardt2021}.
Column (7) -- the association to Galaxy region (GR): bulge (BL), thin disk (TN), thick disk (TH), and halo (HL).
Columns (8), (9) and (10) -- the characteristics of GCs orbits such as {\tt per}, {\tt ecc}, $z_{\rm max}$ according to their shapes taking from  web-page of the project \footnote{Orbits for all 159 GCs in TNG-TVP potentials presented on a web page of the project \url{https://sites.google.com/view/mw-type-sub-halos-from-illustr/GC-TNG}}.
Column (11) -- possible progenitors according to \cite{Malhan2022}.
}
\label{tab:phis-gc}
\end{table*} 
%-------------------------------------------------------------------------%

\begin{figure*}[htbp!]
\centering
\includegraphics[width=0.9\linewidth]{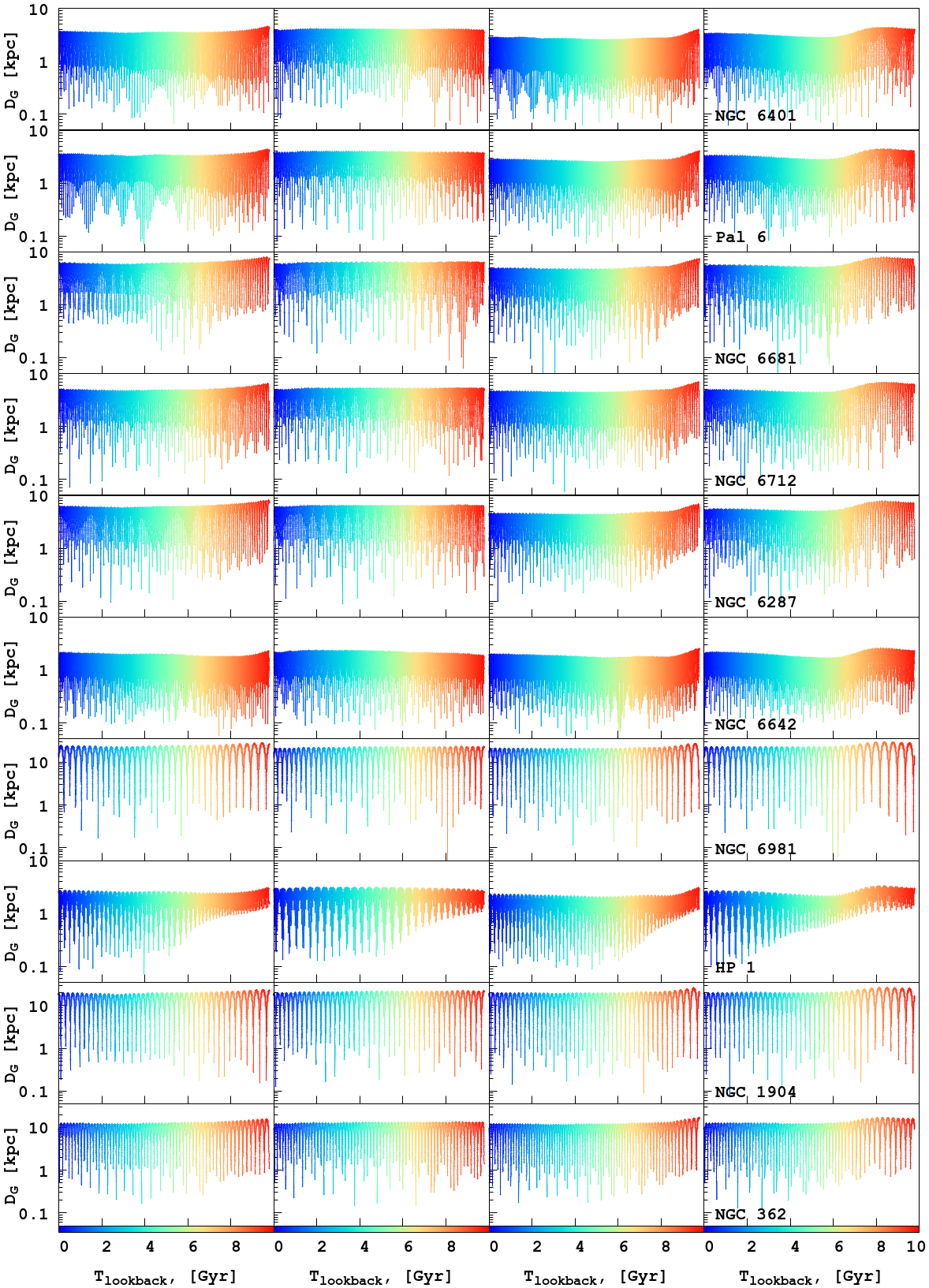}
\caption{3D distance from the Galactic centre for selected 10 GCs for all four TNG-TVPs \textit{from left to right}: {\tt \#411321}, {\tt \#441327}, {\tt \#451323} and {\tt \#462077}.}
\label{fig:3d-dist}
\end{figure*}

In Fig.~\ref{fig:3d-dist} we present the Galactocentric distances for the selected GCs (Table~\ref{tab:bh-phis-param}) for all four TNG-TVP external potentials {\bf (with selected time resolution 1 Myr)}. As we can see, for these 10 objects the close approaches to the GalC happen over almost all the integration time. From Fig.~\ref{fig:3d-dist} we can conclude, that the main changes in the orbital evolution of selected GCs mainly happen after $\sim8$~Gyr lookback time (reddish colour). This behaviour (i.e. the constant increasing of the GCs orbits apocentre) can be easily understood as a consequence of the global mass change of our TNG-TVPs, see for detail our online {\bf figures}\footnote{Milky Way-like TNG-TVP potentials \url{https://sites.google.com/view/mw-type-sub-halos-from-illustr/TNG-MWl}}.

%-------------------------------------------------------------------------%
%-------------------------------------------------------------------------%

\begin{figure*}[htbp!]
\centering
\includegraphics[width=0.90\linewidth]{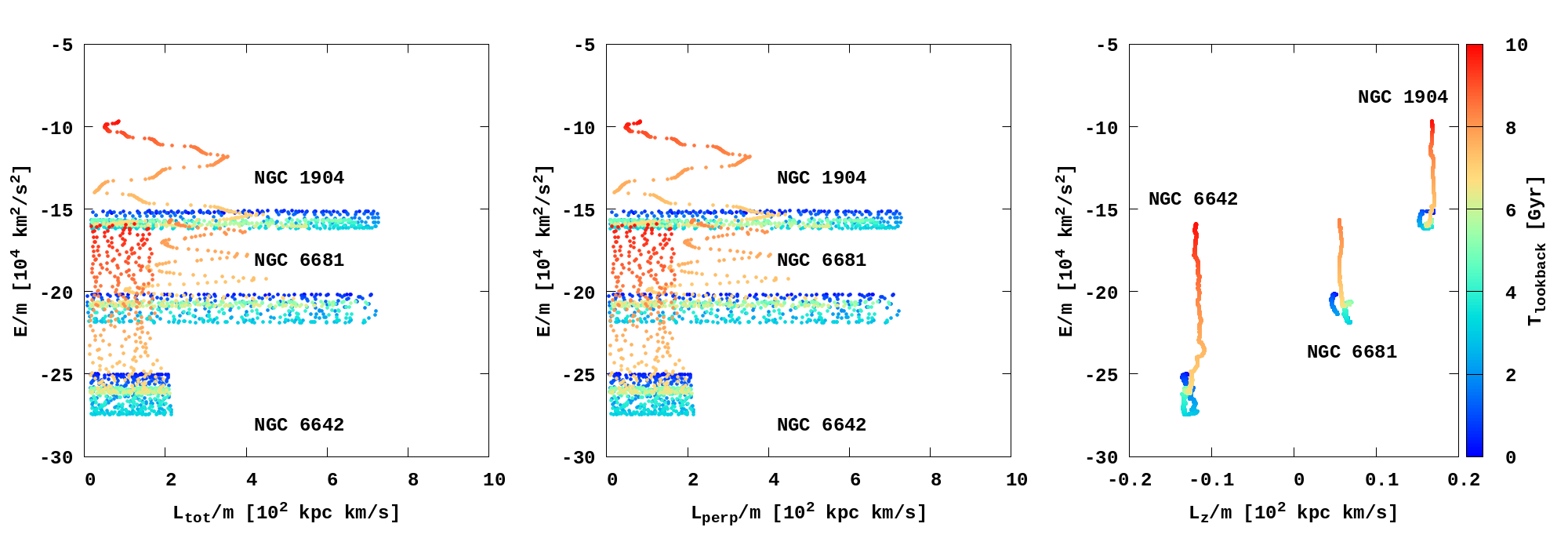}
\caption{{\it From left to right:} total angular momentum ($L_{\rm tot}/m$) versus total energy ($E/m$), the perpendicular component of the angular momentum ($L_{\rm perp}/m$) versus total energy and $z$-th component of the angular momentum ($L_{\rm z}/m$) versus total energy for GCs in {\tt \#411321} TNG-TVP.}
\label{fig:to-moms}
\end{figure*}

We analysed the phase-space distribution of 10 selected GCs in three combinations of the coordinates: total specific angular momentum~($L_{\rm tot}/m$) versus total specific energy ($E/m$), the perpendicular component of the specific angular momentum~($L_{\rm perp}/m$) versus $E/m$ and the $z$-component of the specific angular momentum~($L_{\rm z}/m$) versus $E/m$. In the Fig. \ref{fig:to-moms}, we present the three typical cases for NGC 6642, NGC 1904 and NGC 6681 in the {\tt \#411321} TNG-TVP.

In Fig.~\ref{fig:to-moms}, we can see the quite predictable phase space behaviour of our selected GCs. As expected, based on the generally axisymmetric nature of our TNG-TVP, the $L_{\rm z}/m$ should conserve over the integration time. The $E/m$ generally changes according to the mass grow/loss of the model Galactic potential. Both $L_{\rm tot}/m$ and $L_{\rm perp}/m$ are non-conserving quantities, as we can see in the figures. The small relative changes of the $L_{\rm z}/m$ values of our objects over time can be explained by the small deviation from axial symmetry due to the mass component changes of the TNG-TVP over the orbital period of the individual GCs. 

%-------------------------------------------------------------------------%
%-------------------------------------------------------------------------%

%%%%%%%%%%%%%%%%%%%%%%%%%%%%%%%%%%%%%%%%%%%%%%%%%%%%%%%%%%%%%%%%%%%%%%%%%%%%%%%%
\section{Conclusions} \label{sec:conc}
%%%%%%%%%%%%%%%%%%%%%%%%%%%%%%%%%%%%%%%%%%%%%%%%%%%%%%%%%%%%%%%%%%%%%%%%%%%%%%%%

In this study, we investigated the orbits of 147 Milky Way globular clusters on a cosmological timescale using data from \textit{Gaia} Data Release 3, after excluding GCs with large errors in their proper motions and radial velocities. We employed time-varying external potentials from IllustrisTNG-100 simulations, selecting four potentials that best match the Milky Way, and generated 1000 realisations for each of them. The GCs were backwards integrated using high-order parallel dynamical \textit{N}-body code $\varphi$-GPU. Our main objective was to identify GCs with close passages near the Galactic centre and gain insights into the evolution of the Milky Way's GC subsystem. The main results of our study are summarised below.

\begin{itemize}

\item We found 98 GCs with close passages near the GalC in all four TNG-TVPs.

\item The number of interactions of GCs with the Galactic centre per Gyr varies from 3-4 at distances less than 50~pc to 5-6 at 80~pc for each of the TNG-TVP external potentials.

\item Analysing the same interaction rates with time binning of 1 Gyr, we found almost constant interaction rates in the 1-8 Gyr lookback time interval, which decrease after 8 Gyr due to changes in the masses of the disk and halo components and their scale parameters.

\item We identified 10 GCs, including NGC 6401, Pal 6, NGC 6681, NGC 6712, NGC 6287, NGC 6642, NGC 6981, HP 1, NGC 1904, and NGC 362, with a high probability of close passages near the Galactic centre in all four TNG-TVPs, particularly the first six with a probability of around 100\%.

\end{itemize}

Our main results indicate that the maximum interaction frequency of GCs with the Galactic centre in the Milky Way is a few dozen passages per Gyr within the 100 pc central zone. However, this low interaction frequency cannot fully explain the relatively high mass (of the order of 10$^7$ M$_\odot$) of the Milky Way's NSC, if we only consider periodic capturing of stars from close-passing GCs. Therefore, we need to consider other scenarios as well, such as the full tidal destruction of some of the early GCs during interaction with the forming NSC and GalC. Our study provides valuable insights into the evolution of the Milky Way's GC subsystem by identifying a dozen GCs that are the most likely candidates to approach the Galactic centre.

%%%%%%%%%%%%%%%%%%%%%%%%%%%%%%%%%%%%%%%%%%%%%%%%%%%%%%%%%%%%%%%%%%%%%
\begin{acknowledgements}
%%%%%%%%%%%%%%%%%%%%%%%%%%%%%%%%%%%%%%%%%%%%%%%%%%%%%%%%%%%%%%%%%%%%%

The authors thank the anonymous referee for a very constructive report and suggestions that helped significantly improve the quality of the manuscript.

The work of MI, DK, TP and PB was funded by the Science Committee of the Ministry of Education and Science of the Republic of Kazakhstan (Grant No. AP14870501). 

MI acknowledges the Europlanet 2024 RI project funded by the European Union's Horizon 2020 Research and Innovation Programme (Grant agreement No. 871149).

The work of MI, MS and PB was also supported by the Volkswagen Foundation under the Trilateral Partnerships grant No.~97778 and special stipend No.~9B870 (for PB). 

MI, MS and PB acknowledge the support from the ACIISI, Consejer\'{i}a de Econom\'{i}a, Conocimiento y Empleo del Gobierno de Canarias and the European
Regional Development Fund (ERDF) under grant with reference PROID2021010044.

%The work of PB, MI and MS was supported under the special program of the NRF of Ukraine Leading and Young Scientists Research Support - ”Astrophysical Relativistic Galactic Objects (ARGO): life cycle of active nucleus”, No.~2020.02/0346.

MI and PB acknowledges the support by Ministry of Education and Science of Ukraine under the collaborative grant M/32-23.05.2022.

This work has made use of data from the European Space Agency (ESA) mission GAIA (\url{https://www.cosmos.esa.int/gaia}), processed by the GAIA Data Processing and Analysis Consortium (DPAC, \url{https://www.cosmos.esa.int/web/gaia/dpac/consortium}). Funding for the DPAC has been provided by national institutions, in particular the institutions participating in the GAIA Multilateral Agreement. TP acknowledges the support by the Science and Technology Facilities Council (STFC) Grant Number ST/W000903/1.
\end{acknowledgements}

%%%%%%%%%%%%%%%%%%%%%%%%%%%%%%%%%%%%%%%%%%%%%%%%%%%%%%%%%%%%%%%%%%%%%
\bibliographystyle{mnras}  % style aa.bst
\bibliography{gc-coll}   % your references Yourfile.bib
%%%%%%%%%%%%%%%%%%%%%%%%%%%%%%%%%%%%%%%%%%%%%%%%%%%%%%%%%%%%%%%%%%%%%

%%%%%%%%%%%%%%%%%%%%%%%%%%%%%%%%%%%%%%%%%%%%%%%%%%%%%%%%%%%%%%%%%%%%%
\begin{appendix}
%%%%%%%%%%%%%%%%%%%%%%%%%%%%%%%%%%%%%%%%%%%%%%%%%%%%%%%%%%%%%%%%%%%%%
\section{Visualisation of GCs orbits that have close interaction with Galactic centre.}\label{app:det-orb}
%%%%%%%%%%%%%%%%%%%%%%%%%%%%%%%%%%%%%%%%%%%%%%%%%%%%%%%%%%%%%%%%%%%%%
We present orbital evolution for selected 10 GCs from Table~\ref{tab:bh-prob}. Each GC is represented in four TNG-TVP external potentials {\tt \#411321}, {\tt \#441327}, {\tt \#451323} and {\tt \#462077}. 
The {\tt \#411321-m} TNG-TVP where we added the SMBH influence we see only very insignificant changes in the GCs global orbits. We present here the three GCs with the most visible orbital changes, namely: NGC 362, HP 1 and NGC 6401. The orbital evolution is presented in three planes (X, Y), (X, Z) and (R, Z) (where R is the planar Galactocentric radius). The total time of integration is 10 Gyr lookback time and is shown by the colour line.

%-------------------------------------------------------------------------%
\begin{figure*}[ht]
\centering
\includegraphics[width=0.75\linewidth]{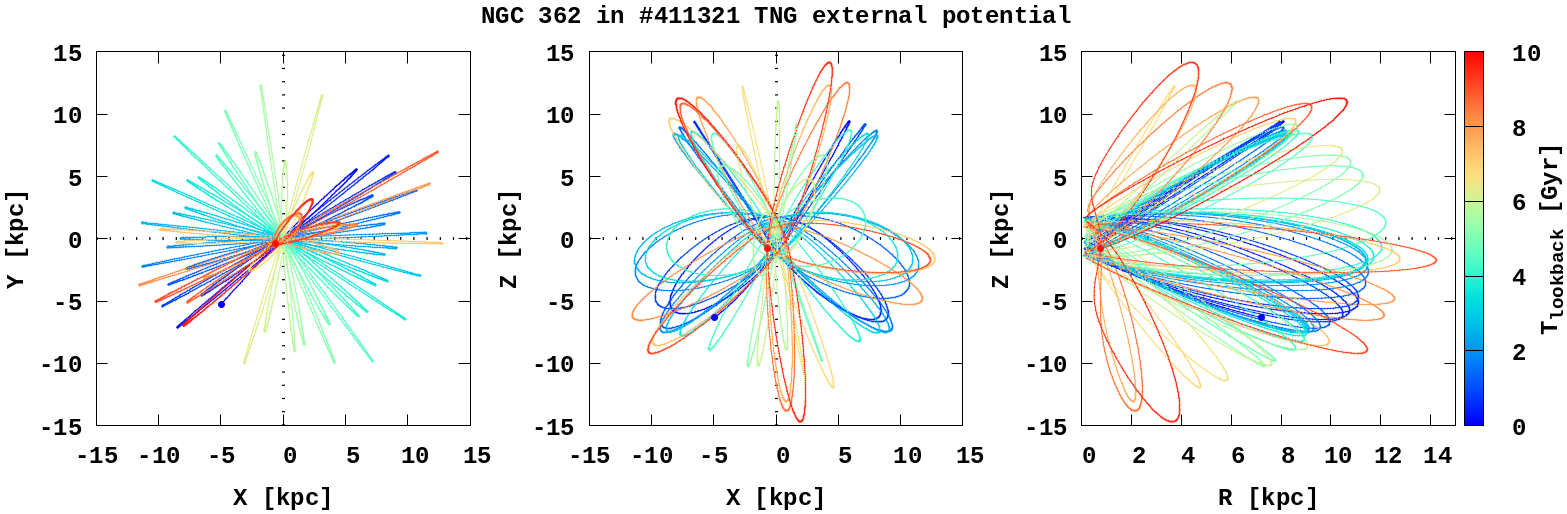}
\includegraphics[width=0.75\linewidth]{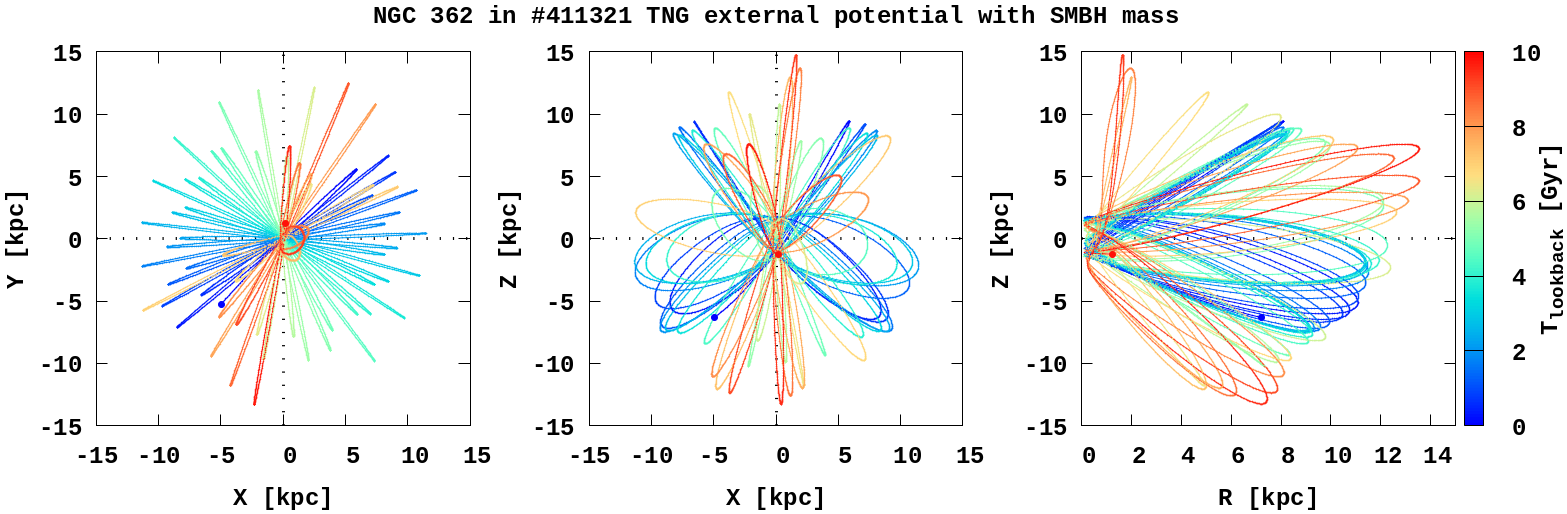}
\includegraphics[width=0.75\linewidth]{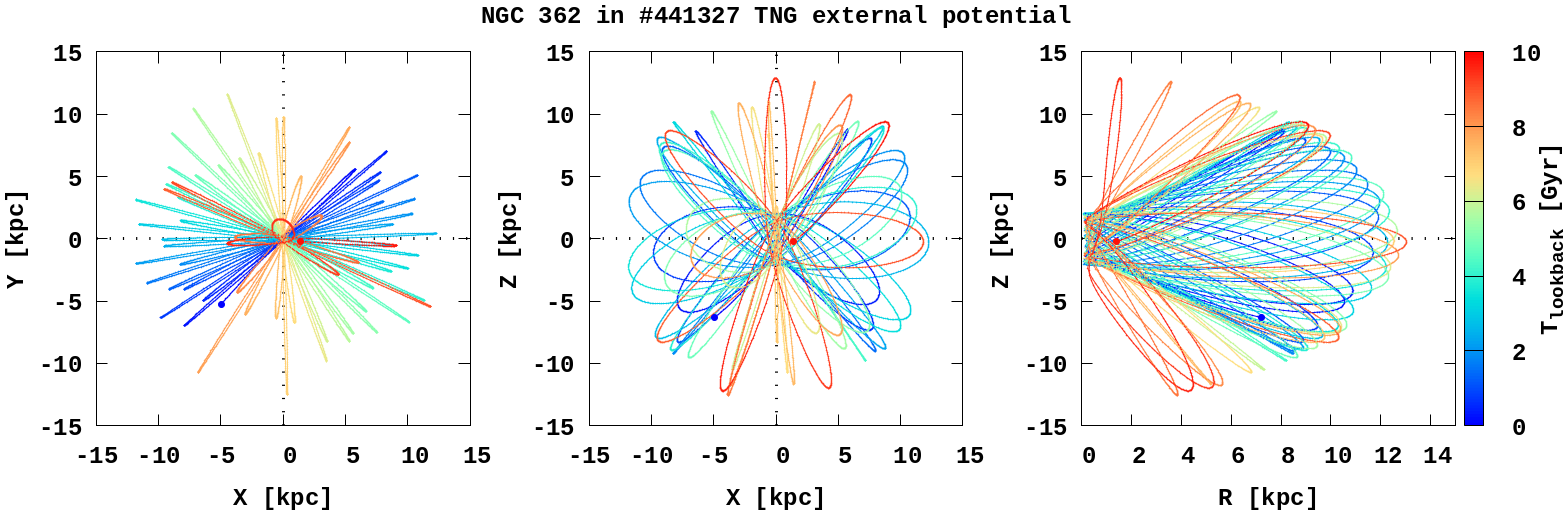}
\includegraphics[width=0.75\linewidth]{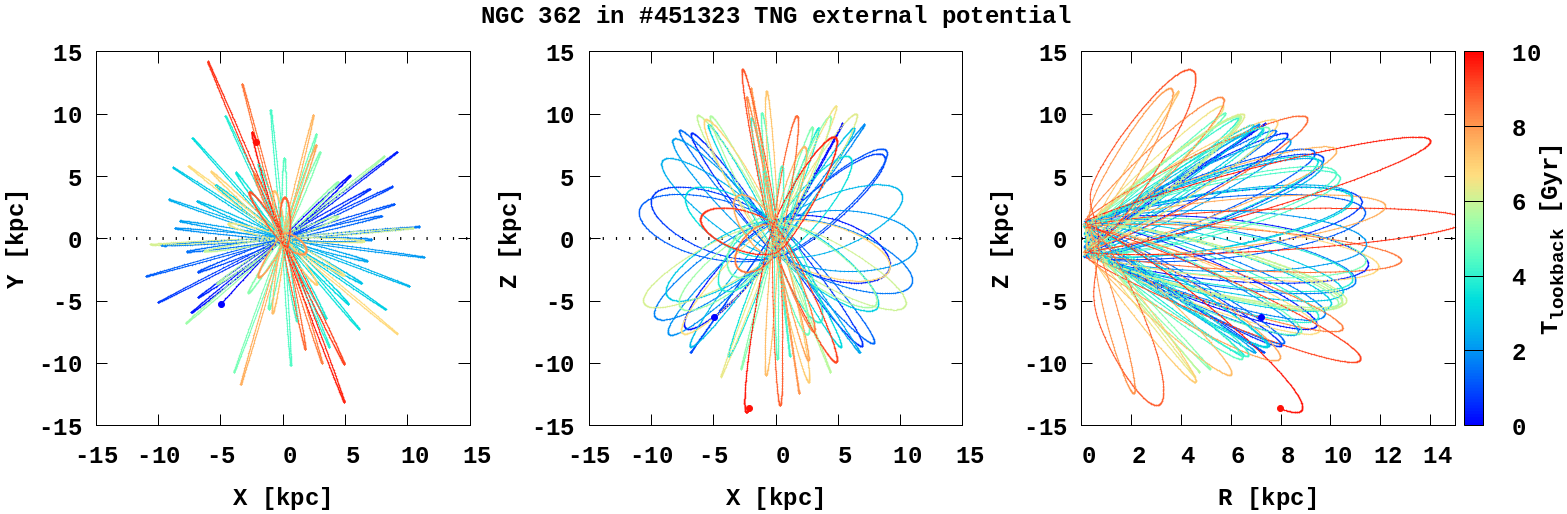}
\includegraphics[width=0.75\linewidth]{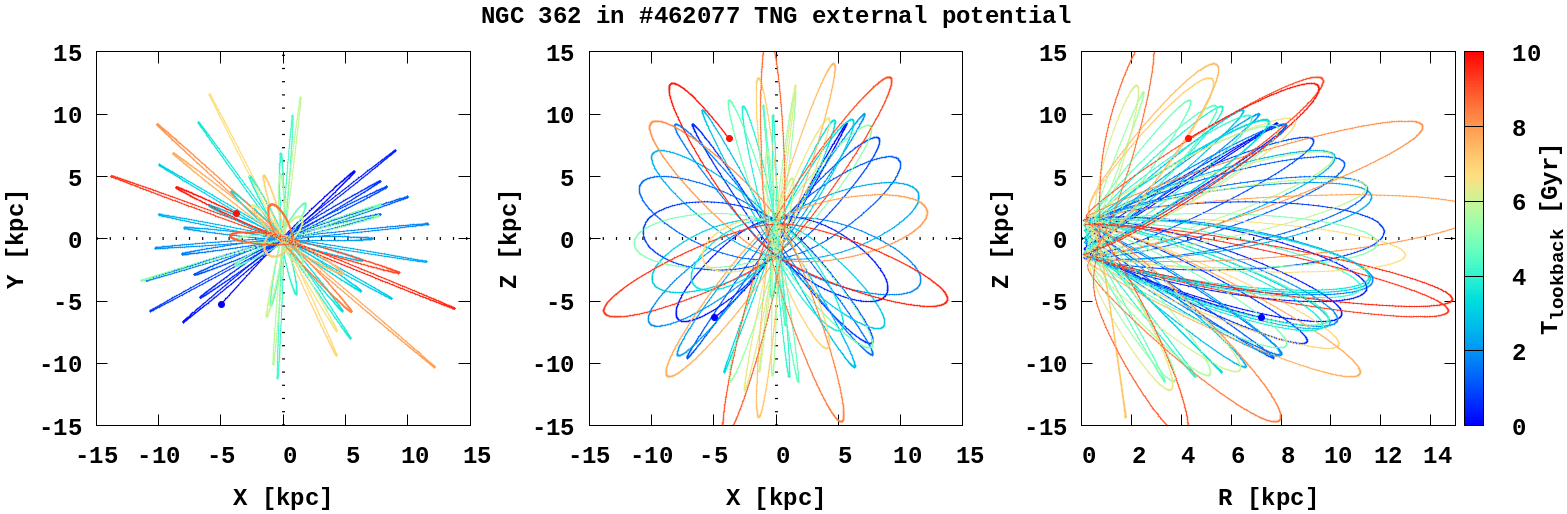}
\caption{NGC~362 orbital changes in four TNG-TVP external potentials and one with SMBH mass. \textit{From top to down}: {\tt \#411321}, {\tt \#411321-m}, {\tt \#441327}, {\tt \#451323} and {\tt \#462077}. The orbital evolution presents in three planes ($X$, $Y$), ($X$, $Z$) and ($R$, $Z$) (where $R$ is the planar Galactocentric radius). The total time of integration is 10 Gyr lookback time represent by colour line.}
\label{fig:orb1}
\end{figure*}
%-------------------------------------------------------------------------%

%-------------------------------------------------------------------------%
\begin{figure*}[ht]
\centering
\includegraphics[width=0.9\linewidth]{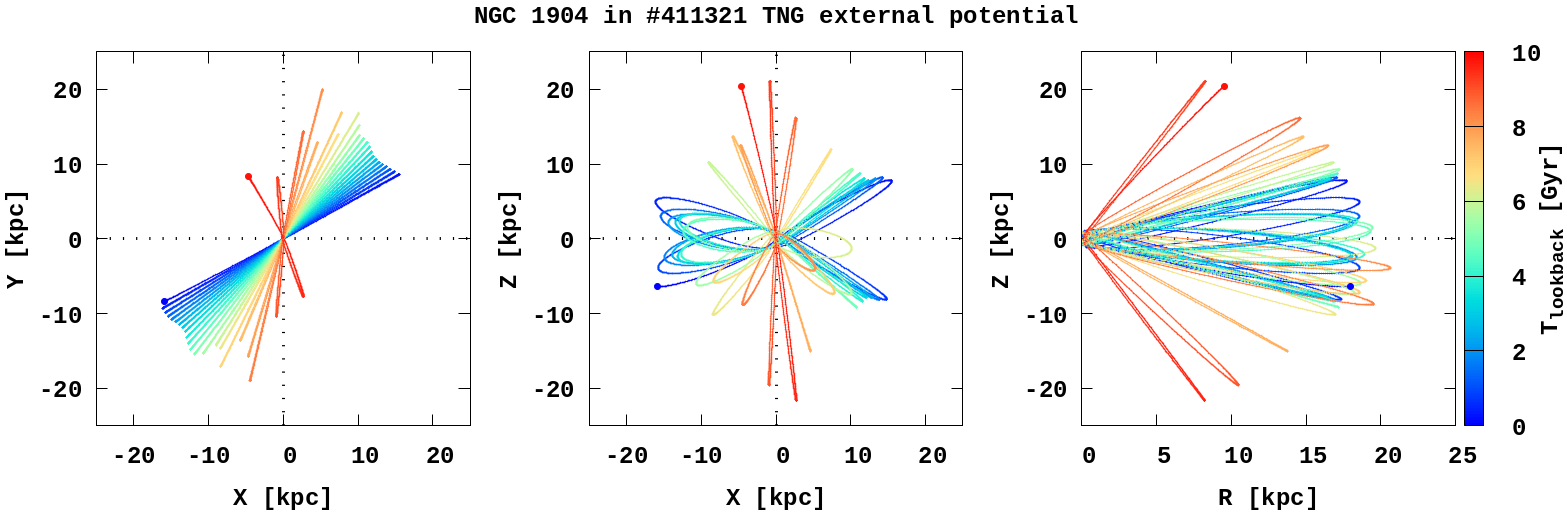}
\includegraphics[width=0.9\linewidth]{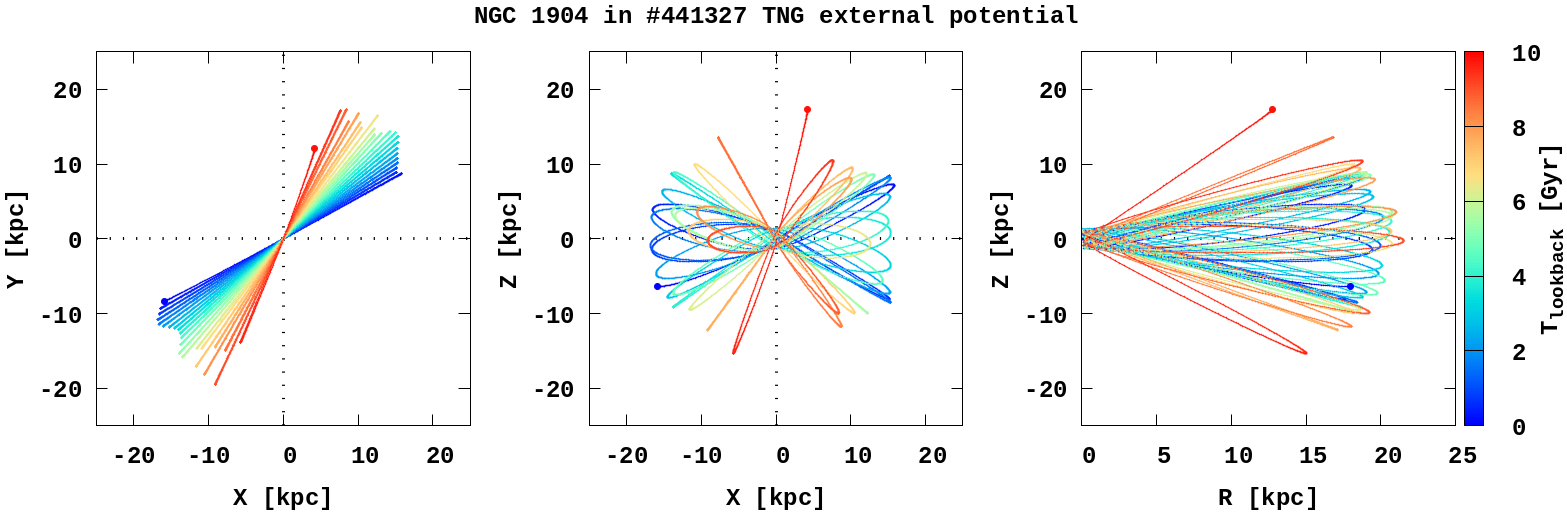}
\includegraphics[width=0.9\linewidth]{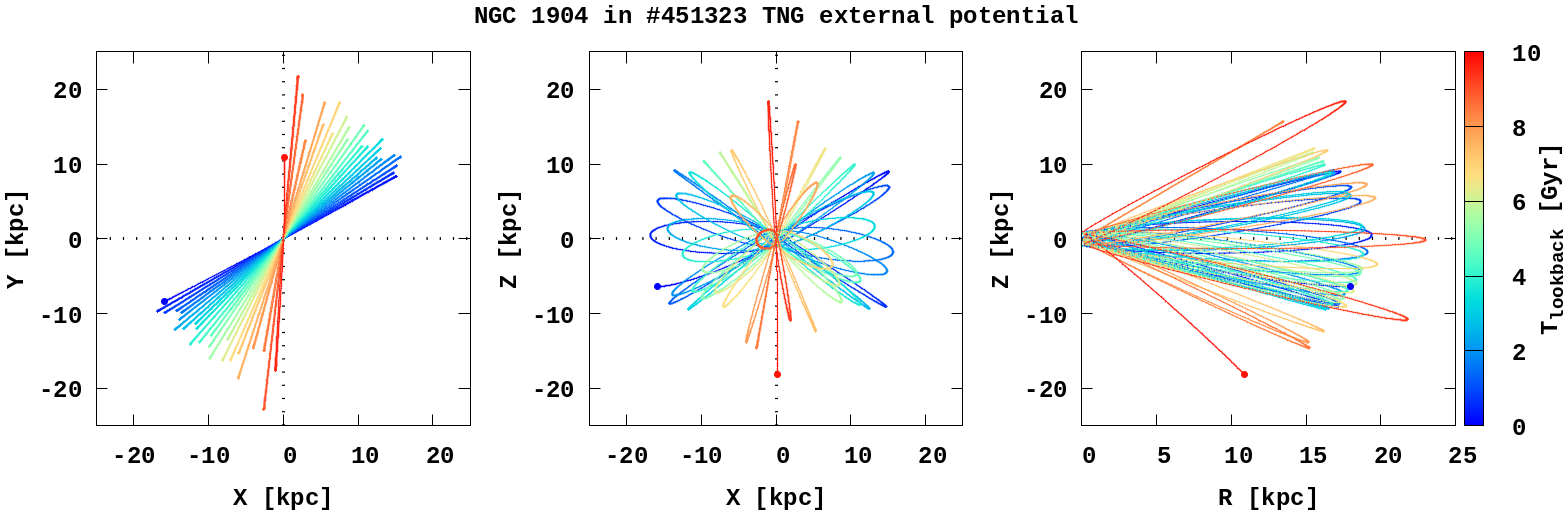}
\includegraphics[width=0.9\linewidth]{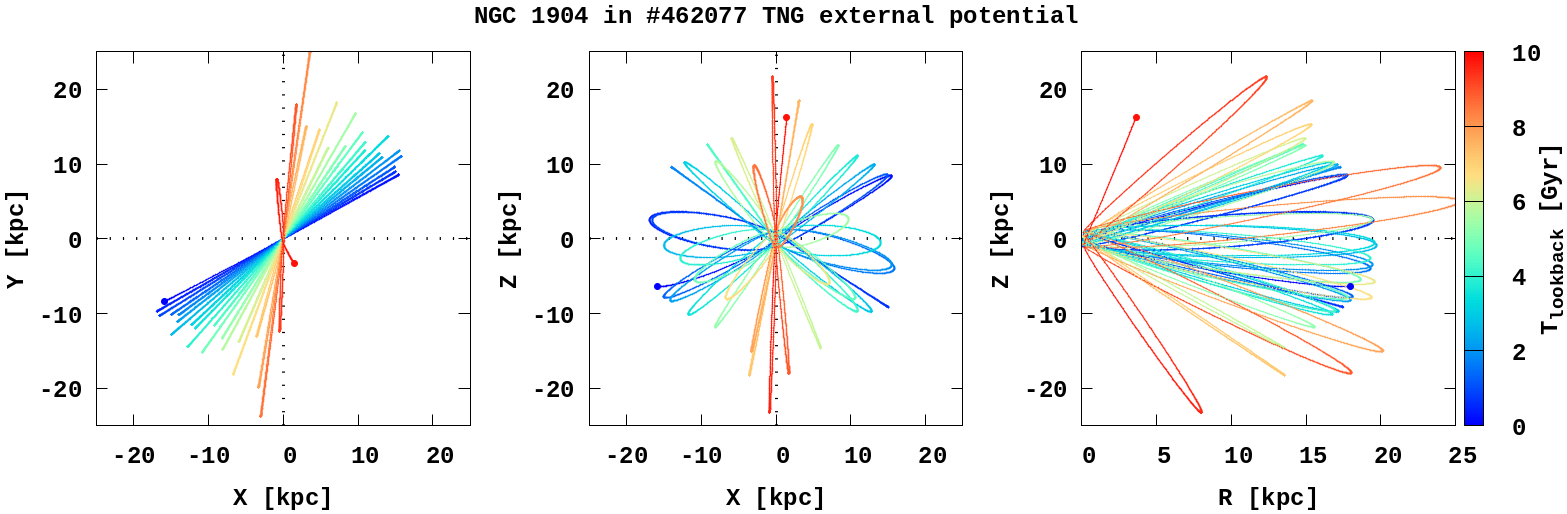}
\caption{The same as in Fig.~\ref{fig:orb1} for NGC 1904 but without {\tt \#411321-m} TNG-TVP.}
\label{fig:orb2}
\end{figure*}
%-------------------------------------------------------------------------%

%-------------------------------------------------------------------------%
\begin{figure*}[ht]
\centering
\includegraphics[width=0.9\linewidth]{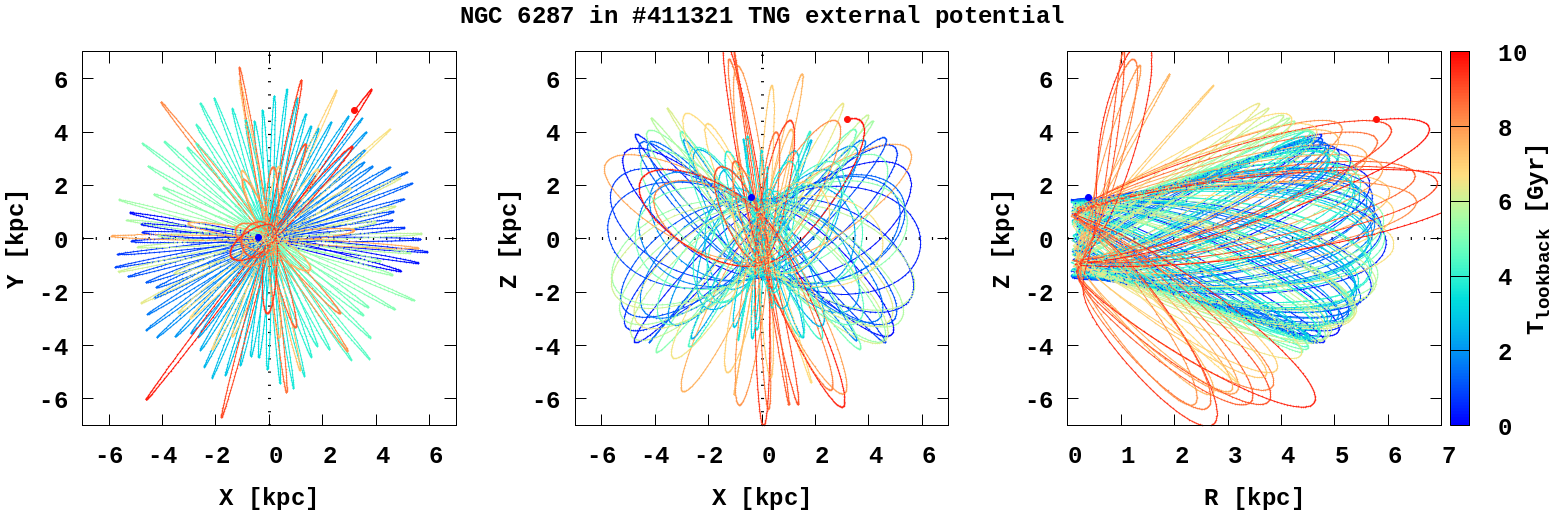}
\includegraphics[width=0.9\linewidth]{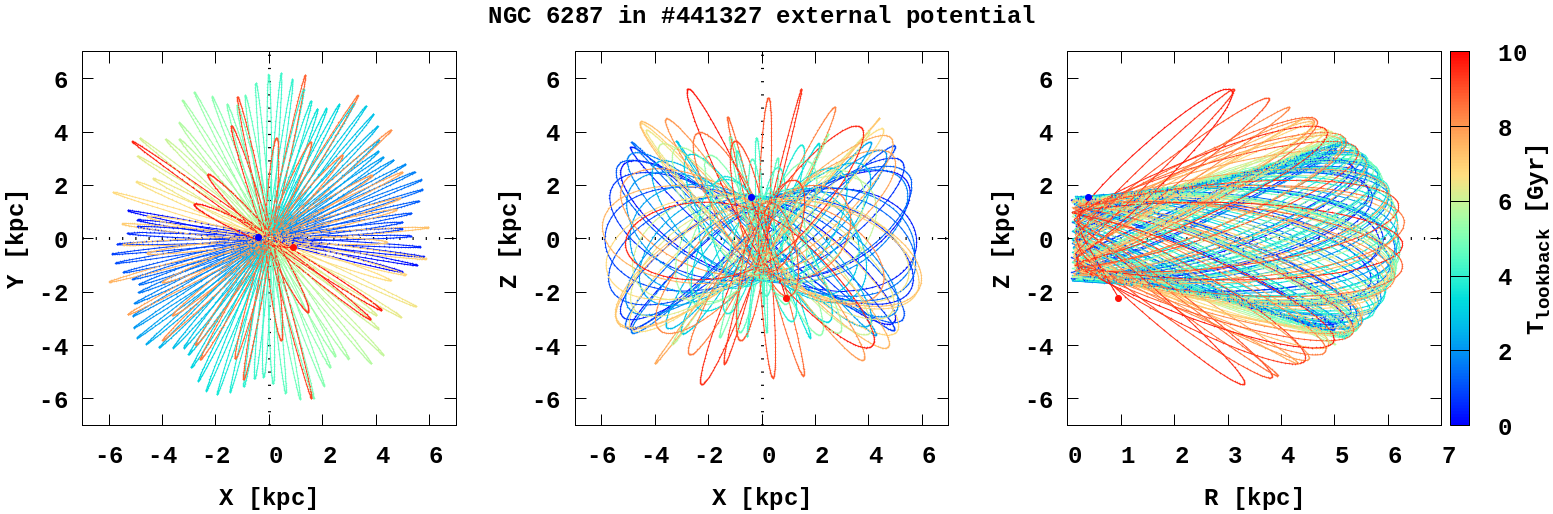}
\includegraphics[width=0.9\linewidth]{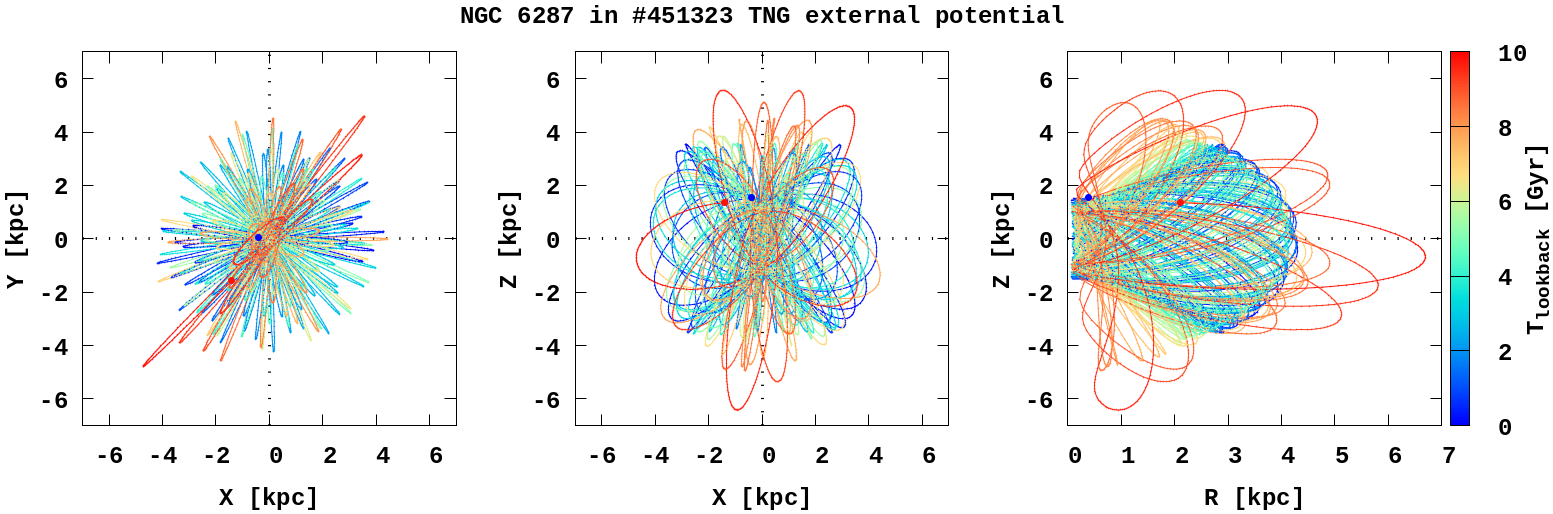}
\includegraphics[width=0.9\linewidth]{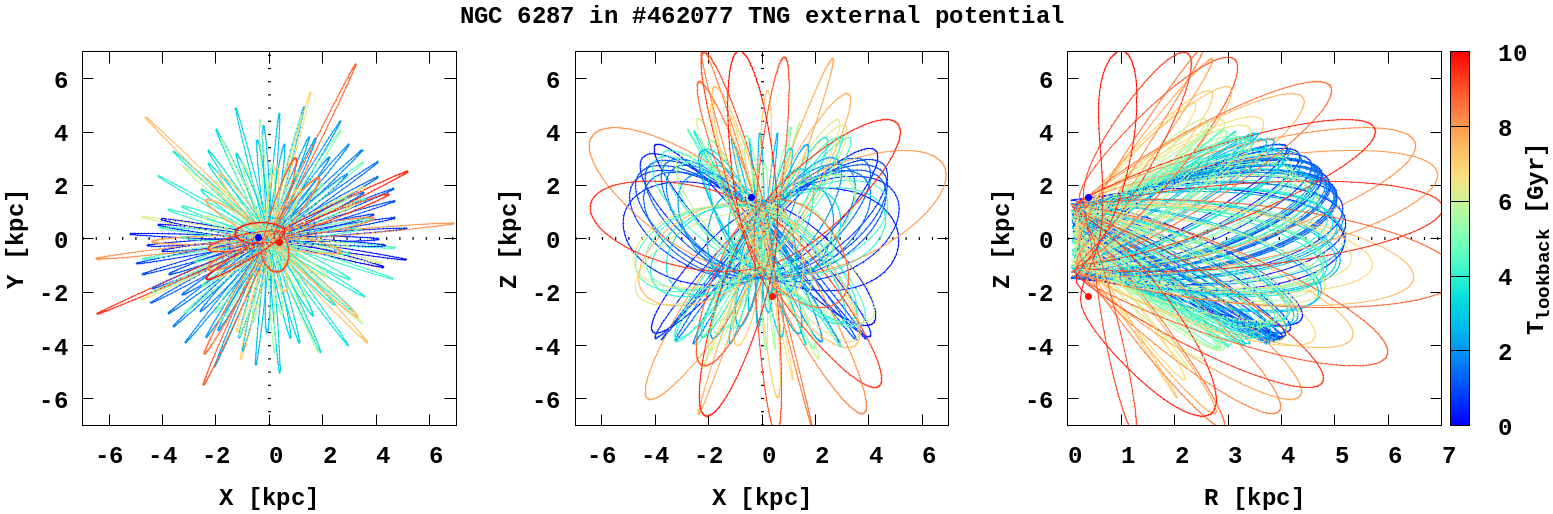}
\caption{The same as in Fig.~\ref{fig:orb1} for NGC 6287 but without {\tt \#411321-m} TNG-TVP.}
\label{fig:orb3}
\end{figure*}
%-------------------------------------------------------------------------%

%-------------------------------------------------------------------------%
\begin{figure*}[ht]
\centering
\includegraphics[width=0.75\linewidth]{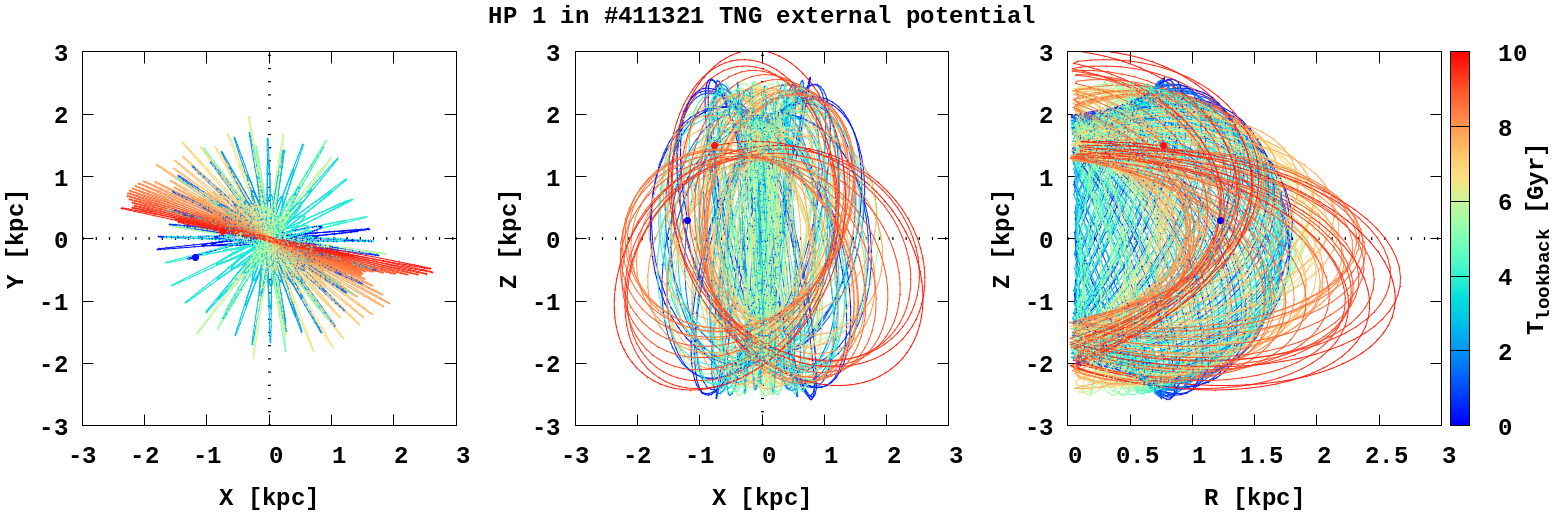}
\includegraphics[width=0.75\linewidth]{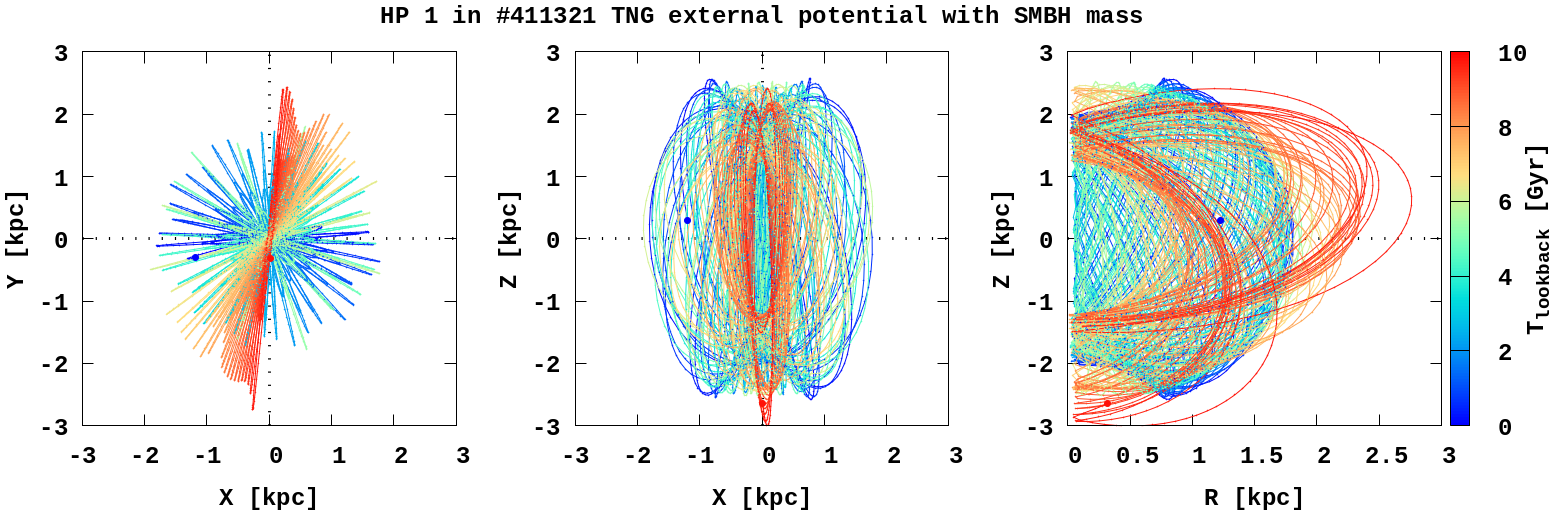}
\includegraphics[width=0.75\linewidth]{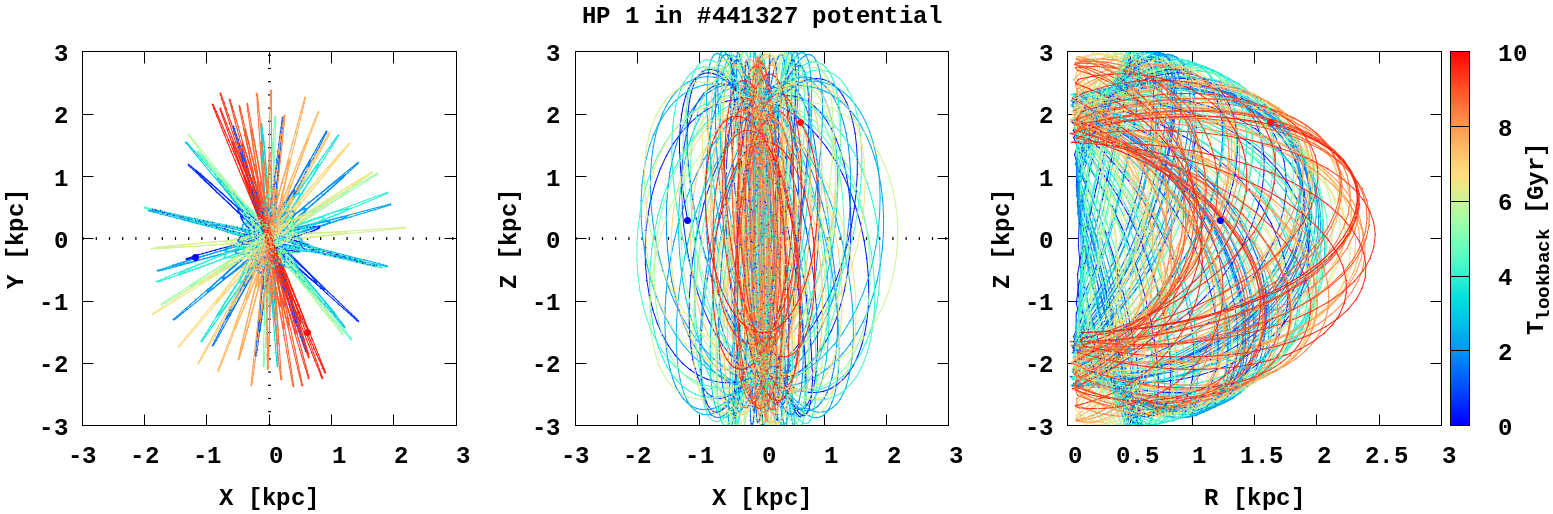}
\includegraphics[width=0.75\linewidth]{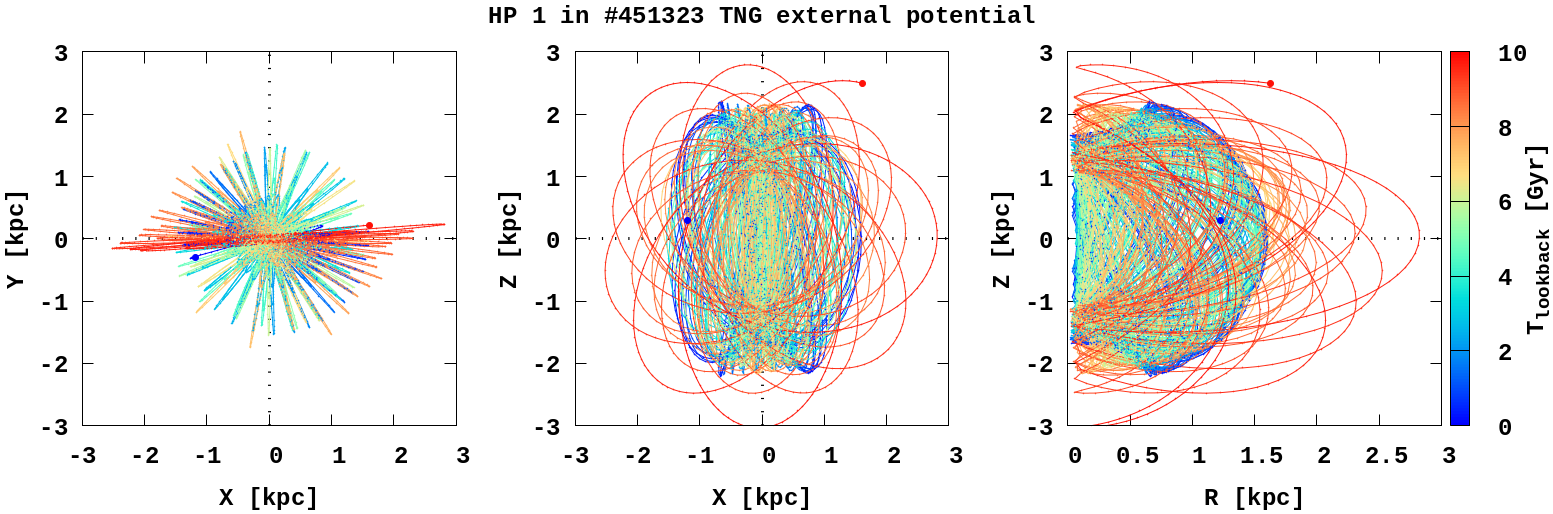}
\includegraphics[width=0.75\linewidth]{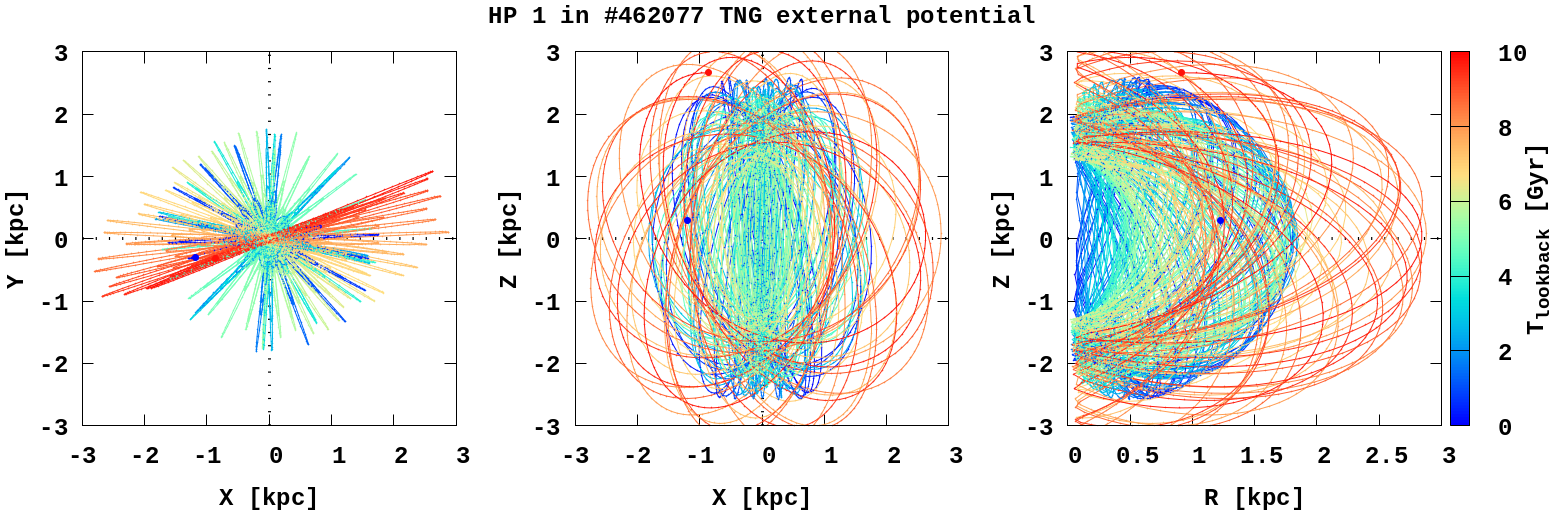}
\caption{The same as in Fig.~\ref{fig:orb1} for HP~1.}
\label{fig:orb4}
\end{figure*}
%-------------------------------------------------------------------------%

%-------------------------------------------------------------------------%
\begin{figure*}[ht]
\centering
\includegraphics[width=0.8\linewidth]{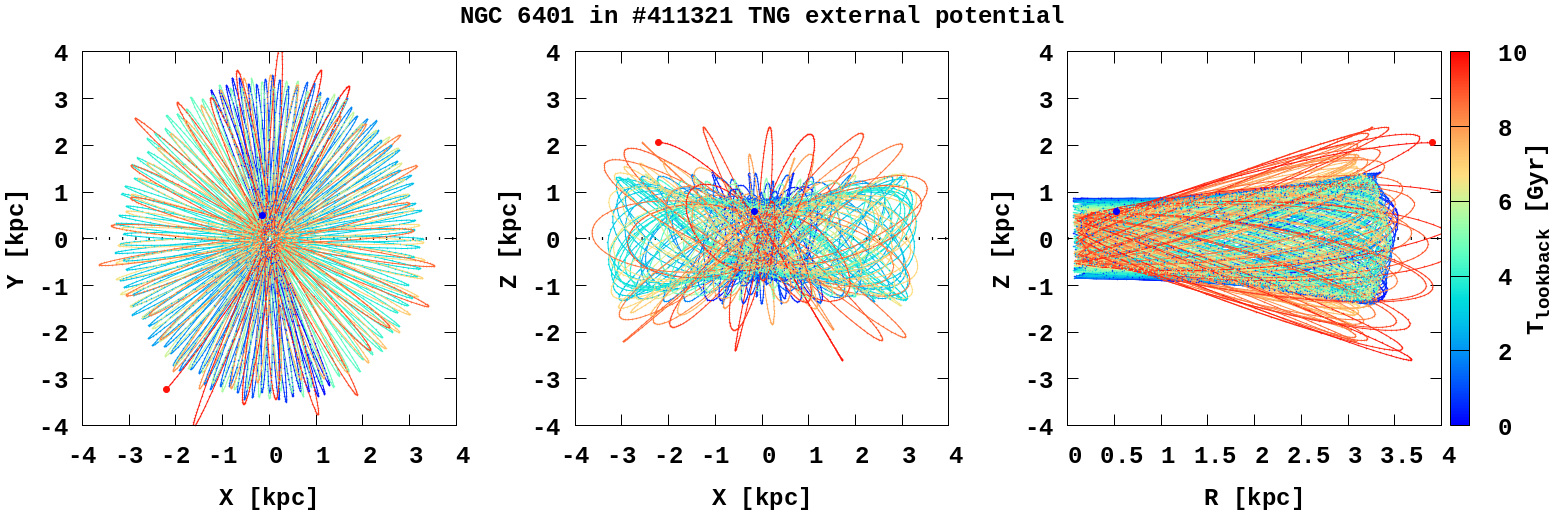}
\includegraphics[width=0.8\linewidth]{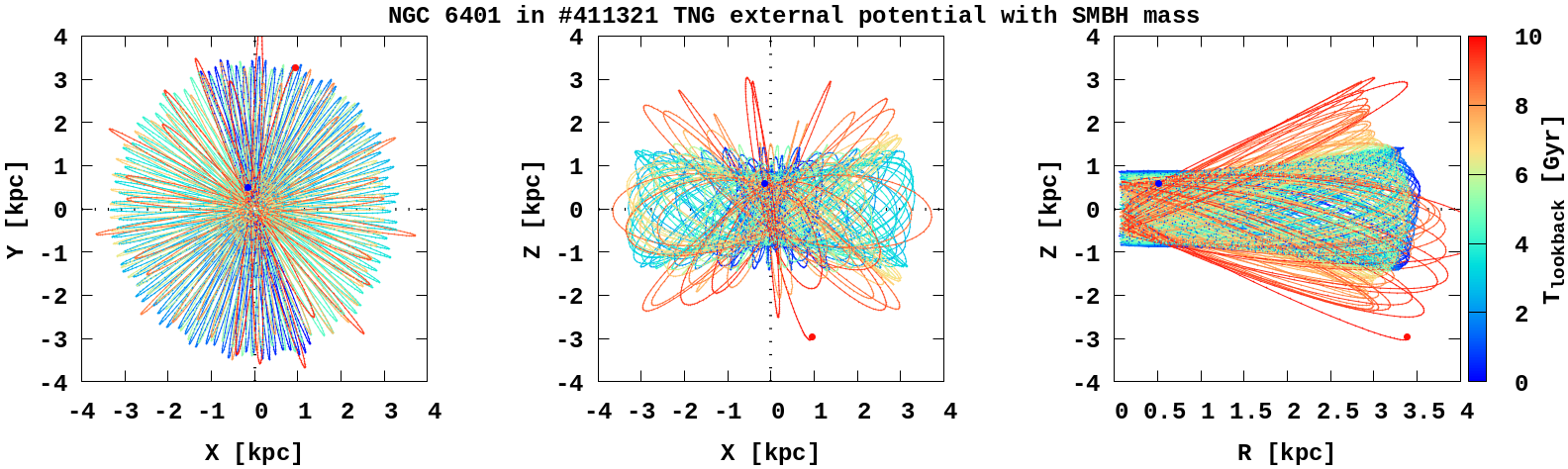}
\includegraphics[width=0.8\linewidth]{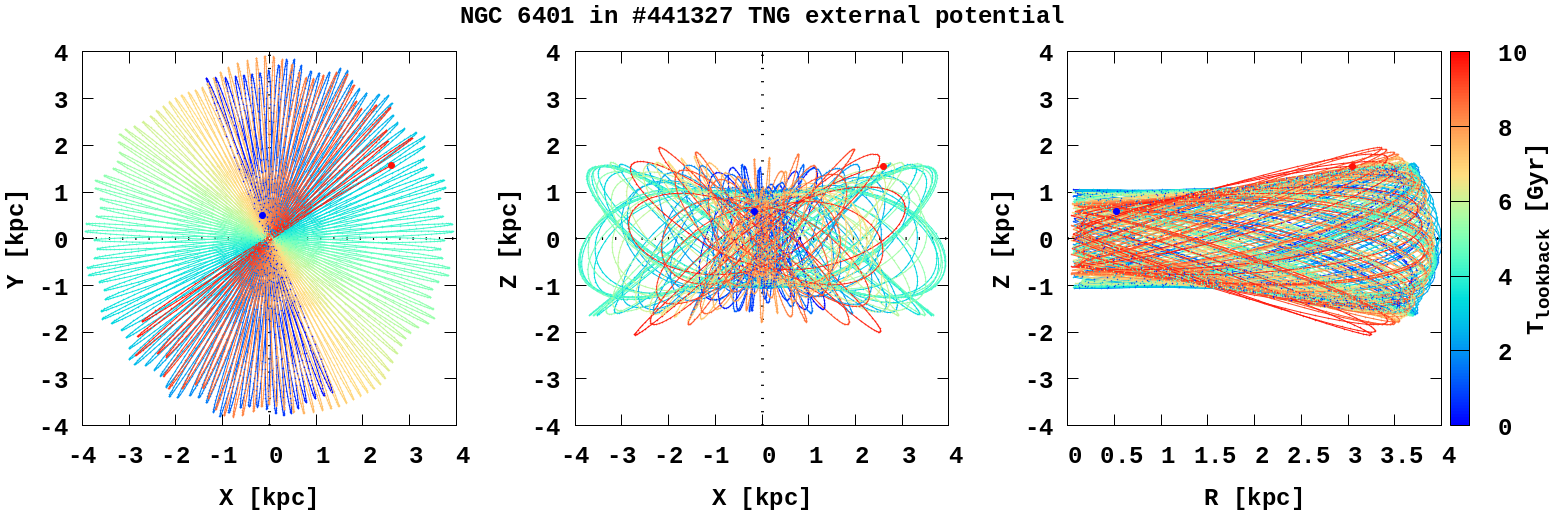}
\includegraphics[width=0.8\linewidth]{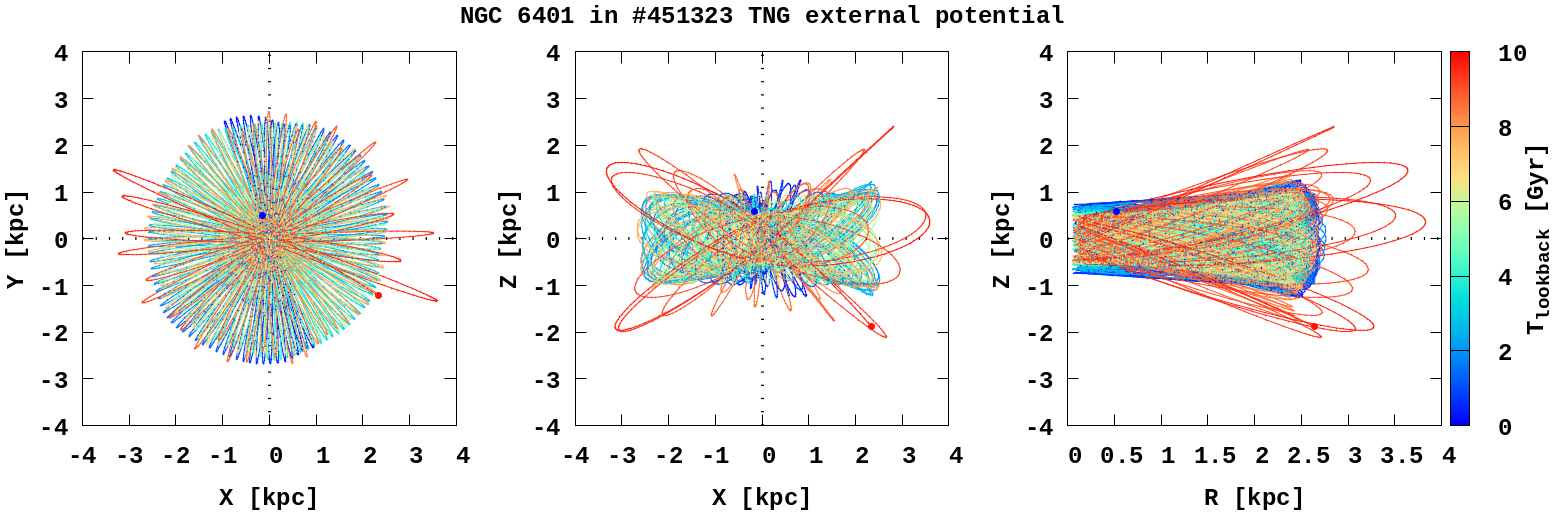}
\includegraphics[width=0.8\linewidth]{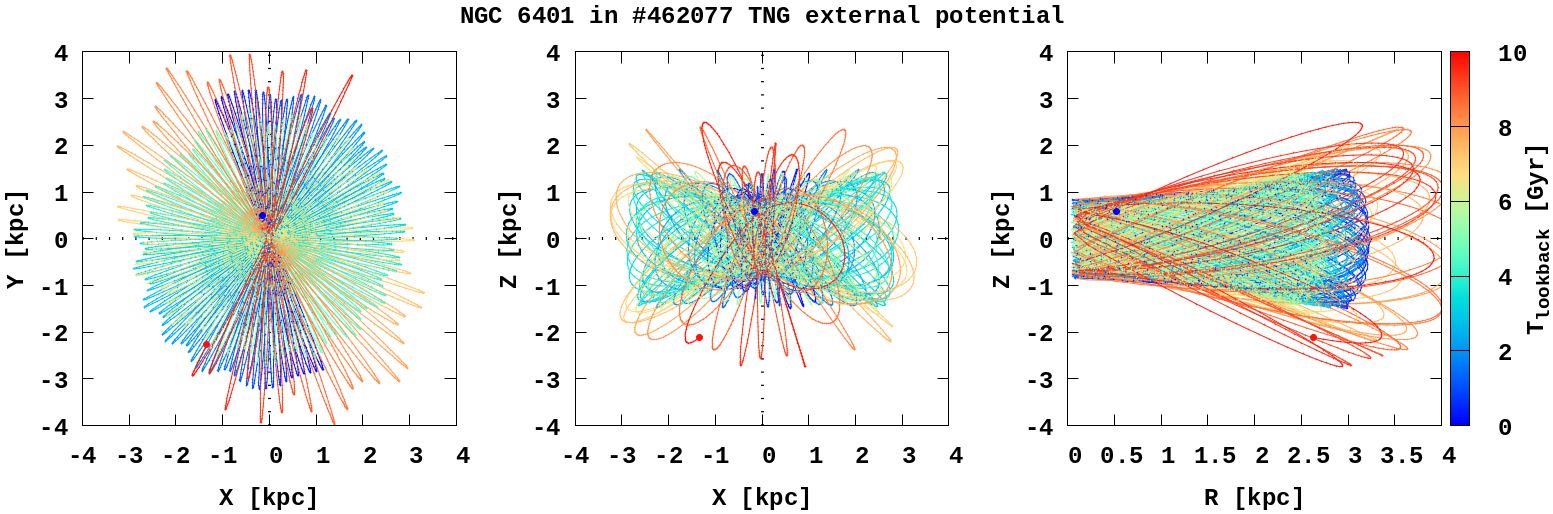}
\caption{The same as in Fig.~\ref{fig:orb1} for NGC 6401.}
\label{fig:orb5}
\end{figure*}
%-------------------------------------------------------------------------%

%-------------------------------------------------------------------------%
\begin{figure*}[ht]
\centering
\includegraphics[width=0.9\linewidth]{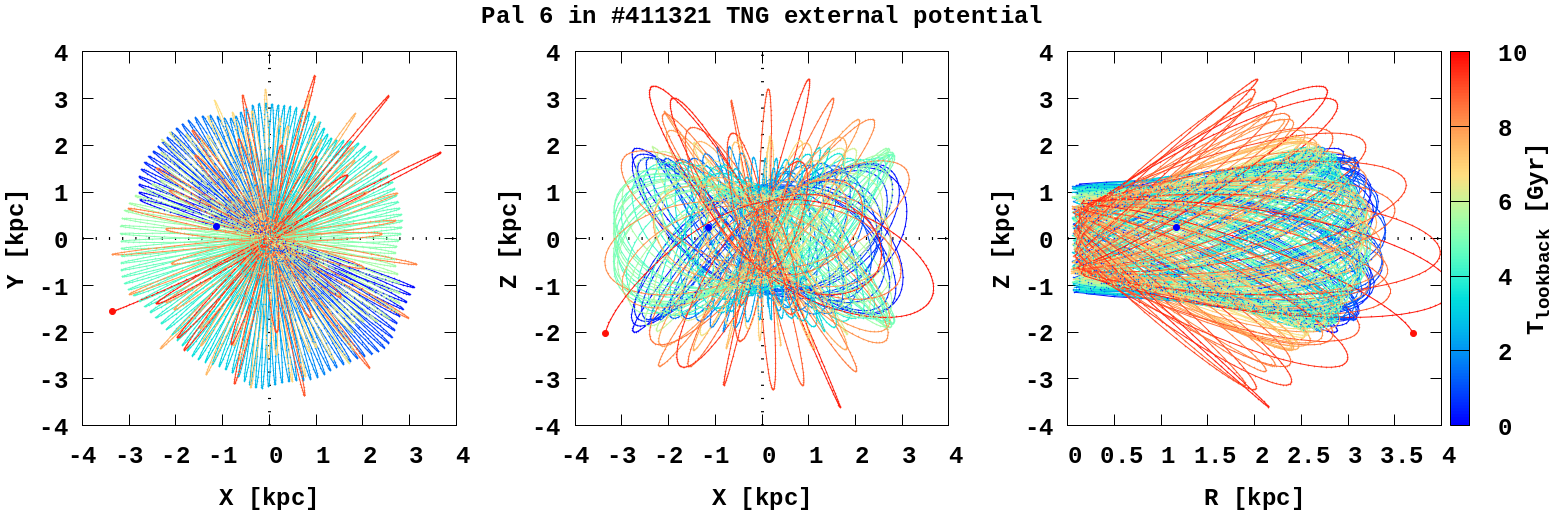}
\includegraphics[width=0.9\linewidth]{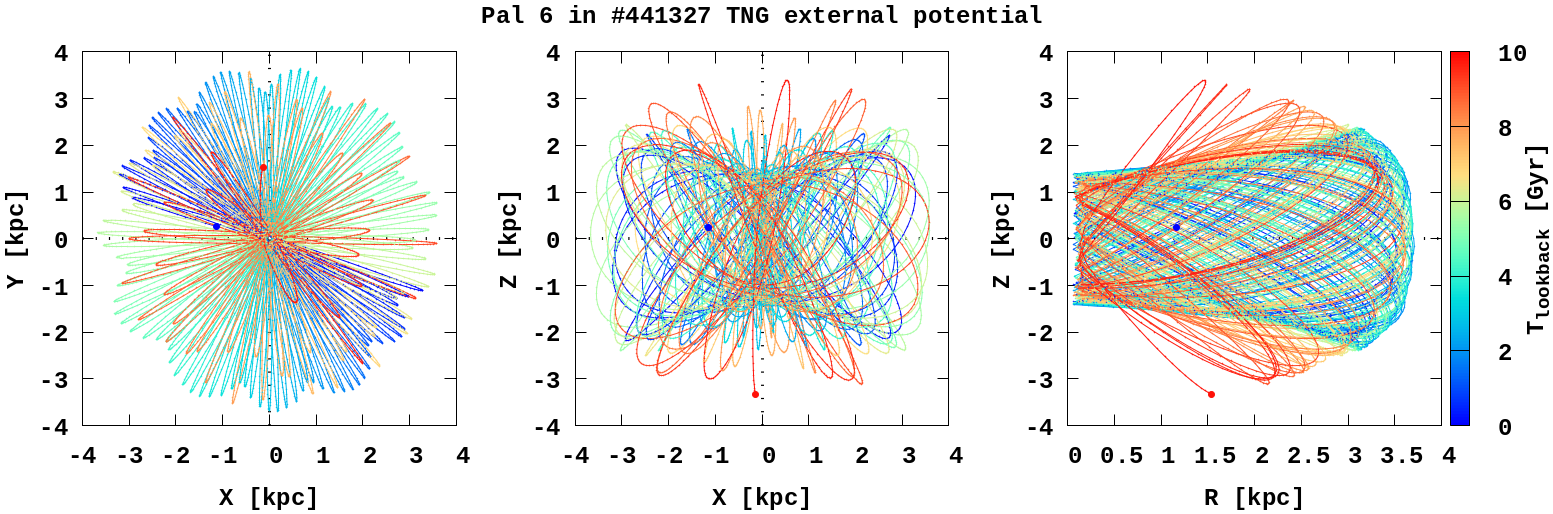}
\includegraphics[width=0.9\linewidth]{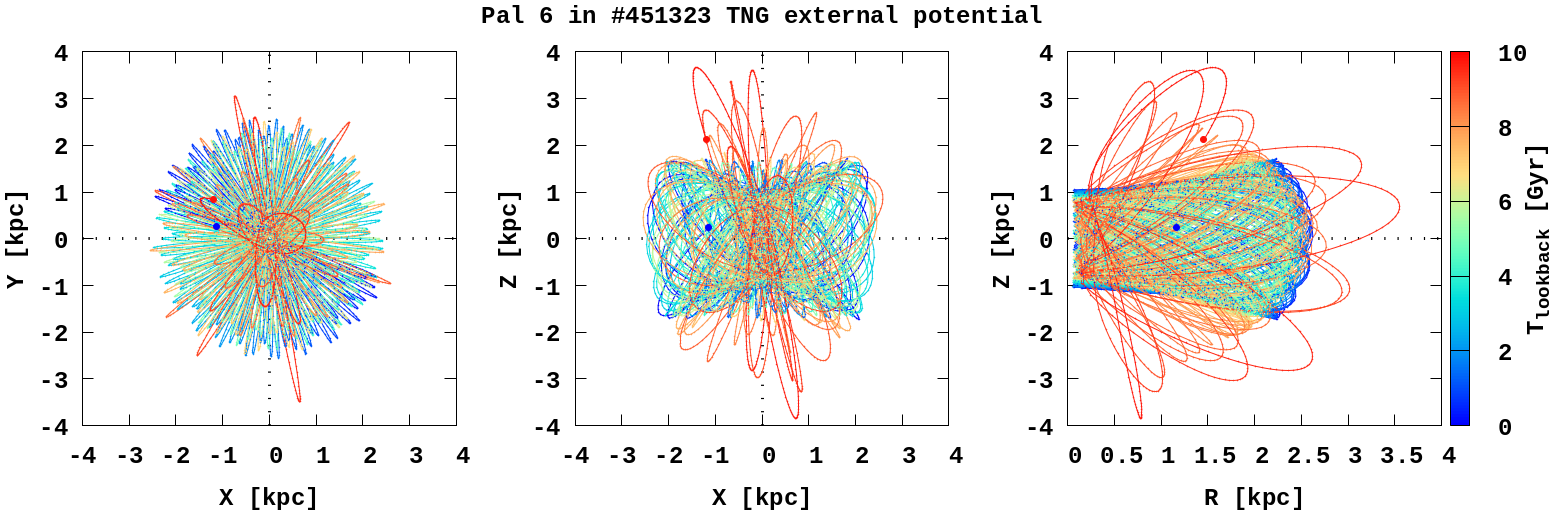}
\includegraphics[width=0.9\linewidth]{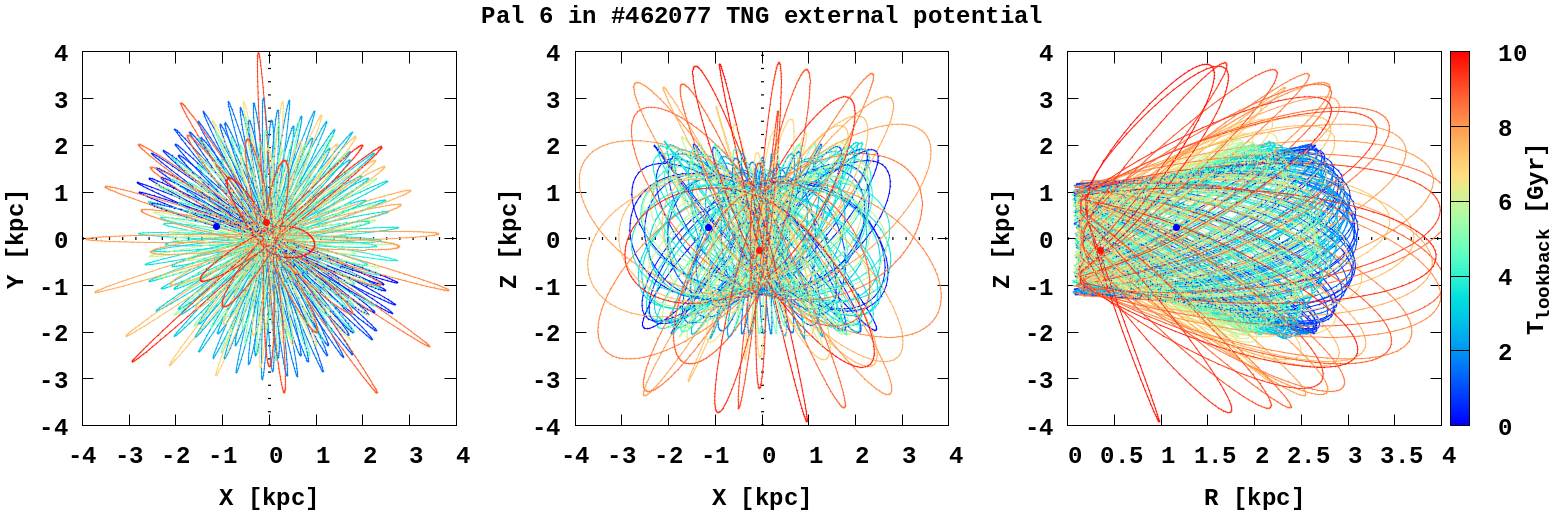}
\caption{The same as in Fig.~\ref{fig:orb1} for Pal 6 but without {\tt \#411321-m} TNG-TVP.}
\label{fig:orb6}
\end{figure*}
%-------------------------------------------------------------------------%

%-------------------------------------------------------------------------%
\begin{figure*}[ht]
\centering
\includegraphics[width=0.9\linewidth]{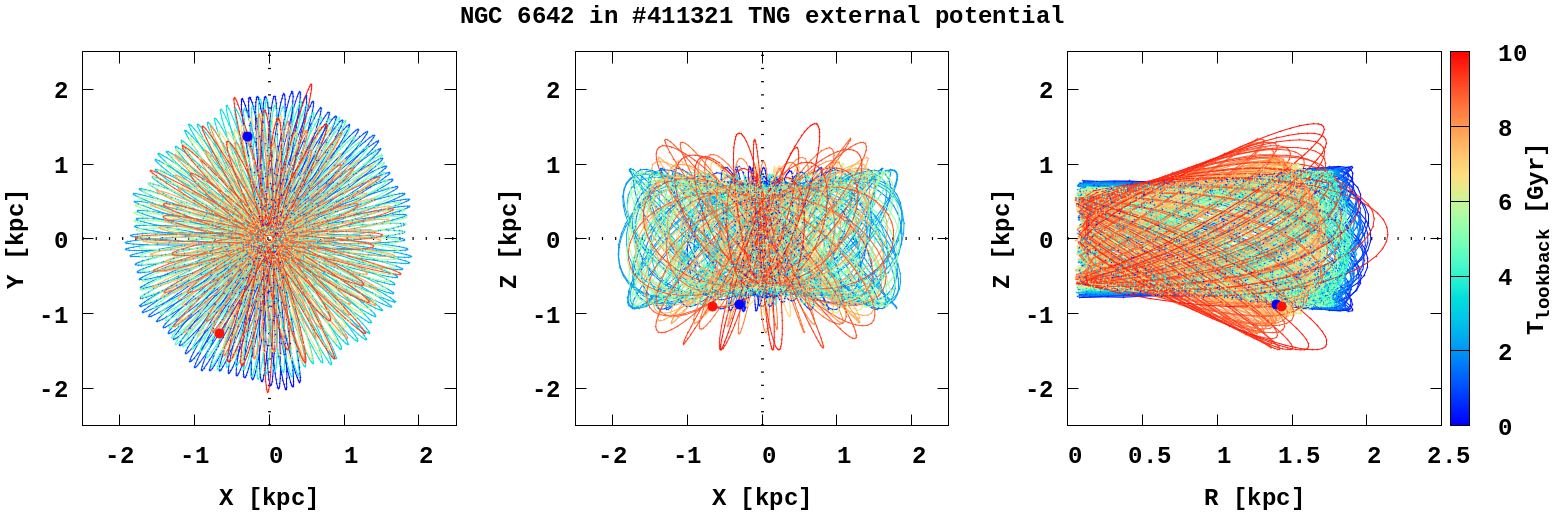}
\includegraphics[width=0.9\linewidth]{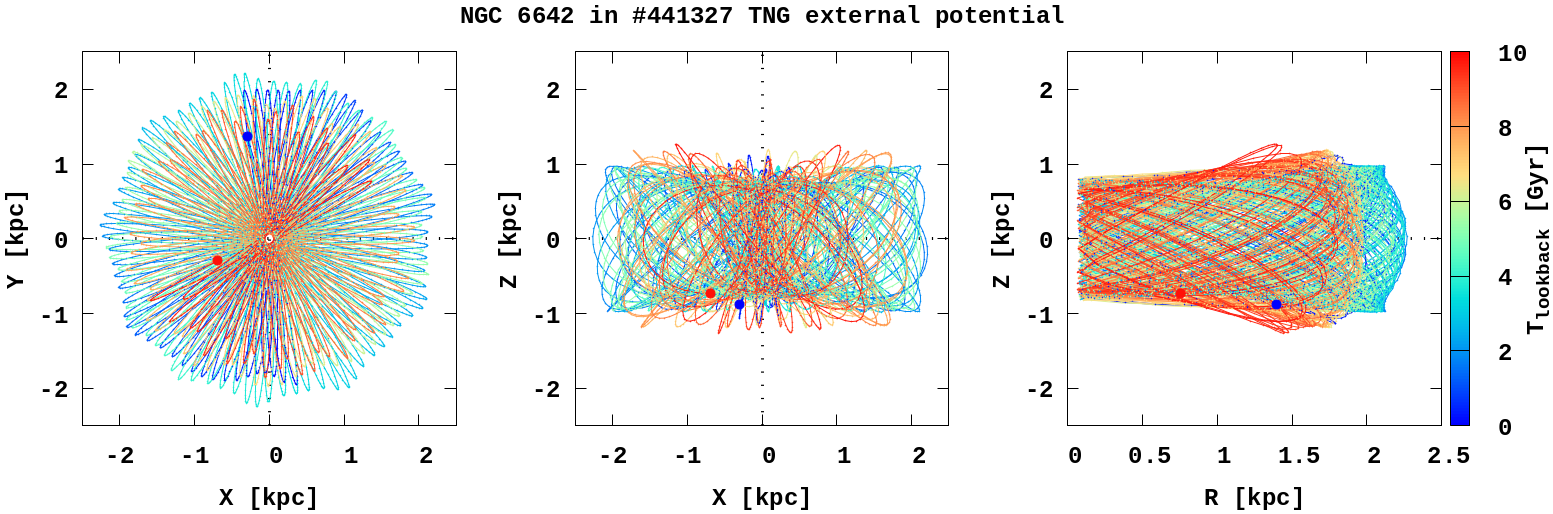}
\includegraphics[width=0.9\linewidth]{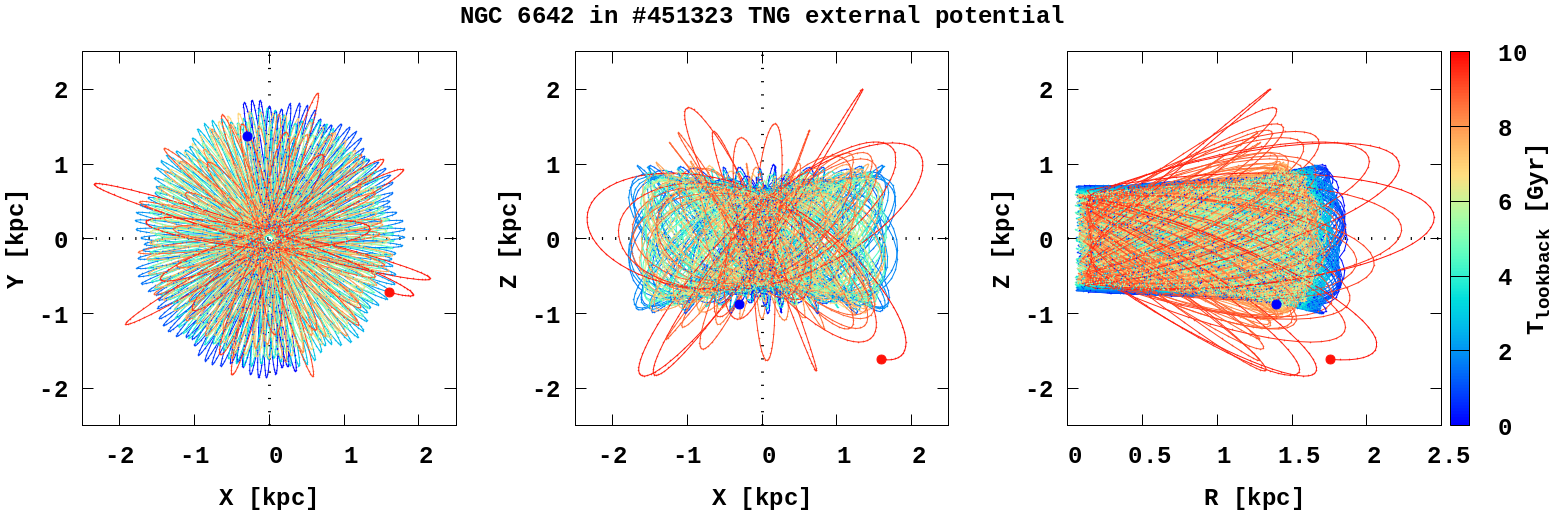}
\includegraphics[width=0.9\linewidth]{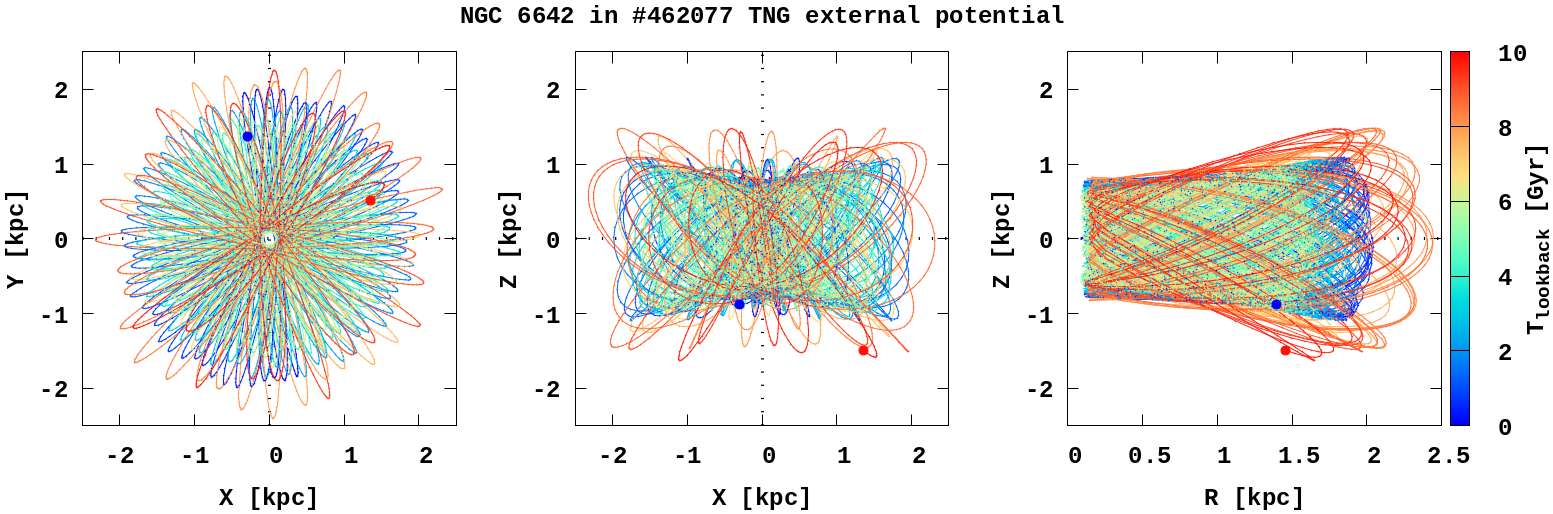}
\caption{The same as in Fig.~\ref{fig:orb1} for NGC 6642 but without {\tt \#411321-m} TNG-TVP.}
\label{fig:orb7}
\end{figure*}
%-------------------------------------------------------------------------%

%-------------------------------------------------------------------------%
\begin{figure*}[ht]
\centering
\includegraphics[width=0.9\linewidth]{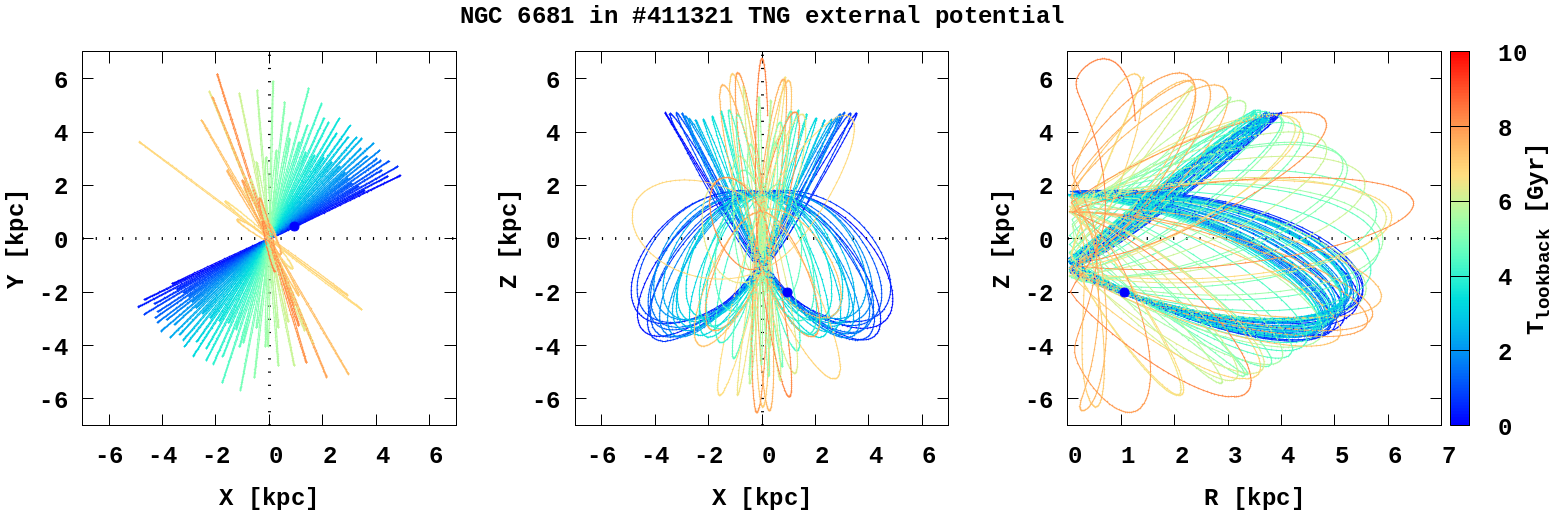}
\includegraphics[width=0.9\linewidth]{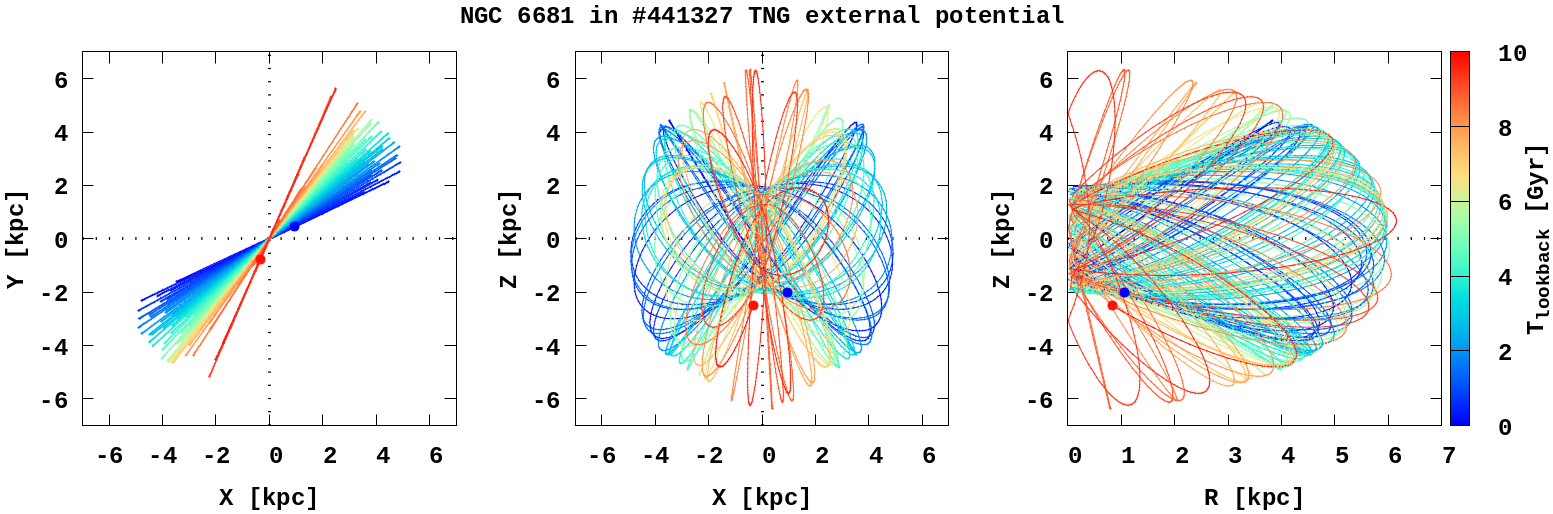}
\includegraphics[width=0.9\linewidth]{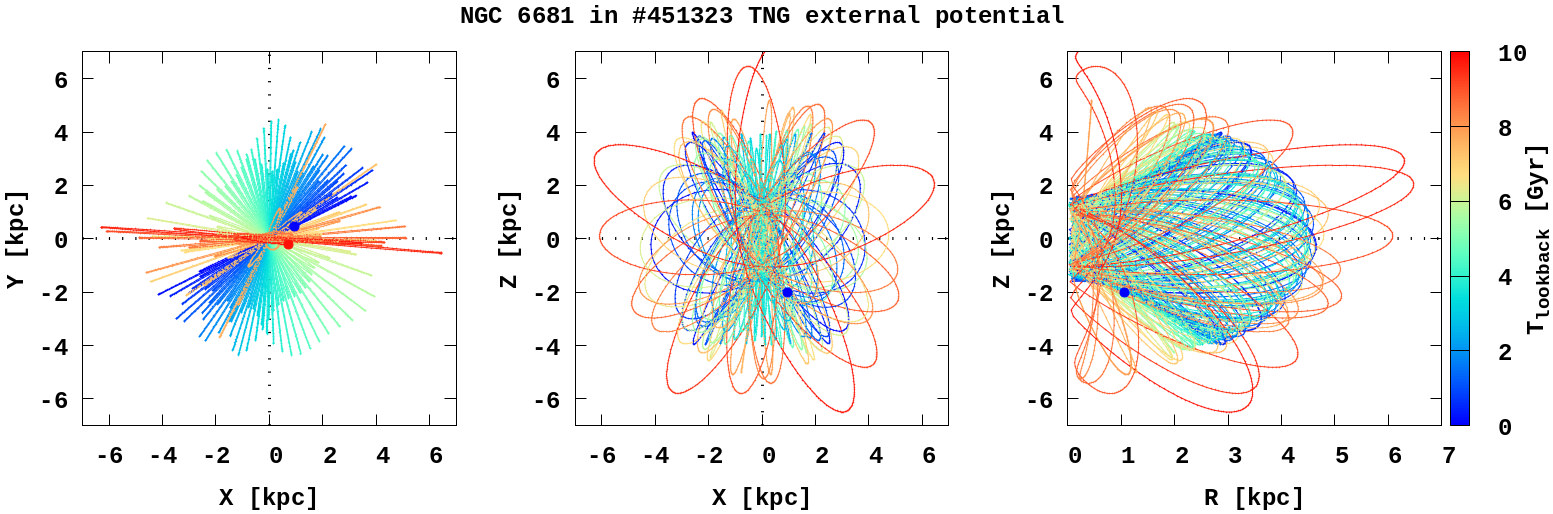}
\includegraphics[width=0.9\linewidth]{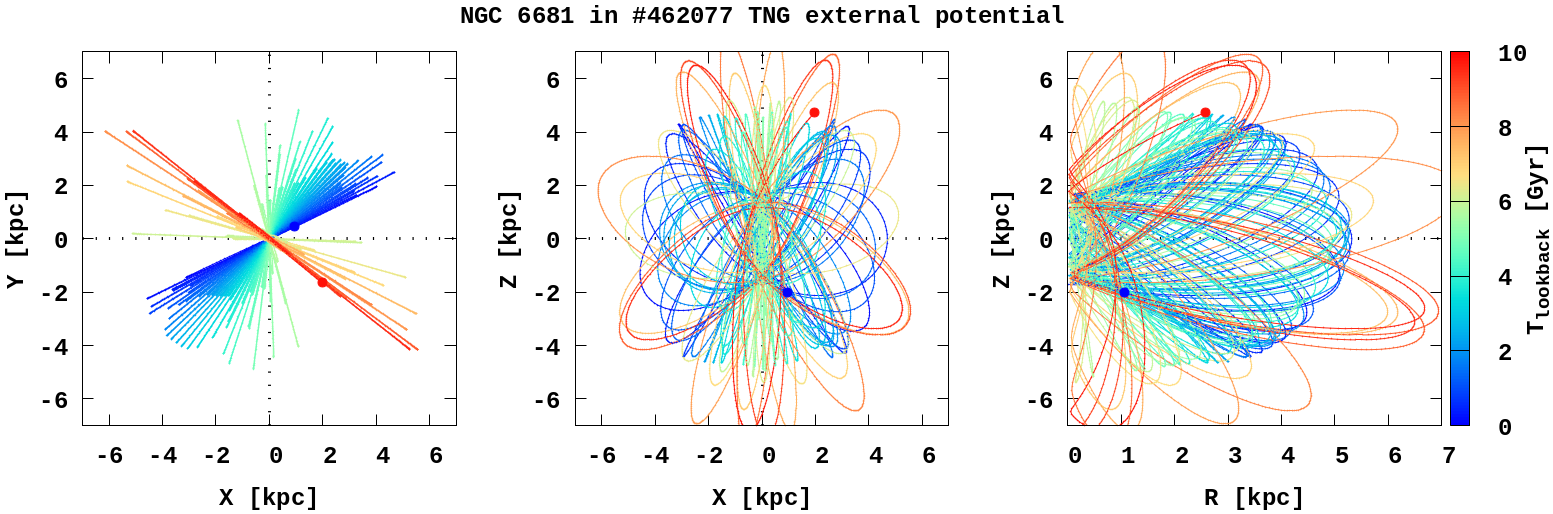}
\caption{The same as in Fig.~\ref{fig:orb1} for NGC 6681 but without {\tt \#411321-m} TNG-TVP.}
\label{fig:orb8}
\end{figure*}
%-------------------------------------------------------------------------%

%-------------------------------------------------------------------------%
\begin{figure*}[ht]
\centering
\includegraphics[width=0.9\linewidth]{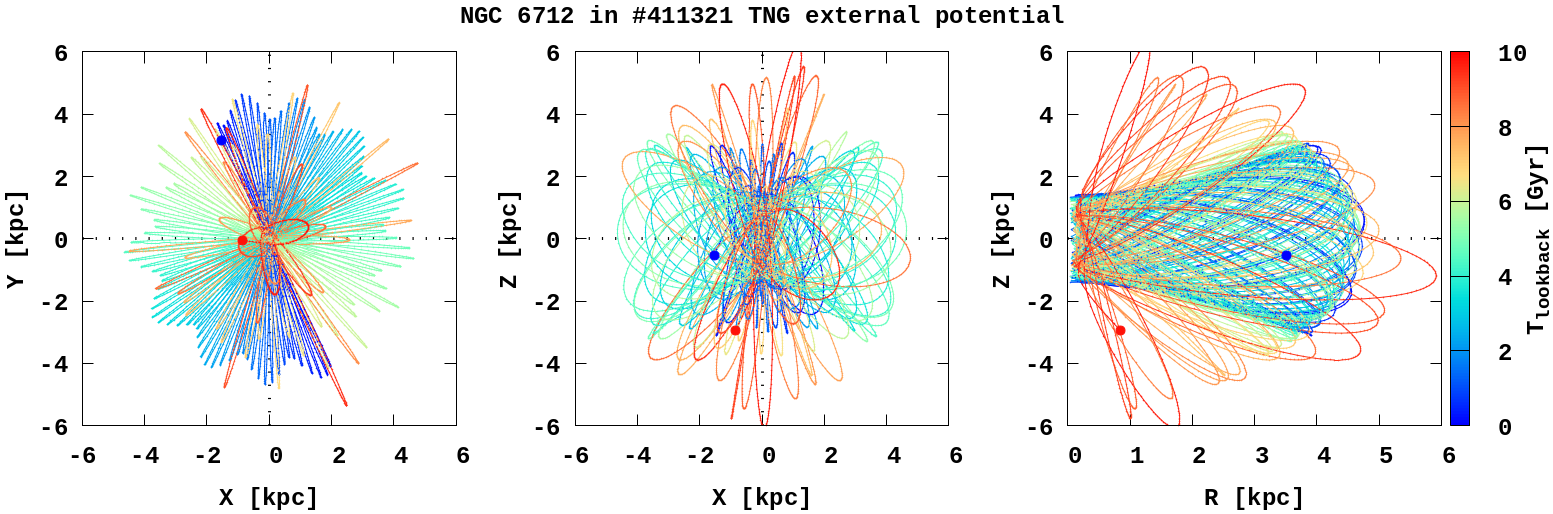}
\includegraphics[width=0.9\linewidth]{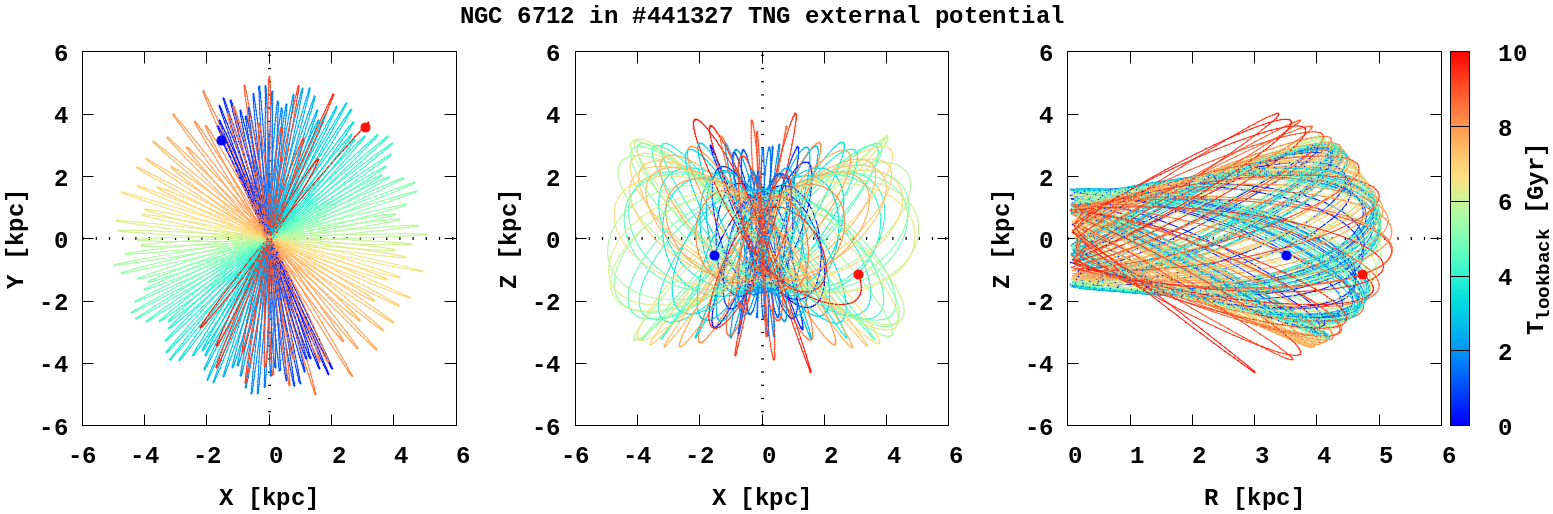}
\includegraphics[width=0.9\linewidth]{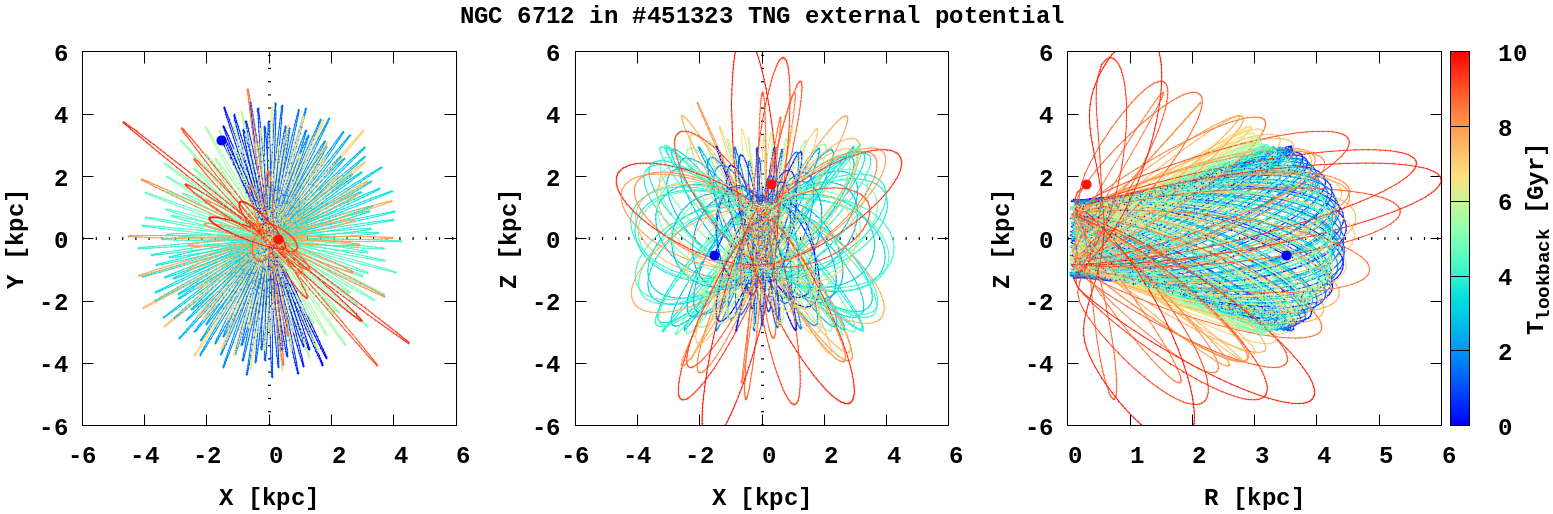}
\includegraphics[width=0.9\linewidth]{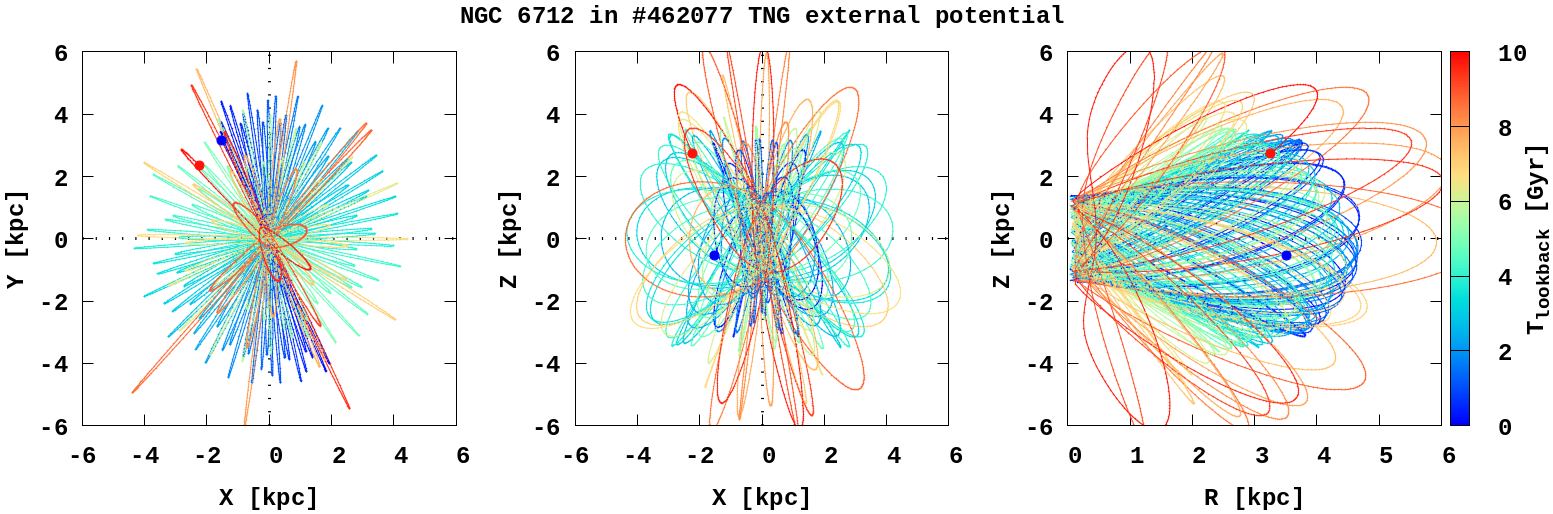}
\caption{The same as in Fig.~\ref{fig:orb1} for NGC 6712 but without {\tt \#411321-m} TNG-TVP.}
\label{fig:orb9}
\end{figure*}
%-------------------------------------------------------------------------%

%-------------------------------------------------------------------------%
\begin{figure*}[ht]
\centering
\includegraphics[width=0.9\linewidth]{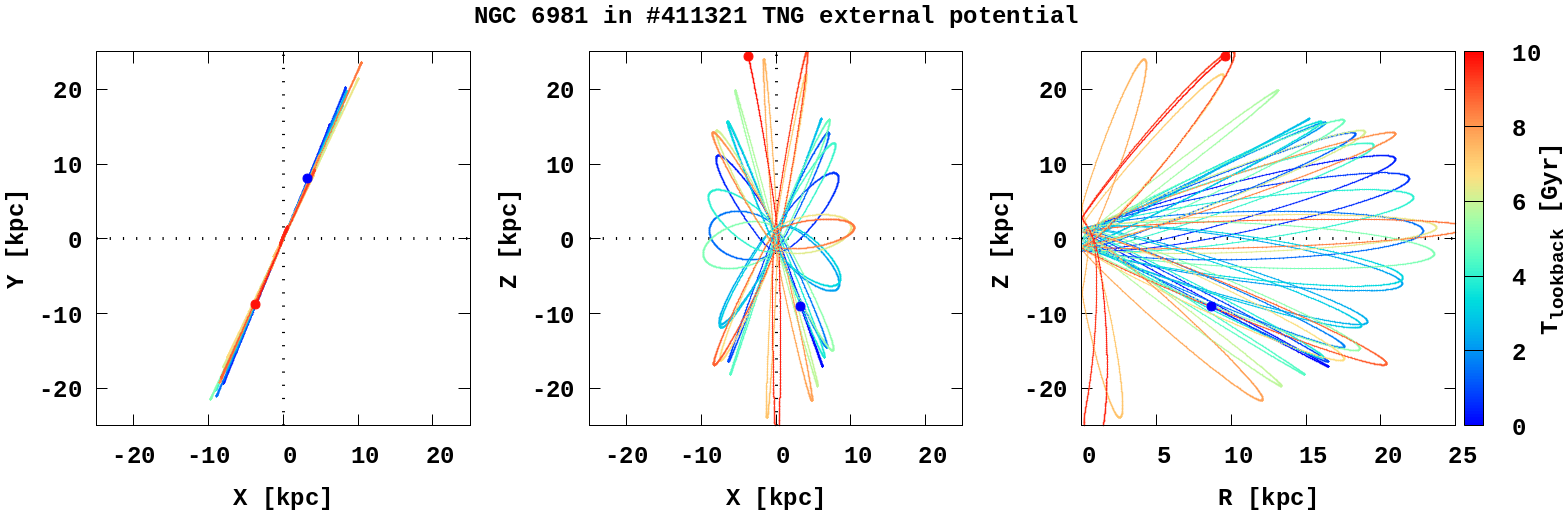}
\includegraphics[width=0.9\linewidth]{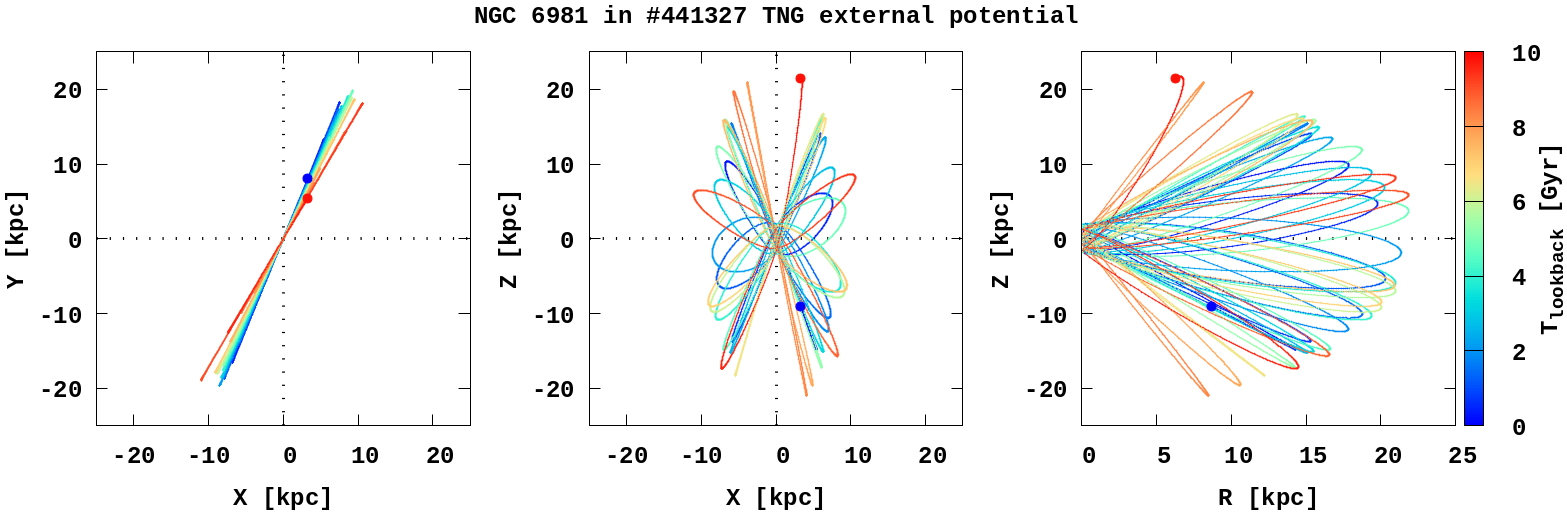}
\includegraphics[width=0.9\linewidth]{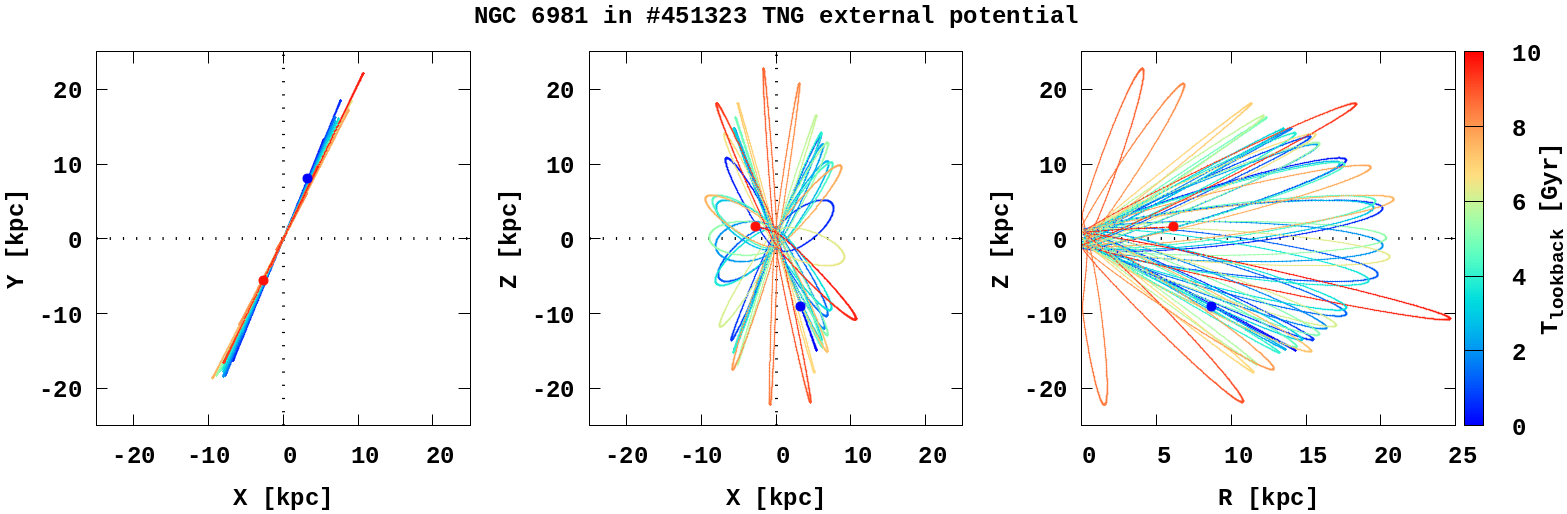}
\includegraphics[width=0.9\linewidth]{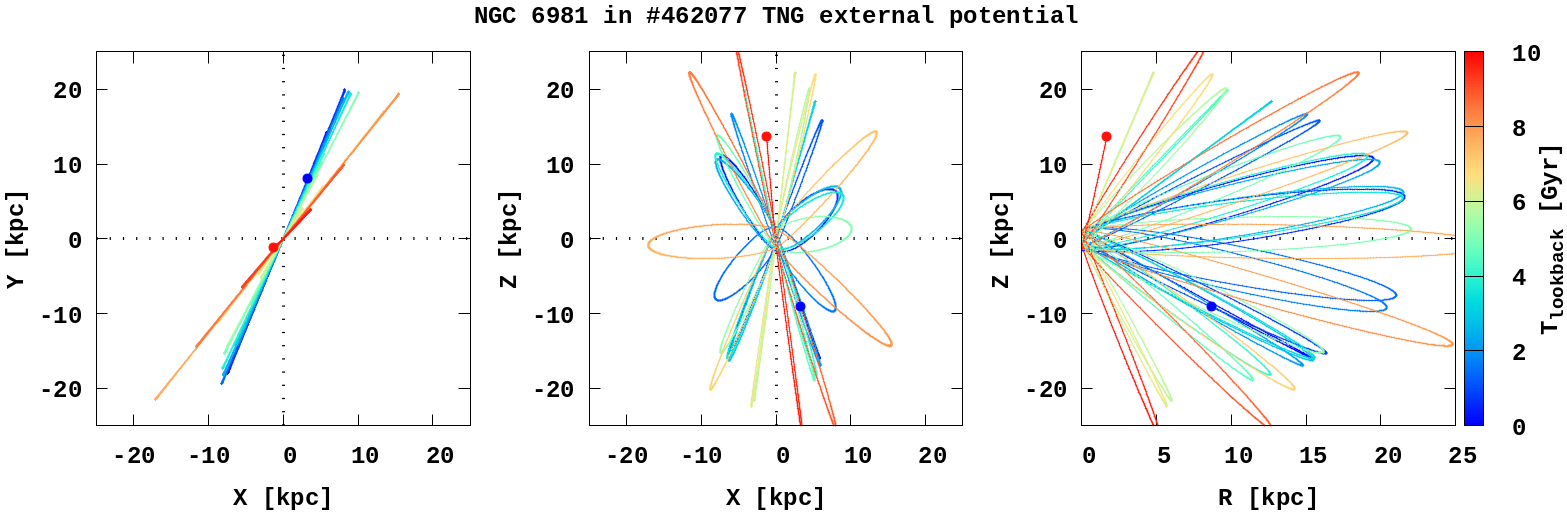}
\caption{The same as in Fig.~\ref{fig:orb1} for NGC 6981 but without {\tt \#411321-m} TNG-TVP.}
\label{fig:orb10}
\end{figure*}
%-------------------------------------------------------------------------%

%%%%%%%%%%%%%%%%%%%%%%%%%%%%%%%%%%%%%%%%%%%%%%%%%%%%%%%%%%%%%%%%%%%%%
\section{Detailed trajectories of Globular Clusters near the Galactic centre.} \label{app:rand-orb-box}
%%%%%%%%%%%%%%%%%%%%%%%%%%%%%%%%%%%%%%%%%%%%%%%%%%%%%%%%%%%%%%%%%%%%%
We present detailed orbital trajectories for GCs that have close passages near the Galactic centre inside the box 100$\times$100 pc in {\tt \#411321} TNG-TVP external potential. The orbital evolution is presented in three planes (X, Y), (X, Z) and (R, Z) (where R is the planar Galactocentric radius). The total time of integration is 10 Gyr lookback time and is represented by a colour line. Grey lines -- taking into account the measurement errors in the orbital shapes.

%-------------------------------------------------------------------------%
\begin{figure*}[htbp!]
\centering
\includegraphics[width=0.96\linewidth]{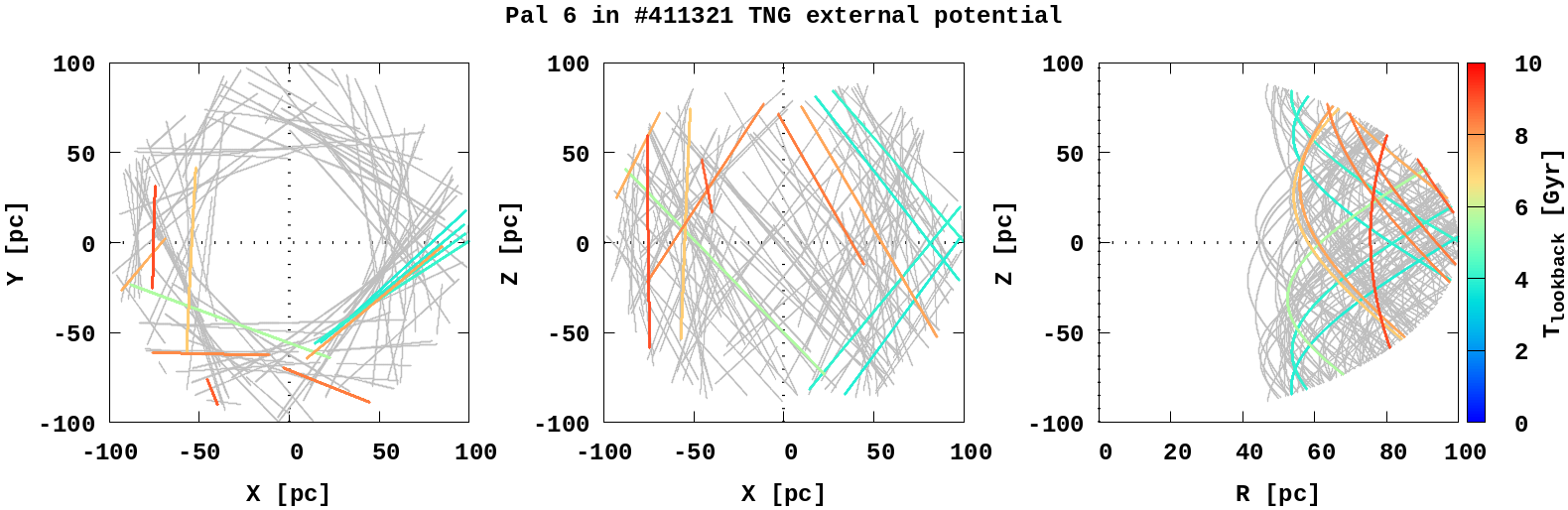}
\includegraphics[width=0.96\linewidth]{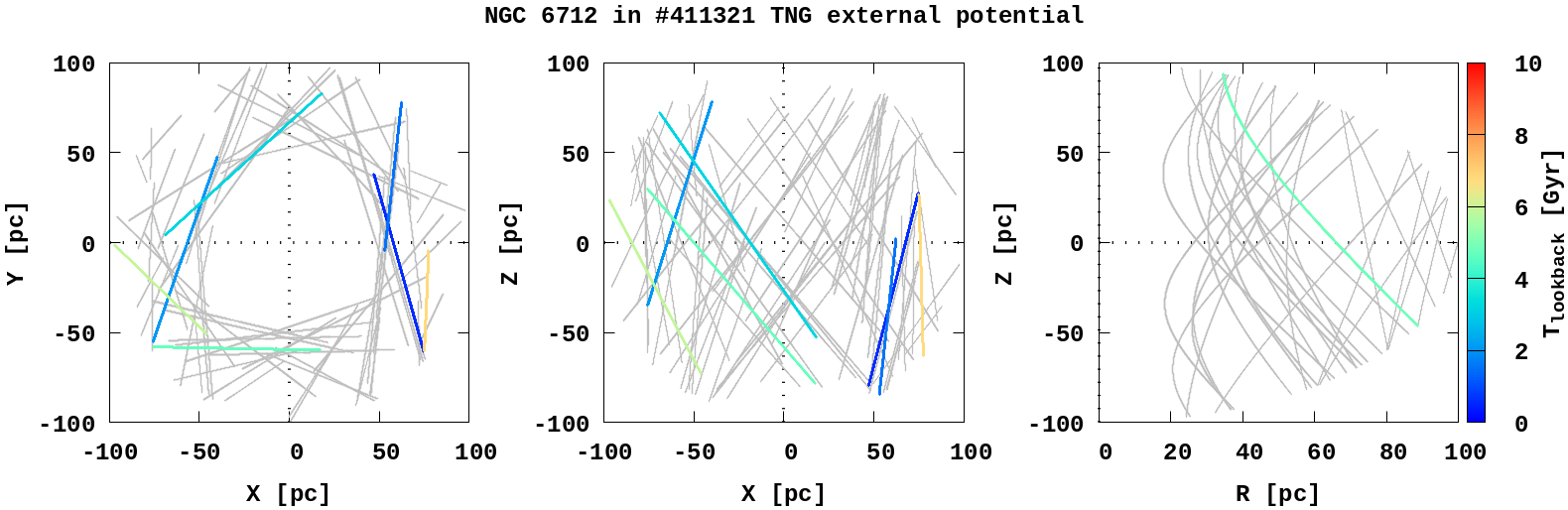}
\includegraphics[width=0.96\linewidth]{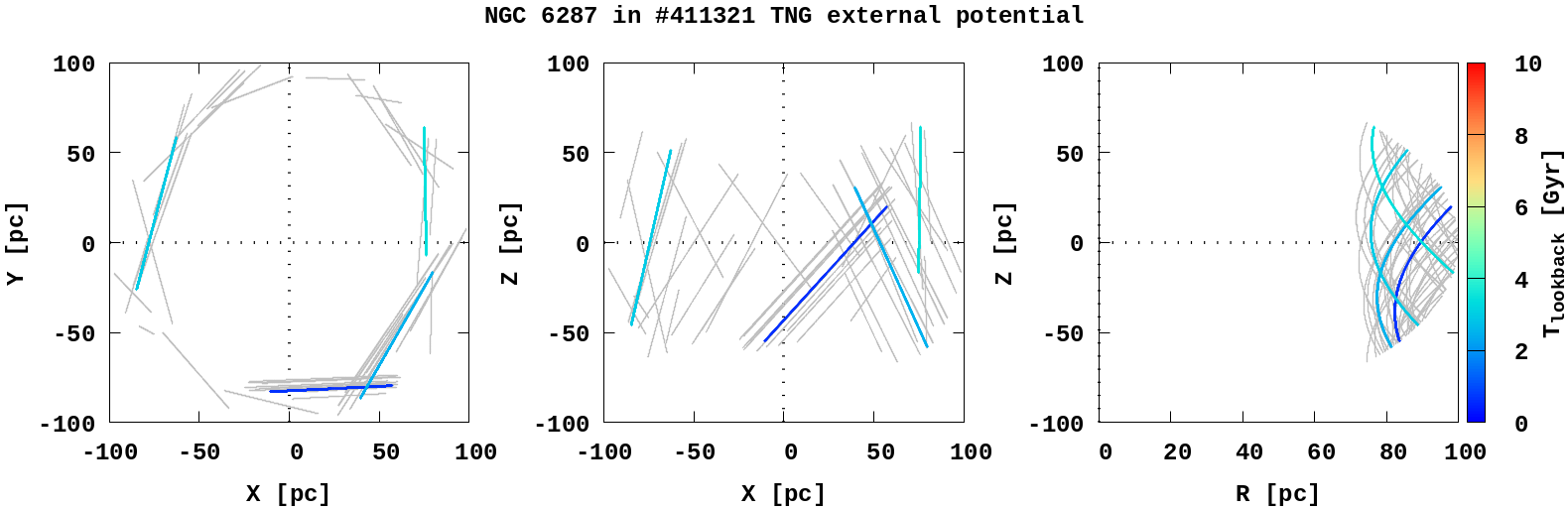}
\includegraphics[width=0.96\linewidth]{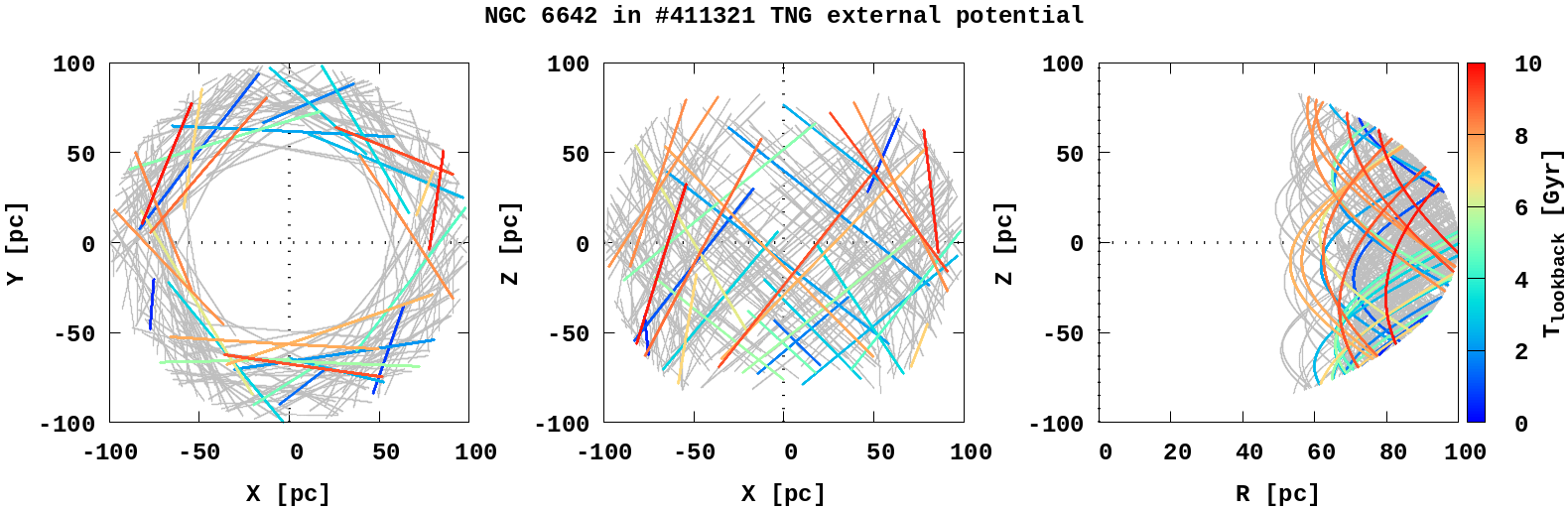}
\caption{The detailed type of orbits that have interactions with the Galactic centre in three planes ($X$, $Y$), ($X$, $Z$) and ($R$, $Z$), where $R$ is the planar Galactocentric radius in box 100$\times$100 pc. The GCs \textit{from top to bottom panels}: Pal 6, NGC 6712, NGC 6287 and NGC 6642 in {\tt \#411321} TNG time-variable potential. The colour line presents the trajectory based on the data from the catalogues. Grey lines -- taking into account the measurement errors.}
\label{fig:rand-orb-box}
\end{figure*}
%-------------------------------------------------------------------------%
%-------------------------------------------------------------------------%
\begin{figure*}[htbp!]
\centering
\includegraphics[width=0.98\linewidth]{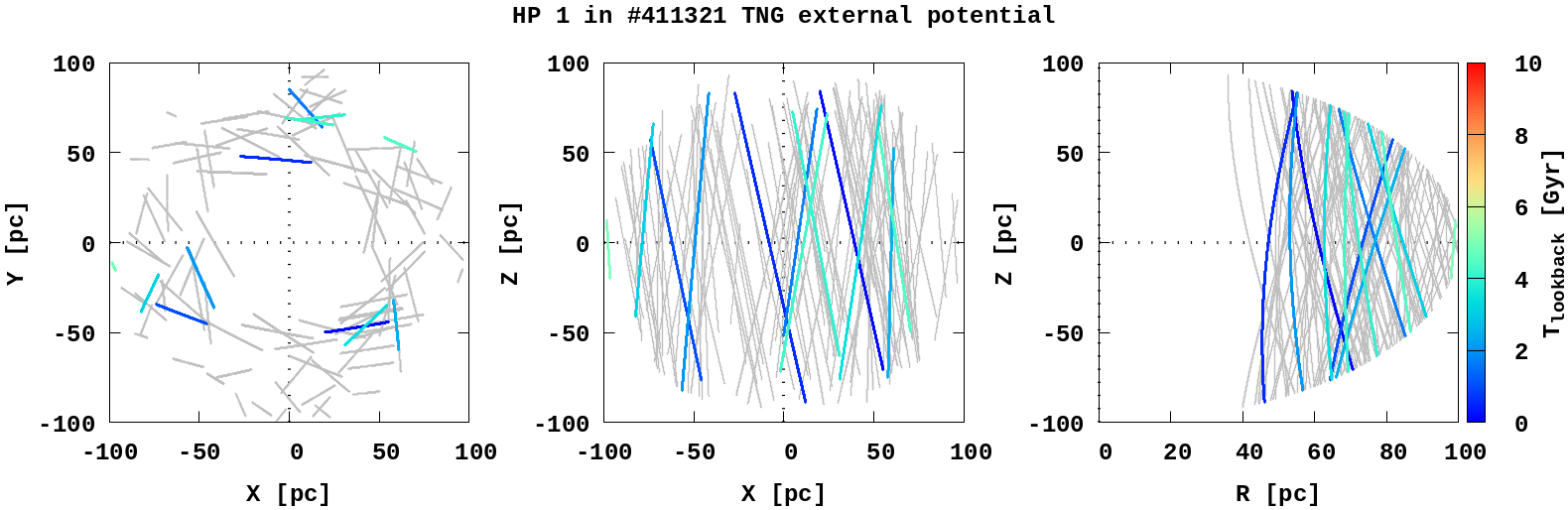}
\includegraphics[width=0.98\linewidth]{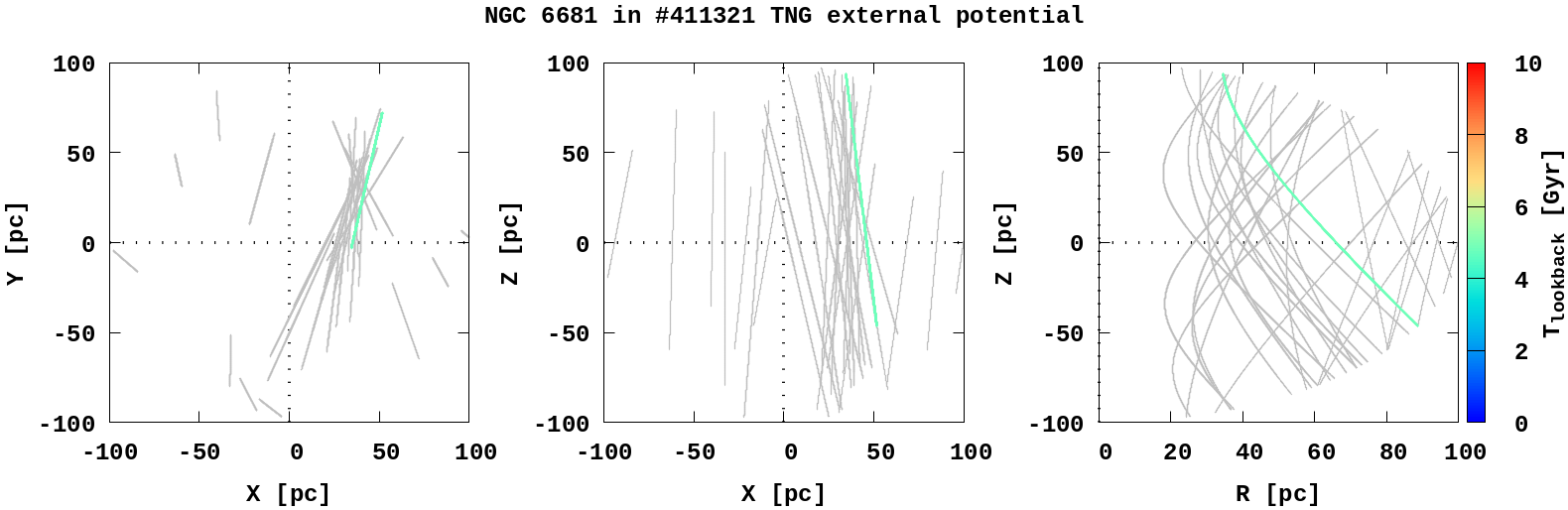}
\includegraphics[width=0.98\linewidth]{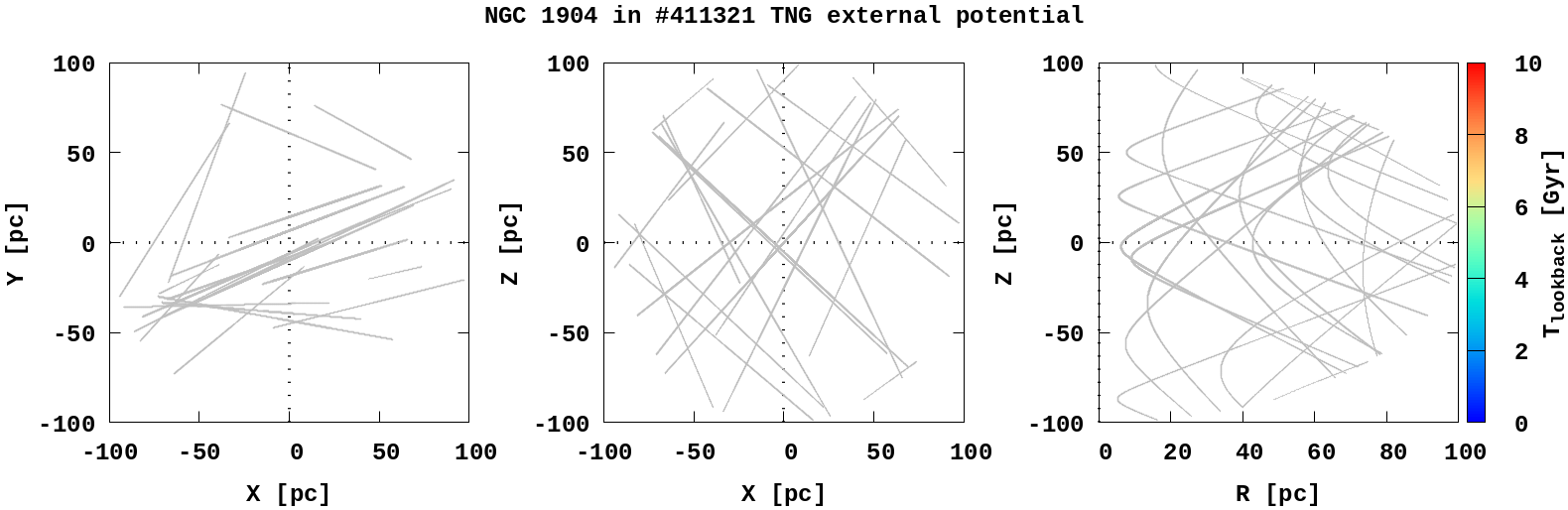}
\caption{As in Fig.~\ref{fig:rand-orb-box} \textit{from top to bottom panels}: HP 1, NGC 6681 and NGC 1904.}
\label{fig:rand-orb-box2}
\end{figure*}
%-------------------------------------------------------------------------%

%%%%%%%%%%%%%%%%%%%%%%%%%%%%%%%%%%%%%%%%%%%%%%%%%%%%%%%%%%%%%%%%%%%%%
\section{Evolution of the Globular Clusters interaction rates.} \label{app:hist-gc-bh}
%%%%%%%%%%%%%%%%%%%%%%%%%%%%%%%%%%%%%%%%%%%%%%%%%%%%%%%%%%%%%%%%%%%%%
We present the interaction rate of GCs with GalC as a function of the relative distance from the centre in different time intervals (colour dashed lines) for TNG-TVPs: {\tt \#441327}, {\tt \#451323}, {\tt \#462077}. Each time interval has a bin of 1~Gyr. The solid black line is a global close passages rate for a whole time interval 10~Gyr. The grey solid lines are a result of 1000 simulations with different initial data random realisations. Also, we demonstrate the contribution of individual GCs to global collision rate at different time ranges for the listed above TNG-TVPs.

%-------------------------------------------------------------------------%
\begin{figure*}[ht]
\centering
\includegraphics[width=0.43\linewidth]{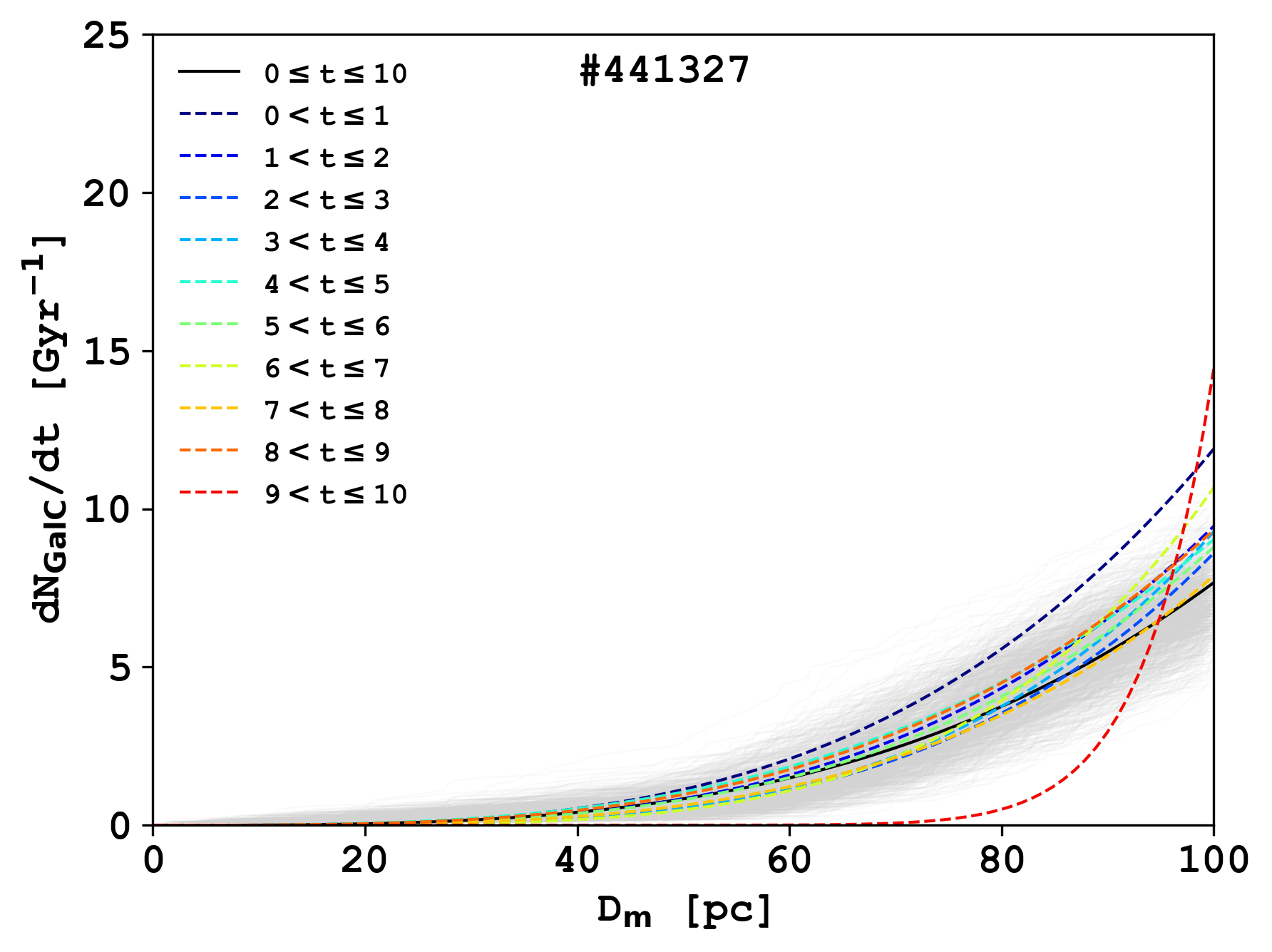}
\includegraphics[width=0.46\linewidth]{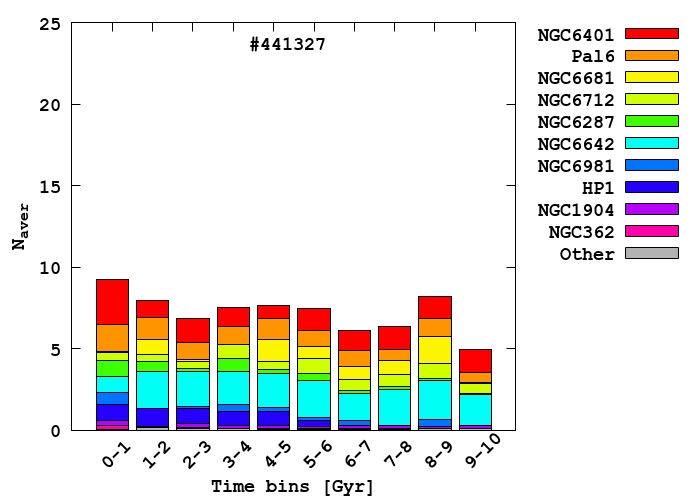}\linebreak
\includegraphics[width=0.43\linewidth]{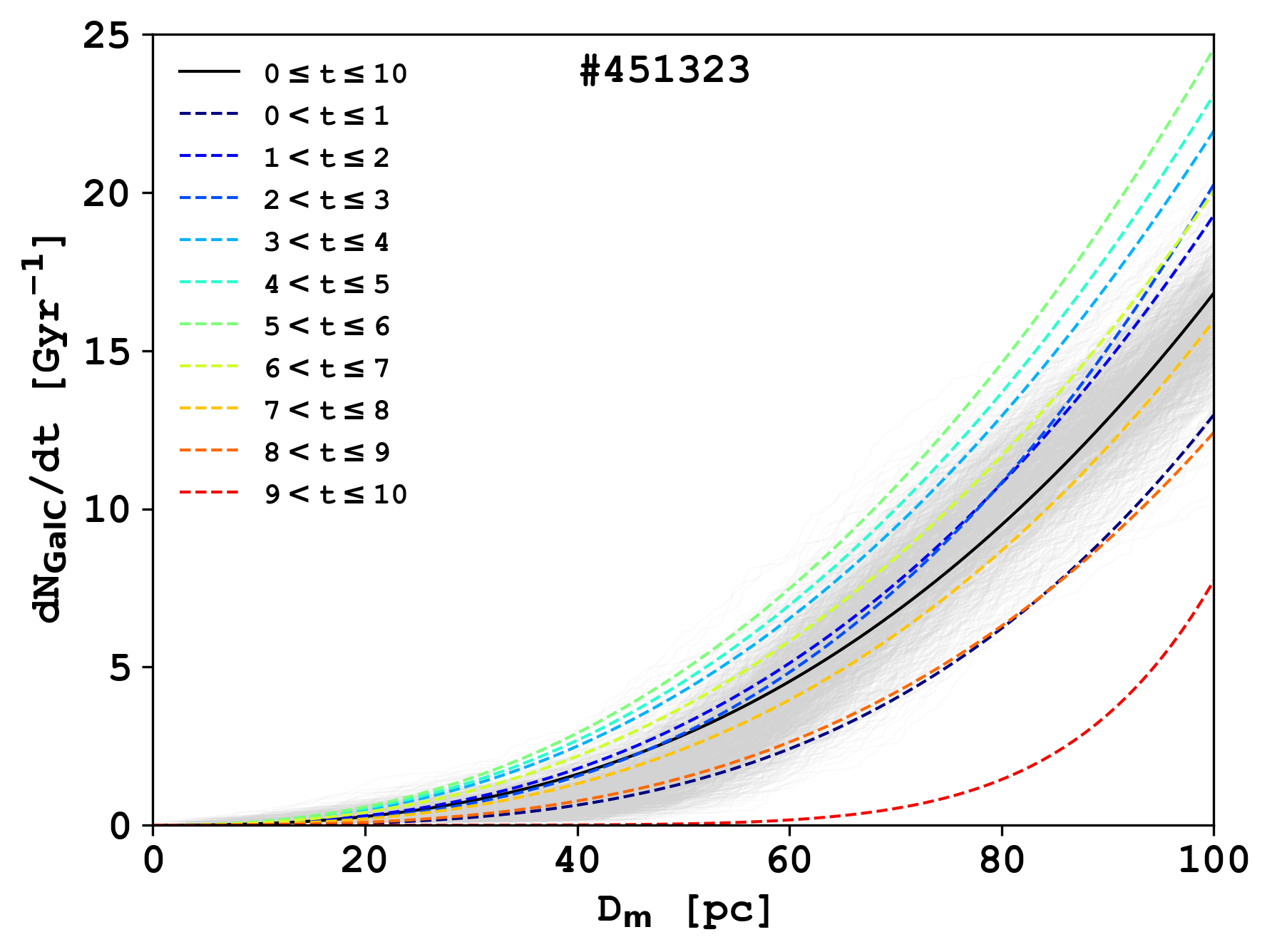}
\includegraphics[width=0.46\linewidth]{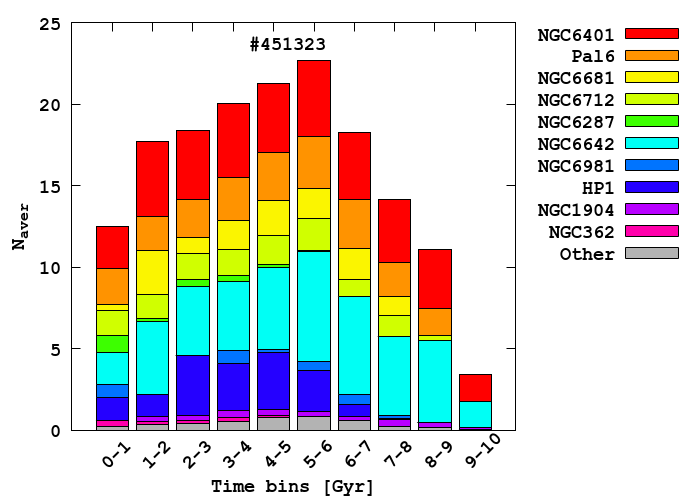}\linebreak
\includegraphics[width=0.43\linewidth]{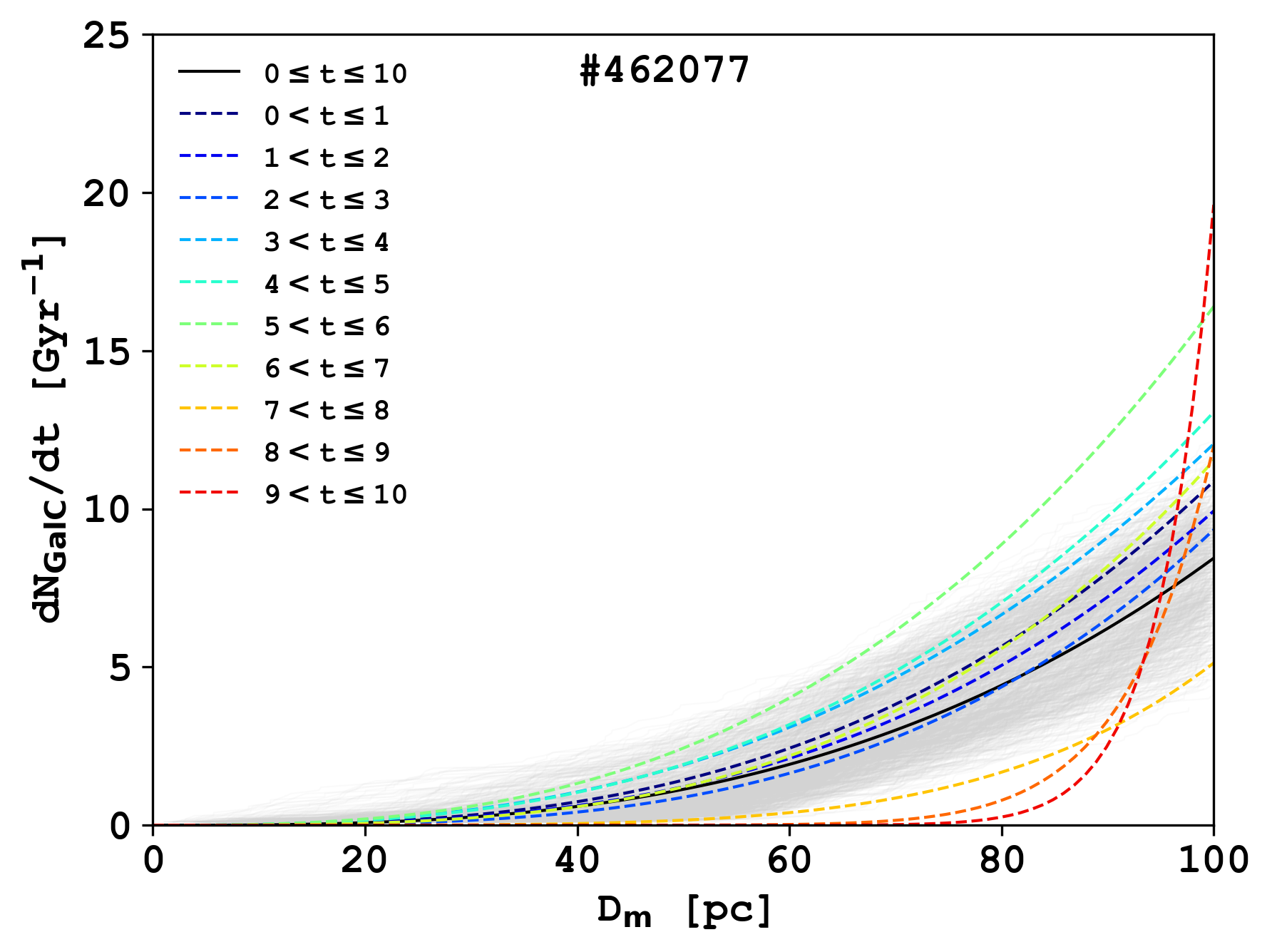}
\includegraphics[width=0.46\linewidth]{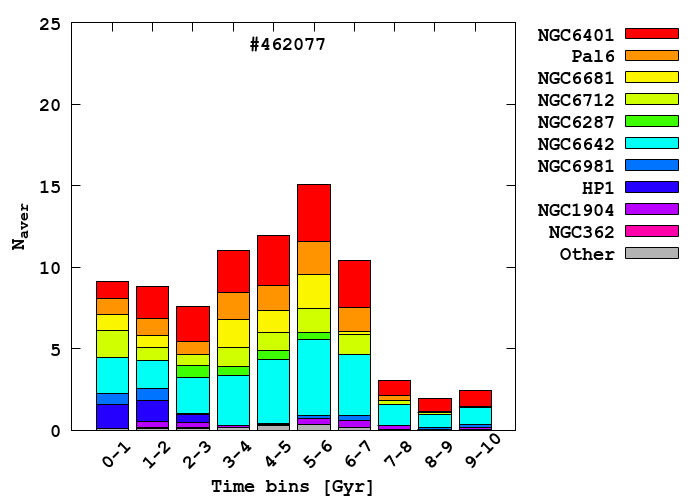}
\caption{\textit{Left:} Interaction rate of GCs with GalC as a function of the relative distance in different time intervals (colour dashed lines) for TNG-TVPs \textit{from top to bottom}: {\tt \#441327}, {\tt \#451323}, {\tt \#462077}. Each time interval has a length of 1~Gyr. The solid black line is a global close passages rate for a whole time interval 10~Gyr. The grey solid lines is a results for 1000 simulations with different random realisations. Right:} Contribution of individual GCs into global collision rate at different time ranges for listed above TNG-TVPs.
\label{fig:hist-gc-bh}
\end{figure*}
%-------------------------------------------------------------------------%

\clearpage

\end{appendix}

%%%%%%%%%%%%%%%%%%%%%%%%%%%%%%%%%%%%%%%%%%%%%%%%%%%%%%%%%%%%%%%%%%%%%
\end{document}